\documentclass[journal]{IEEEtran}
\ifCLASSINFOpdf
\else
\fi
\usepackage{siunitx}
\usepackage{amsthm}
\usepackage{amssymb}
\usepackage{bm}
\usepackage[linesnumbered,ruled]{algorithm2e}
\input  alphabet.tex
\DeclareMathOperator\arccosh{arccosh}
\usepackage{graphicx}
\graphicspath{{figures/}}
\usepackage{bbm}
\usepackage{xcolor}

\usepackage{url}

\begin{document}

\title{A unified joint reconstruction approach in structured illumination microscopy using unknown speckle patterns}

\author{Penghuan~Liu \IEEEmembership{}

\thanks{ Penghuan Liu 
    got the PhD degree in \'Ecole Centrale de Nantes at the Laboratoire des Sciences du Num\'erique de Nantes (LS2N),  
    France. He is now an independent researcher in Foliu village, 061802 Wuqiao county, Hebei province, China.
    E-mail: liu.penghuan@outlook.com}
}

\maketitle

\begin{abstract}

The structured illumination microscopy using unknown speckle patterns (blind-speckleSIM) has shown the capacity to surpass the Abbe’s diffraction barrier \cite{mudry2012structured}, giving the possibility to design cheap and versatile SIM devices.  However, the state-of-the-art joint reconstruction methods in blind-speckleSIM has a relatively low contrast in super-resolution part in comparison to conventional SIM \cite{mudry2012structured}\cite{labouesse2017joint} and the hyperparameter in this model is not easy to tune. In this paper, a unified joint reconstruction approach is proposed with the hyperparameter proportional to the noise level. The performance of different regularizations could be evaluated under the same model. Numerical simulations show that the $\ell_{2,1}$ regularizer gives more satisfactory results than the previously used regularizer terms in joint reconstruction approach. Moreover, the degradation entailed by out-of-focus light in conventional SIM could be solved easily in blind-speckleSIM setup. 

\end{abstract}


\IEEEpeerreviewmaketitle

\section{Introduction}

\subsection{Super-resolution fluorescence microscopy}
\IEEEPARstart{T}{he} conventional optical microscopy is a diffraction limited system whose spatial resolution is limited by diffraction effect (often modeled as a low-pass filter in Fourier domain).  In the past twenty years, numerous super-resolution techniques has been proposed in fluorescence microscopy to surpass this limitation, enabling resolution about 10 to \SI{100}{\nano\metre}. Stochastic optical reconstruction microscopy (STORM) \cite{rust2006sub} or photo activated localization microscopy (PALM) \cite{hess2006ultra} achieve super-resolution by sequential activation of photo switchable fluorophores. In each imaging cycle, only a fraction of fluorophores are activated at any given moment, such that the position of each fluorophore can be determined with high precision. To build the final super-resolution image, thousands of exposures are required and the fluorophore activation and deactivation process are quite time consuming. Stimulated emission depletion fluorescence scanning microscopy (STED) creates super-resolution image by reducing the diffraction spot with the help of an additional STED pulse. The  scanning process strongly limits the data acquisition time for large size object, even with parallelizing STED \cite{bergermann20152000}.

Structured illumination microscopy (SIM) retrieves super-resolution by illuminating the object $\rho$ with a few structured patterns $I_m$. In the linear regime, the measured dataset $\{ y_m\}_{m=1}^M$ has relation with the sample $\rho$ by \cite{Goodman05}:
\begin{equation}
\label{eq.SIM}
y_m = h\ast (\rho I_m) + \epsilon_m
\end{equation}
where $h$ is the point spread function (PSF) of the system and $\ast$ denotes convolution operator. The production of the sample $\rho$ with structured patterns $I_m$ transfers otherwise unobservable high-frequency information about the sample into a lower-frequency region, thus pass through the optical system \cite{Gustafsson2000}. As a wide-field imaging technique, SIM acquisitions are much faster, and super-resolution imaging in living samples has been demonstrated with SIM \cite{Kner09}. 

In standard SIM, the object is illuminated by a set of harmonic patterns with designed spatial frequencies and phases. However, generating a perfect known harmonic illumination is a difficult task and strong distortions of the light grid can be induced
within the investigated volume by the sample \cite{jost2015optical}. The blurring in illumination will reduce the SR capacity in SIM  and introduce strong artifacts\cite{Ayuk13}. 

One way to tackle this problem is using unknown speckle patterns as a substitute for the harmonic illumination in SIM (blind-speckleSIM).  Compared with harmonic illumination, the speckle patterns are easier to generate while the super-resolution is still attainable\cite{mudry2012structured}. Several reconstruction methods has been proposed in blind-speckleSIM, as shown in \cite{mudry2012structured}\cite{labouesse2017joint}\cite{min2013fluorescent}\cite{yeh2017structured}\cite{idier2017super}. In reference \cite{idier2017super}, a marginal approach is reported and the super-resolution capacity of blind-speckleSIM has been demonstrated as good as classic SIM by taking advantage of the second-order statistics of the data in asymptotic condition when the Fourier support of speckle is identified with the OTF of system. However, the computational complexity of the methods presented in \cite{idier2017super} is $\Oc(N^3)$, which is too high for realistic size images. Possible solutions to reduce the computational burden in marginal approach are out of the scope in this paper. On the other hand, the methods shown in \cite{mudry2012structured}\cite{labouesse2017joint}\cite{min2013fluorescent} share similar framework, i.e.\ they are trying to reconstruct the object by minimizing a data fidelity term plus a regularizer:
\begin{equation}
\label{eq.modelUnconstrained}
\arg \min_\Qv \frac{1}{2}\lVert\Yv-\Hv\Qv \rVert_F^2 + \mu \Phi(\Qv)
\end{equation}
where $\lVert \cdot \rVert_F$ denotes Frobenius norm, $\Qv= [\qb_1,\cdots,\qb_M] = [\rhob\circ \Iv_1, \cdots,\rhob\circ\Iv_M]$ and $\Phi(\Qv)$ is the regularizer term that enforces the priori knowledge of $\Qv$, such as positivity constraint \cite{mudry2012structured}, positivity and sparsity constraint \cite{labouesse2017joint}, or joint sparsity constraint \cite{min2013fluorescent}. In this model, the super-resolution is induced by the regularizer term, while the data fidelity term gives no super-resolution information if only the first order statistics of speckle is used \cite{labouesse2017joint}.

\subsection{Contribution of this paper}
In this paper, the super-resolution information is retrieved with $\ell_{p,q}$ norm regularizer to enforce joint sparsity of matrix $\Qv$, with $p\geq 1, 0<q\leq 1$ (for the definition of $\ell_{p,q}$ norm, please see in section \ref{sec.lpq}). The mathematical analysis indicates that the joint sparsity of $\Qv$ is equivalent with the sparsity of object $\rhob$ when $p=1$ or $2$. Moreover, Other prior information on the object could be easily incorporated into the model without major changes of the associated primal-dual algorithm, such as the positivity constraint and total variation (TV) regularization. 

To tackle the hyperparameter tuning problem in \cite{mudry2012structured} \cite{min2013fluorescent}\cite{labouesse2017joint}, the unconstrained minimization model \eqref{eq.modelUnconstrained} is transformed to a constrained  one:
\begin{equation}
\label{eq.modelConstrained}
\arg \min_\Qv  \;\Phi(\Qv)  \qquad \text{s.t.} \quad \lVert\Yv-\Hv\Qv \rVert_F \leq \xi
\end{equation}
Although model \eqref{eq.modelConstrained} and \eqref{eq.modelUnconstrained} are equivalent in the sense that for any $\xi \geq 0$, the solution of \eqref{eq.modelConstrained} is either the null vector, or it is a minimizer of \eqref{eq.modelUnconstrained}, for some $\mu >0$ (see \cite{rockafellar2015convex}), $\xi$ in \eqref{eq.modelConstrained} is much easier to set because it has a clear meaning which is proportional to the standard variance of the noise, while tuning the hyperparameter $\mu$ in \eqref{eq.modelUnconstrained} is an non-trivial task (please note here that the hyperparameter $\mu$ does not influence the performance of the positivity constraint presented in \cite{mudry2012structured}). The idea to solve the constrained problem \eqref{eq.modelConstrained} is firstly transforming it to an unconstrained one with the help of indicator function, and then solve the unconstrained optimization problem by the primal-dual splitting method  \cite{condat2013primal}.

The nonmodulated background signal from the out-of-focus light will degrade the image quality in conventional SIM dramatically \cite{SIMprotocol2017nature}. Instead of modeling the background with a smooth function \cite{Orieux2012Bayesian} or carrying on background subtraction heuristically \cite{Lal2016Structured}, a much easier and natural strategy in blind-speckleSIM setup is to estimate the object by Eq. \eqref{eq.variance_qm}, taking advantage of the information that speckle patterns are second-order stationary random process. 
Numerical simulations show that this estimator could remove the fixed background signal.

The organization of this paper is as follows. In Section~\ref{sec.lpq}, the constrained $\ell_{p,q}$ reconstruction model is presented together with the primal-dual algorithm to solve it. The simulation results of the proposed method and reconstructions from experimental data are presented in Section~\ref{sec.simu} and conclusions are drawn in Section~\ref{sec.conclusion}.

\section{Problem Formulation}\label{sec.lpq}
The discretized form of \eqref{eq.SIM}, where each 2D quantity is displayed by a column vector is:
\begin{equation}
\label{eq.obs}
\yb_m = \Hv (\rhob \circ \Iv_m) + \epsilonb_m
\end{equation}
in which $\yb_m \in \mathbb{R}^L$ is the recorded raw image, $\Hv \in \mathbb{R}^{L\times N}$ is the discrete convolution matrix built from the discretized PSF and $\circ$ denotes the element-wise product.  $\rhob \in \mathbb{R}^N$ stands for the discretized fluorescence density, $\Iv_m \in \mathbb{R}^N$ is the $m$-th illumination with homogeneous intensity mean $I_0$, and $\epsilon_m \in \mathbb{R}^L$ is the noise in imaging process. By introducing an auxiliary variable $\qb_m = \rhob \circ \Iv_m$, the image formation model \eqref{eq.obs} can be written  in matrix form as:

\begin{equation}
\Yv = \Hv \Qv + \epsilonb
\end{equation}
with $\Yv=[\yb_1,\cdots,\yb_M] \in \mathbb{R}^{L\times M}$ and $\Qv=[\qb_1,\cdots,\qb_M] \in \mathbb{R}^{N\times M}$.  Now our task is to estimate matrix $\Qv$ from the measurement matrix $\Yv$. Once $\Qv$ is obtained,  $\rhob$ can be retrieved either by the mean of $\qb_m$:
\begin{equation}
\label{eq.rhoQmean}
\rhob = \frac{1}{I_0} \bar{\qb}
\end{equation}
or by their standard deviation:

\begin{equation}
\label{eq.variance_qm}
\rhob \propto \sqrt{\frac{1}{M}\sum_m\big(\qb_m-\bar{\qb}\big)^2}
\end{equation} 
where $\bar{\qb} = \frac{1}{M} \sum_{m=1}^M \qb_m$. The relation \eqref{eq.variance_qm} holds because speckle patterns are second-order stationary. The differences between estimators \eqref{eq.rhoQmean} and \eqref{eq.variance_qm} are explored in Section~\ref{sec.simu}.

To introduce the prior information of $\Qv$, the following regularizer term, i.e.\ $q$-th power of $\lVert \Qv \rVert_{pq}$  
plus total variation of $\rhob$, is chosen: 

\begin{equation}
\begin{aligned}
\arg \min_{\Qv} \quad  \lVert \Qv \rVert_{pq}^q + \mu  \lVert \rhob \rVert _{TV}\\
\text{subject to}\qquad \big\lVert \Hv\Qv - \Yv \big\rVert_F \leq \xi 
\end{aligned}
\label{eq.JSP2}
\end{equation}
where  $\lVert \Qv \rVert_{pq}$ denotes the $\ell_{p,q}$ norm of matrix $\Qv$:
\begin{equation}
\lVert \Qv \rVert_{pq} = \Big( \sum_{n=1}^N \lVert\qb^n\rVert_p^q \Big)^{1/q}, \quad p \geq 1,\; 0 \leq q \leq 1
\end{equation}
with $\qb^n$ the $n$-th row of $\Qv$. The TV norm is introduced as a normal regularizer for the “natural” images to enforce the piece-wise smooth property of the object. Given a $N_1\times N_2$ object $\rho$, the \emph{isotropic} TV is defined as:
\begin{equation}
\begin{aligned}
&TV(\rho) = \sum_{n_1 =1}^{N_1} \sum_{n_2 =1}^{N_2} \\
&\sqrt{(\rho[n_1+1,n_2]-\rho[n_1,n_2])^2+(\rho[n_1,n_2+1]-\rho[n_1,n_2])^2}
\end{aligned}
\end{equation}

\subsection{Relationship between joint sparsity and prior information of object $\rho$} 
In this section, the relation between the joint sparsity of $\Qv$ and the prior of object $\rho$ is analyzed. For the $n$-th row of matrix $\Qv$, its $\ell_p$ norm is given by:
\begin{equation}
\lVert\qb^n\rVert_p = \Big[\lvert \rho_nI_{1n} \rvert^p + \cdots +\lvert \rho_n I_{Mn} \rvert^p \Big]^{1/p}
\end{equation}
when $p=1$,
\begin{equation}
\label{eq.jointSparsityrhoSparsity1}
\lVert\qb^n\rVert_p = \rho_n (\lvert I_{1n}\rvert + \cdots + \lvert I_{Mn}\rvert) = MI_0\rho_n
\end{equation}
when $p=2$
\begin{equation}
\label{eq.jointSparsityrhoSparsity2}
\lVert\qb^n\rVert_p = \rho_n \big(I_{1n}^2 + \cdots + I_{Mn}^2\big)^{1/2} = \sqrt{ M k}\rho_n
\end{equation}
where $k$ is a constant for the fully developed speckle patterns \cite[Chapter~7]{goodman2015statistical}. So the sparsity of $\lVert\qb^n\rVert_p$ is equivalent with the sparsity of $\rho_n$  when $p=1$ or $2$. 

\subsection{Expressing TV norm of $\rhob$ as a function of $\Qv$}

Since the constraint in \eqref{eq.JSP2} is on variable $\Qv$ rather directly on $\rhob$, we need to transform $\lVert \rhob \rVert_{TV}$ to a function of $\Qv$. Eq. \eqref{eq.rhoQmean} can be written in matrix form as:
\begin{equation}
\rhob = \Ac \bm{\mathfrak{q}}
\end{equation}
with 
\begin{equation}
\begin{aligned}
\Ac = \left[
\begin{array}{ccc}
\frac{1}{MI_0}\mathbbm{1}_N,\cdots,\frac{1}{MI_0}\mathbbm{1}_N
\end{array}
\right], \; \bm{\mathfrak{q}} 
= \left[
\begin{array}{c}
\qb_1  \\
\qb_2\\
\vdots  \\
\qb_M \\
\end{array}
\right]
\end{aligned}
\end{equation}
where $\mathbbm{1}_N$ is the $N\times N$ identity matrix. Now we have:
\begin{equation}
\lVert \rhob \rVert_{TV} = \lVert\Ac\bm{\mathfrak{q}}\lVert_{TV}
\end{equation}
\subsection{The equivalently unconstrained form}
To simplify the notation, let us define:
\begin{equation}
\begin{aligned}
\Hc = \mathbbm{1}_M \otimes \Hv = \left[
\begin{array}{cccc}
\Hv & & & \\
& \Hv& &\\
& & \ddots & \\
& & & \Hv \\
\end{array}
\right], \;
 \bm{\mathfrak{y}} 
= \left[
\begin{array}{c}
\yb_1  \\
\yb_2\\
\vdots  \\
\yb_M \\
\end{array}
\right]
\end{aligned}
\end{equation}
where $\mathbbm{1}_M \in \mathbb{R}^{M\times M}$ is the identity matrix and $\otimes$ denotes Kronecker product. Let us partition $\bm{\mathfrak{q}}$ into $N$ groups $\{ \bm{\mathfrak{q}}_{\Gc_1},\cdots,\bm{\mathfrak{q}}_{\Gc_N} \}$ with $\bm{\mathfrak{q}}_{\Gc_n} \in \mathbb{R}^M$ corresponds to the $n$-th row $\qb^n$ in matrix $\Qv$. Then the $\ell_{p,q}$ norm of matrix $\Qv$ is equivalent with the group $\ell_{\Gc,p,q}$ norm of vector $\bm{\mathfrak{q}}$:
\begin{equation}
\lVert \bm{\mathfrak{q}} \rVert_{\Gc pq} = \Big( \sum_{n=1}^N \lVert \bm{\mathfrak{q}}_{\Gc_n} \rVert_p^q \Big)^{1/q} 
\label{eq.groupSparsity1}
\end{equation}

According to the definition of TV norm and $\ell_{\Gc,p,q}$ norm, we can easily verify that the TV norm of $\rhob$ can be seen as the $\ell_{\Gc,2,1}$ norm of $\Cv\rhob$, with $\Cv = (\Cv_1;\Cv_2) \in \mathbb{R}^{2N \times N}$ and $\Cv_1, \Cv_2$ the first-order horizontal and vertical finite difference operators, and the $n$-th group is given by $(\Cv\rhob)_{\Gc_n} = [(\Cv_1\rhob)_n; (\Cv_2\rhob)_n] \in \mathbb{R}^2$. So now 
\begin{equation}
\lVert \rhob \rVert_{TV} = \lVert\Cv\Ac\bm{\mathfrak{q}}\lVert_{\Gc 21}
\end{equation}

Let us define $\Dc = \Cv\Ac$ and now the problem \eqref{eq.JSP2} can be expressed in vector form:
\begin{equation}
  \arg \min_{\bm{\mathfrak{q}}} \;    \lVert \bm{\mathfrak{q}} \rVert_{\Gc pq}^q + \mu \lVert \Dc \bm{\mathfrak{q}} \rVert_{\Gc 21} \quad \text{s.t.}\;\;  \lVert \Hc\bm{\mathfrak{q}} - \bm{\mathfrak{y}} \rVert \leq \xi
\label{eq.blindSIMJSPvec}
\end{equation}

Inspired by the so-called C-SALSA algorithm \cite{afonso2011augmented}, I first transform problem \eqref{eq.blindSIMJSPvec} to an unconstrained optimization problem by introducing an indicator function and then solve it with primal-dual splitting algorithm. The feasible set $\Ev(\xi,\Hc,\bm{\mathfrak{y}})$ is defined as:
\begin{equation}
\Ev(\xi,\Hc,\bm{\mathfrak{y}}) =\big\{ \bm{\mathfrak{q}} \in \mathbb{R}^{MN} \mid \lVert \Hc \bm{\mathfrak{q}} -  \bm{\mathfrak{y}} \rVert_2 \leq \xi \big\}
\end{equation}
which is possible infinite in some directions. Then problem \eqref{eq.blindSIMJSPvec} can be written as an unconstrained problem:
\begin{equation}
\label{eq.blindSIMJSPvecConstrained}
\arg \min_{\bm{\mathfrak{q}}} \;   \lVert \bm{\mathfrak{q}} \rVert_{\Gc pq}^q + \mu \lVert \Dc \bm{\mathfrak{q}} \rVert_{\Gc 21} + \imath_{\Ev(\epsilon,\mathbbm{1},\bm{\mathfrak{y}})}(\Hc\bm{\mathfrak{q}})
\end{equation}
where $\imath_\Sc : \mathbb{R}^{MN} \rightarrow \{\mathbb{R} \cup \infty\}$ is the indicator function of set $\Sc \subset \mathbb{R}^{MN}$ :
\begin{equation}
\imath_\Sc(\sb) =  \begin{cases} 
 0,  &  \sb \in \Sc \\
	\infty,   & \sb \not\in \Sc
	\end{cases}
\end{equation}

\subsection{Primal-dual splitting method}

To solve \eqref{eq.blindSIMJSPvecConstrained} with primal-dual algorithm,  three auxiliary variables $\bm{\mathfrak{d}}, \bm{\mathfrak{p}},\bm{\mathfrak{r}}$ are introduced with $ \bm{\mathfrak{d}} = \bm{\mathfrak{q}}$, $ \bm{\mathfrak{p}}=\Dc  \bm{\mathfrak{q}}$ and  $\bm{\mathfrak{r}}=\Hc  \bm{\mathfrak{q}}$. Then \eqref{eq.blindSIMJSPvecConstrained} can be rewritten as:
\begin{equation}
\begin{aligned}
\arg \min_{\bm{\mathfrak{q}}} \;  f(\bm{\mathfrak{q}}) + g_1(\bm{\mathfrak{d}}) +g_2(\bm{\mathfrak{p}}) + g_3( \bm{\mathfrak{r}})\\
\text{with} \qquad  f(\bm{\mathfrak{q}}) = 0, \quad g_1(\bm{\mathfrak{d}}) =  \lVert \bm{\mathfrak{d}} \rVert_{\Gc pq}^q \\
g_2(\bm{\mathfrak{p}})  = \mu \lVert  \bm{\mathfrak{p}} \rVert_{\Gc 21}, \quad g_3( \bm{\mathfrak{r}}) = \imath_{\Ev(\epsilon,\mathbbm{1},\bm{\mathfrak{y}})}(\bm{\mathfrak{r}})
\end{aligned}
\label{eq.blindSIMJSPvecConstrained2}
\end{equation}
We can check that \eqref{eq.blindSIMJSPvecConstrained2} is a particular case considered in \cite{condat2013primal}, and the associated primal-dual algorithm is presented in Algorithm \ref{alg:blind-SIM-JSPTV2}.

\begin{algorithm}
	\caption{Primal-dual splitting method to solve $\ell_{p,q}$ plus TV norm minimization of form 2}\label{alg:blind-SIM-JSPTV2}
	\SetKwInOut{Input}{Input}
	\SetKwInOut{Output}{Output}
	
	Initialize the parameter $\tau, \sigma, \theta >0$ and $\bm{\mathfrak{q}}_0, \bm{\mathfrak{d}}_0, \bm{\mathfrak{r}}_0$ \\
	Set $k=0$ \\
	\While{stopping criterion is not met}{

      $\bar{\bm{\mathfrak{q}}}_{k+1} = \bm{\mathfrak{q}}_k-\tau\big ( \bm{\mathfrak{d}}_k + \Dc^* \bm{\mathfrak{p}}_k  + \Hc^* \bm{\mathfrak{r}}_k \big)$

		$ \bar{\bm{\mathfrak{d}}}_{k+1} = \text{prox}_{\sigma g_1^*} \big(\bm{\mathfrak{d}}_k+ \sigma (2\bar{\bm{\mathfrak{q}}}_{k+1} - \bm{\mathfrak{q}}_k)\big)$ \;

	$\bar{\bm{\mathfrak{p}}}_{k+1} = \text{prox}_{\sigma g_2^*} \big(\bm{\mathfrak{p}}_k+ \sigma \Dc (2\bar{\bm{\mathfrak{q}}}_{k+1} - \bm{\mathfrak{q}}_k)\big)$ \;

	$\bar{\bm{\mathfrak{r}}}_{k+1} = \text{prox}_{\sigma g_3^*} \big(\bm{\mathfrak{r}}_k+ \sigma \Hc (2\bar{\bm{\mathfrak{q}}}_{k+1} - \bm{\mathfrak{q}}_k)\big)$ \;

               $\bm{\mathfrak{q}}_{k+1} = \theta \bar{\bm{\mathfrak{q}}}_{k+1} + (1-\theta) \bm{\mathfrak{q}}_k$\;
		
		$\bm{\mathfrak{d}}_{k+1} = \theta \bar{\bm{\mathfrak{d}}}_{k+1} + (1-\theta) \bm{\mathfrak{d}}_k  $ \;
$\bm{\mathfrak{p}}_{k+1} = \theta \bar{\bm{\mathfrak{p}}}_{k+1} + (1-\theta) \bm{\mathfrak{p}}_k $ \;
	
$\bm{\mathfrak{r}}_{k+1} = \theta \bar{\bm{\mathfrak{r}}}_{k+1} + (1-\theta) \bm{\mathfrak{r}}_k $ \;

                $k \leftarrow k+1$ \;
		
	}
\end{algorithm}

The $\text{prox}_f(\xb)$ in the iterations denotes the proximal operator of the function $f$, whose definition is given by \cite{combettes2011proximal}:
\begin{equation}
\text{prox}_{\lambda f}(\xb) = \arg \min_{\yb} f(\yb) + \frac{1}{2\lambda}\lVert \xb - \yb\rVert^2
\end{equation}
The proximal operator for $h^*$, the conjugate function of $f$, can be obtained from the relation:
\begin{equation}
\text{prox}_{\lambda f^*}(\xb) = \xb - \lambda \text{prox}_{f/ \lambda }(\xb)
\end{equation}

For the indicator function  $g_3$, its proximal operator can be written analytically \cite{afonso2011augmented}:
\begin{equation}
\text{prox}_{\imath_{\Ev(\xi,\mathbbm{1},\bm{\mathfrak{y}})}/\beta}(\xb) = \bm{\mathfrak{y}} + \begin{cases} 
	\epsilon \frac{ \xb-\bm{\mathfrak{y}} }{ \lVert \xb-\bm{\mathfrak{y}} \rVert_2 },  &  \lVert \xb-\bm{\mathfrak{y}} \rVert \geq \xi \\
	\xb-\bm{\mathfrak{y}},   & \lVert \xb-\bm{\mathfrak{y}} \rVert \leq \xi
	\end{cases}
\end{equation}
While the proximal operator for $\ell_{\Gc,p,q}$ norm of different $(p,q)$ pairs are shown in Appendix \ref{appen.lpqNormProximal}. 

Following \cite[Theorem 5.2]{condat2013primal}, the convergence of primal-dual iteration is granted if $q=1$ and the parameters $(\tau,\sigma,\theta)$ in algorithm \ref{alg:blind-SIM-JSPTV2} satisfy:
\begin{equation}
\begin{aligned}
\tau \sigma \big\lVert \mathbbm{1}_{MN} + \Dc^*\Dc + \Hc^*\Hc  \big\rVert_{op} \leq 1\\
\theta \in ] 0,2 [
\end{aligned}
\end{equation}
where $\Dc^*, \Hc^*$ denotes the conjugate transpose of matrix $\Dc$ and $\Hc$ and $\lVert \cdot \rVert_{op}$ is the operator norm of corresponding matrix. From the definition of operator norm, we have 
\begin{equation}
\big\lVert \mathbbm{1}_{MN} + \Dc^*\Dc  + \Hc^* \Hc \big\rVert_{op} \leq \lVert \mathbbm{1}_{MN}\rVert_{op} + \lVert \Dc^* \Dc \rVert_{op}+\lVert \Hc^* \Hc \rVert_{op}
\end{equation}
with 
\begin{equation}
\begin{aligned}
\lVert \Dc^* \Dc \rVert_{op} = \lVert \Dc \rVert_{op}^2 \leq \lVert \Ac \rVert_{op}^2 \lVert \Cv \rVert_{op}^2\\
\lVert \Hc^* \Hc \rVert_{op} = \lVert \Hc \rVert_{op}^2 = \lVert \mathbbm{1}_M \rVert_{op}^2 \lVert \Hv \rVert_{op}^2
\end{aligned}
\end{equation}
According to the inequality between root-mean square and arithmetic mean, we have:
\begin{equation}
\label{eq.operatorNormConverges2}
\lVert \Ac \bm{\mathfrak{q}} \rVert_2^2 \leq \frac{1}{MI_0} \lVert \bm{\mathfrak{q}} \rVert_2^2
\end{equation}
In addition, $\lVert \Cv \rVert_{op}^2 \leq 8$ \cite{chambolle2011first}, so 
\begin{equation}
\lVert \Dc^* \Dc \rVert_{op} \leq \frac{8}{MI_0}
\end{equation}
Since $\Hv$ is a low-pass convolution operator with symmetric boundary conditions, we have $\lVert \Hv \rVert_{op} = 1$, so that
\begin{equation}
\big\lVert \mathbbm{1}_{MN} + \Dc^* \Dc +\Hc^* \Hc \big\rVert_{op} \leq 2+\frac{8}{MI_0}
\end{equation}

So the primal-dual algorithm will converge as long as $\tau \sigma \leq \frac{1}{2+8/(MI_0)}$ and $\theta \in ]0,2[\,$ in the case $q=1$. When $0\leq q<1$, the algorithm can not give the global minimum any more since the $\ell_q$ norm is not convex function.
 
The computational burden of the primal-dual algorithm mainly lie on $\Hc\bm{\mathfrak{r}}$ and $\Hc^*\bm{\mathfrak{r}}$, for $\bm{\mathfrak{r}} \in \mathbb{R}^{MN}$. Taking advantage of the fast Fourier transform (FFT) algorithm, the computational complexity of the primal-dual algorithm in each iteration is $\Oc(MN\log N)$.

\section{Simulation results and experiments}\label{sec.simu}
To study the numerical performance of the proposed $\ell_{p,q}$ norm model, a 2D 'star-like' simulated target whose fluorescence density in the polar coordinates given by: $\rho(r,\theta) \propto [1+\cos(40\theta)]$ is used as the true object. The top-left quarter of the object is shown in Fig. \ref{fig.lpqnorm}a. One advantage of this object is that its spatial frequencies increases as you move closer to the star center, making it easy to visualize the resolution improvement. The point spread function is chosen as:
\begin{equation}
\label{eq:psfSIMU}
h(r,\theta) = \Big(\frac{J_1(NAk_0r)}{k_0r}\Big)^2\frac{k_0^2}{\pi}
\end{equation}
where  $J_1$ is the first order Bessel function of the first kind, NA is the objective numerical aperture set to 1.49 and $k_0=\frac{2\pi}{\lambda}$ is the free-space wavenumber with $\lambda$ the emission and the excitation wavelengths.  The radius $r$ from the center of the object that a conventional wide-field microscopy can reach could be easily deduced from the relation:

\begin{equation}
\frac{2\pi r}{40} = \frac{\lambda}{2NA}
\end{equation}

The sampling step in the object should be finer than $\lambda/ 8NA$ to observe a SR factor of two. In the simulations a sampling step of $\lambda/ 20$ is adopted so that aliasing does not destroy the attainable SR. For the sampling rate in the raw images, no information is lost as long as it is higher than the Nyquist rate $4NA/\lambda$. In the simulations performed in this section, I set the sampling rate for the raw images the same as the object.

The speckle patterns are generated through the same optical device as the collection of raw images, unless otherwise stated. Under this condition, its frequency support has the same shape as the OTF of the system for the unapodized pupil \cite[Section 7.7]{goodman2015statistical}. The boundary conditions of the object $\rhob$ is assumed to be periodic, thus the convolution matrix $\Hv$ will have a block-circulant with circulant-block (BCCB) structure \cite{hansen2006deblurring} and the matrix vector product $\Hv\vb$ could be performed with fast Fourier transform (FFT) algorithm .  

Firstly, numerical simulations are performed with 300 speckle patterns. The low resolution raw images are corrupted with Gaussian white noise, corresponding SNR 40 dB. In the primal-dual algorithm, I set  $\theta = \sigma =1$ and $\tau = 0.35$, with $\bm{\mathfrak{q}}_0, \bm{\mathfrak{d}}_0, \bm{\mathfrak{r}}_0$ initialized with zeros. $\xi$ is set to its true value $\xi_{\text{real}} = \sqrt{MN}\nu$ ,  where $\nu$ is the standard variance of noise, unless otherwise stated. The hyperparameter for TV regularizer is set to $0$, except in the situation when Poisson noise is considered.

The Wiener deconvolution of the mean of raw images $\bar{\yb} = \frac{1}{M} \sum_m \yb_m$ is shown in Fig. \ref{fig.lpqnorm}(b). As has been expected, we see no super-resolution (patterns inside the green solide line) in the Wiener deconvolution of the wide-field image. The reconstructed image obtained by methods presented in \cite{mudry2012structured} using only positivity constraint is shown in Fig. \ref {fig.lpqnorm}(c). It retrieves partial super-resolution information, however, the modulation contrast in super-resolution part is relatively low, coinciding with the results reported in \cite{mudry2012structured}. Figure \ref {fig.lpqnorm}(d,e) are obtained using $\ell_{2,0}$ norm regularizer with M-SBL algorithm as in \cite{min2013fluorescent} and $\ell_1 + \ell_2$ norm plus positivity regularizer with PPDS algorithm presented in \cite{labouesse2017joint}, respectively. The image reconstructed by M-SBL algorithm do not scale well and there are some artifacts in low resolution part. I stop the M-SBL iterations after a fixed number of iterations (i.e.\ 20) as indicated in reference \cite{min2013fluorescent} and the computational complexity of M-SBL algorithm is in fact $\Oc(N^3)$, as high as the marginal approach \cite{idier2017super} plotted in Fig. \ref{fig.lpqnorm}(f). The $\ell_1$ regularizer in \cite{labouesse2017joint} can be seen as a specific case of the $\ell_{p,q}$ regularizer, i.e.\ $p=q=1$. 

\def\wdth{1.6in}
\def\wdthnar{1.55in}

\begin{figure}[htbp]
	\centering
	\begin{tabular}{cc}

          \includegraphics[width=\wdth]{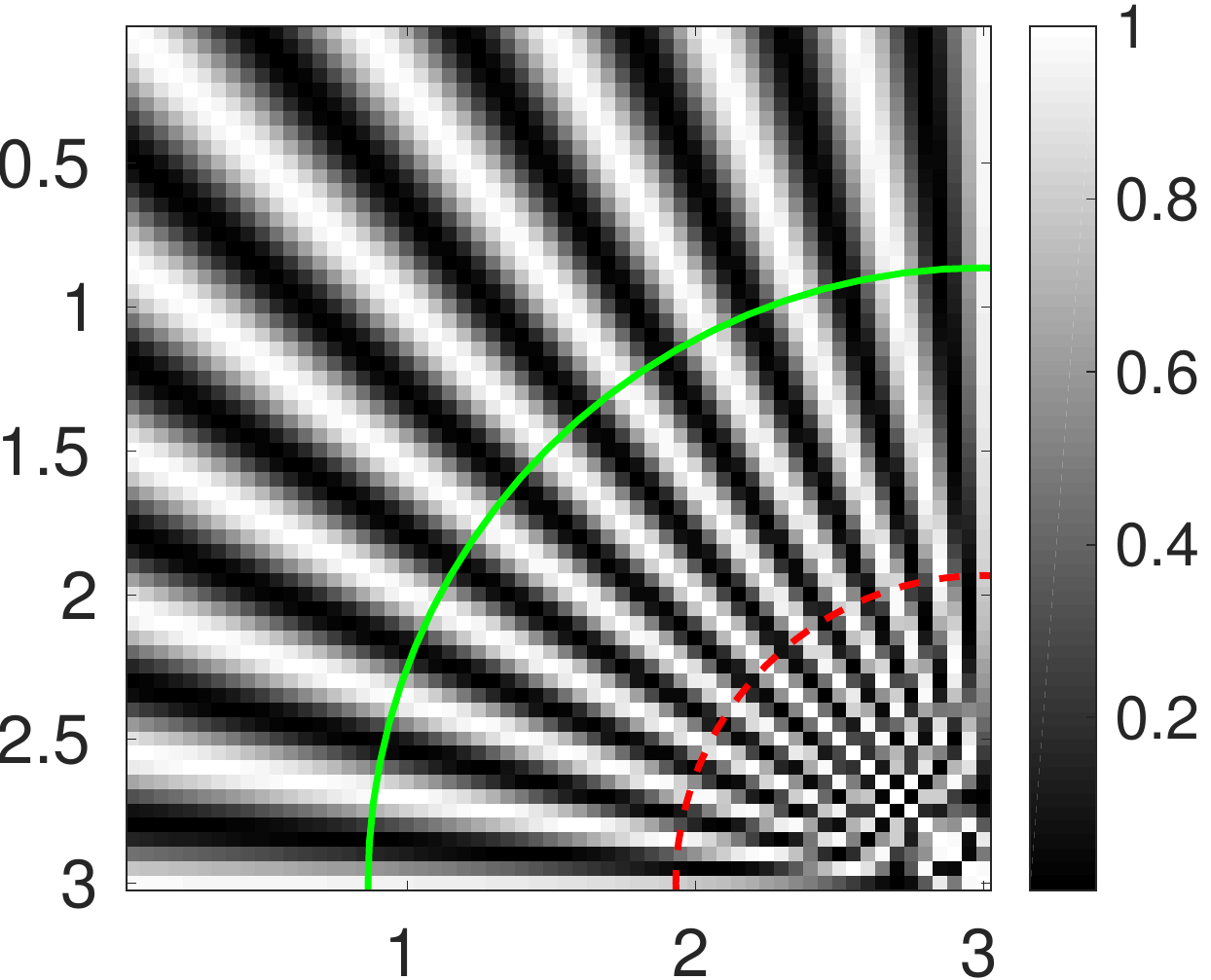}
           & \includegraphics[width=\wdth]{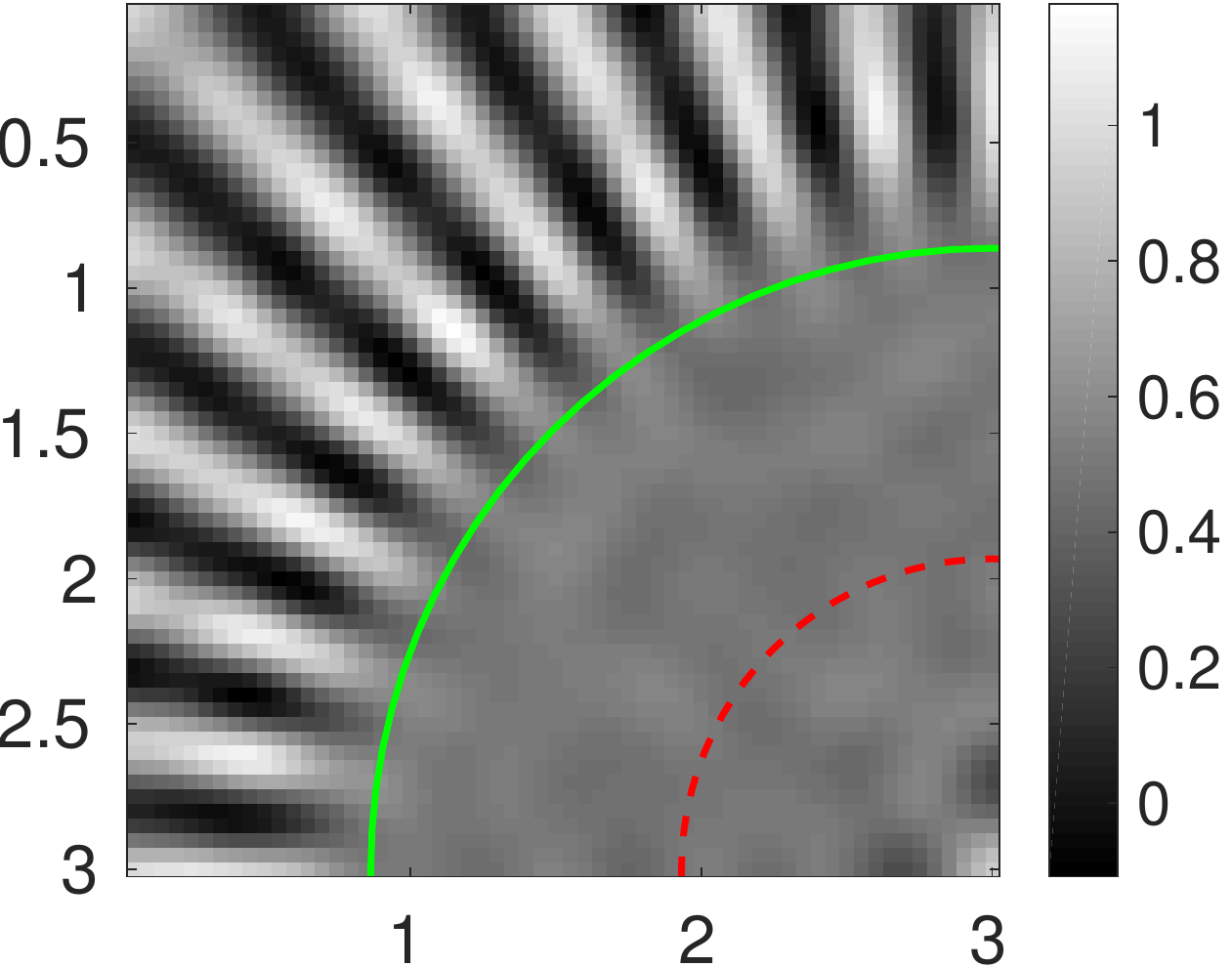}\\
\small a.) True object & b.) Deconvolution of $\bar{\yb}$  \\

\includegraphics[width=\wdthnar]{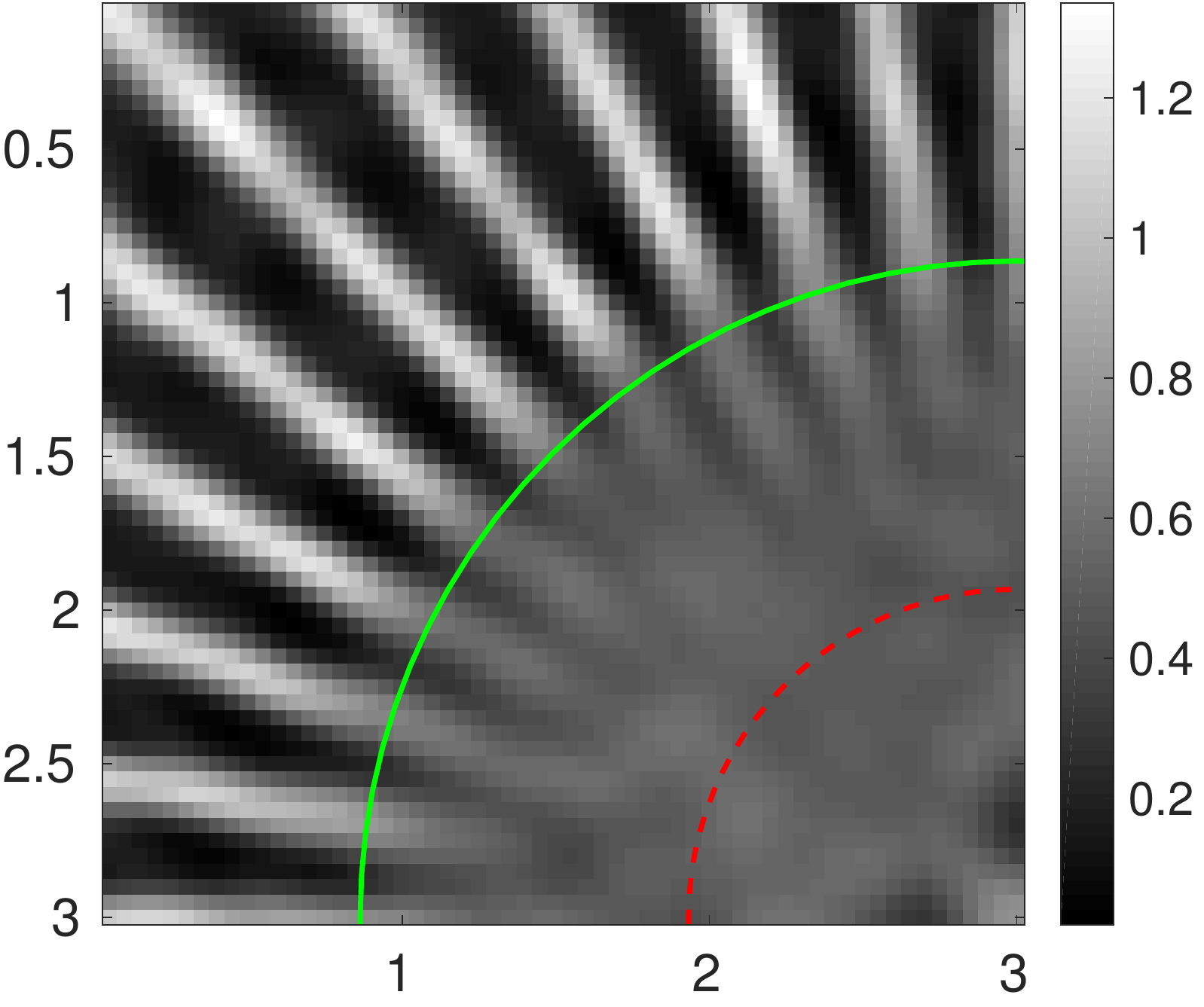}
&\includegraphics[width=\wdth]{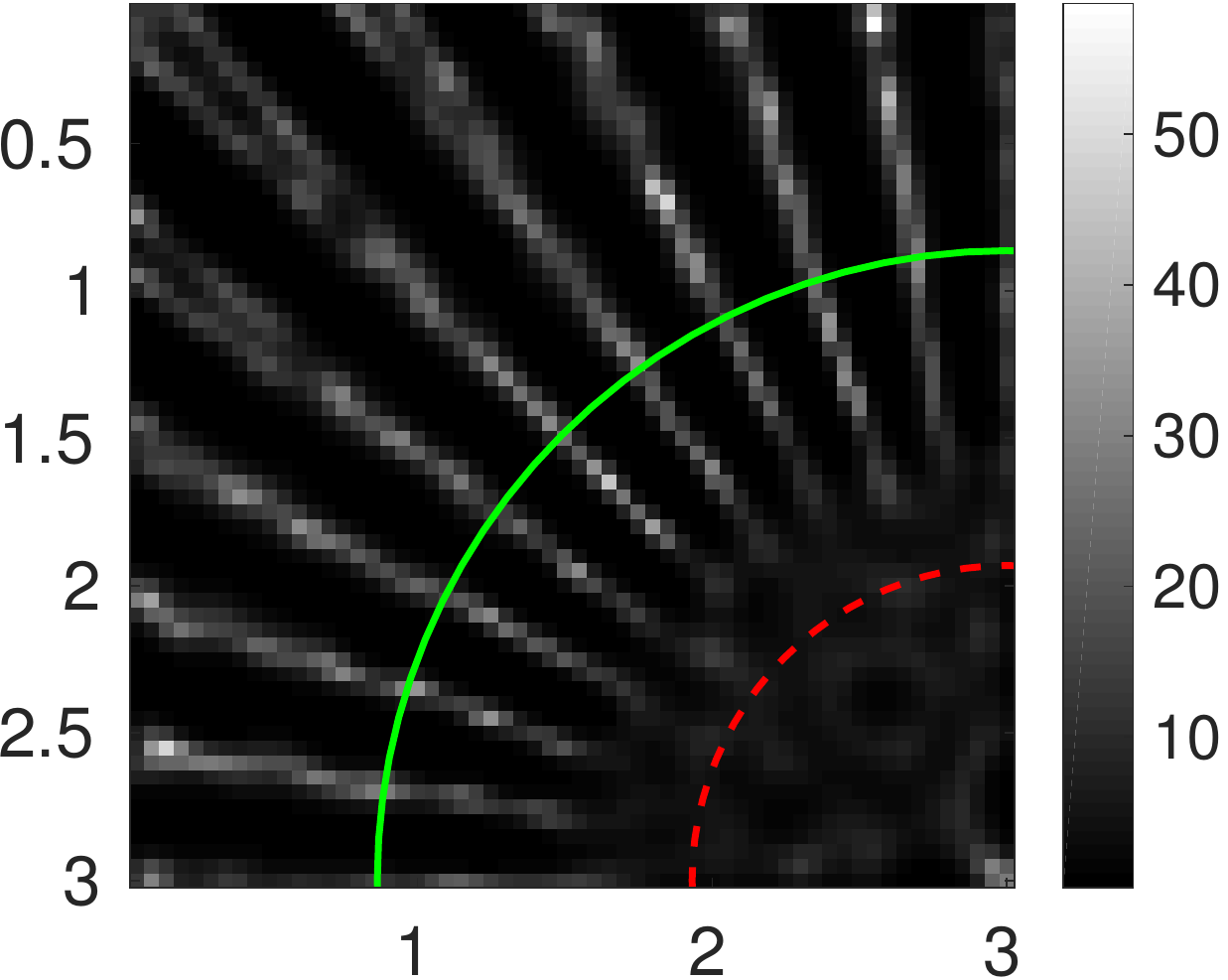}\\
\small c.) Positivity constraint \cite{mudry2012structured} &  d.) M-SBL for $\ell_{2,0}$ norm \cite{min2013fluorescent}  \\

\includegraphics[width=\wdth]{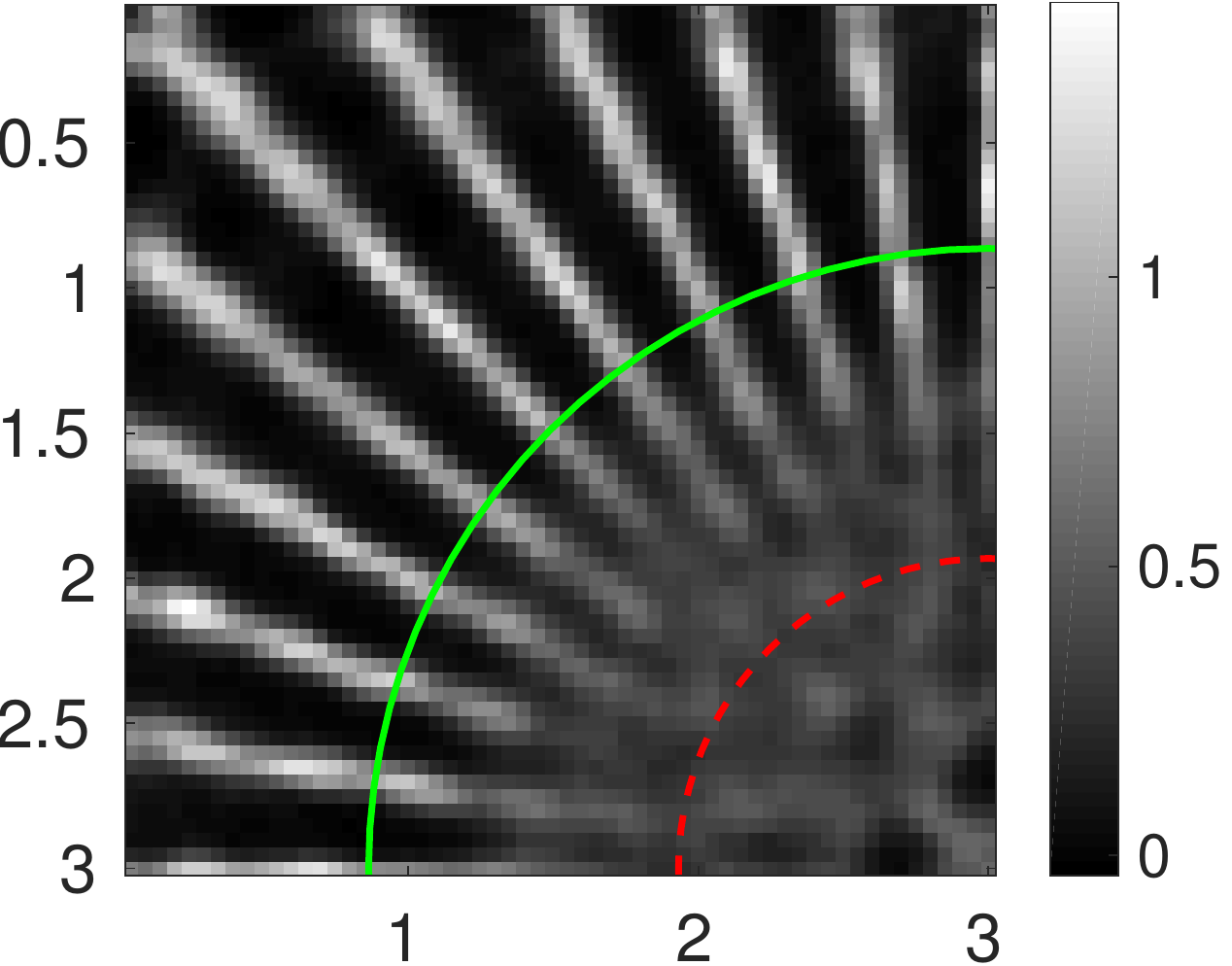}
           & \includegraphics[width=\wdthnar]{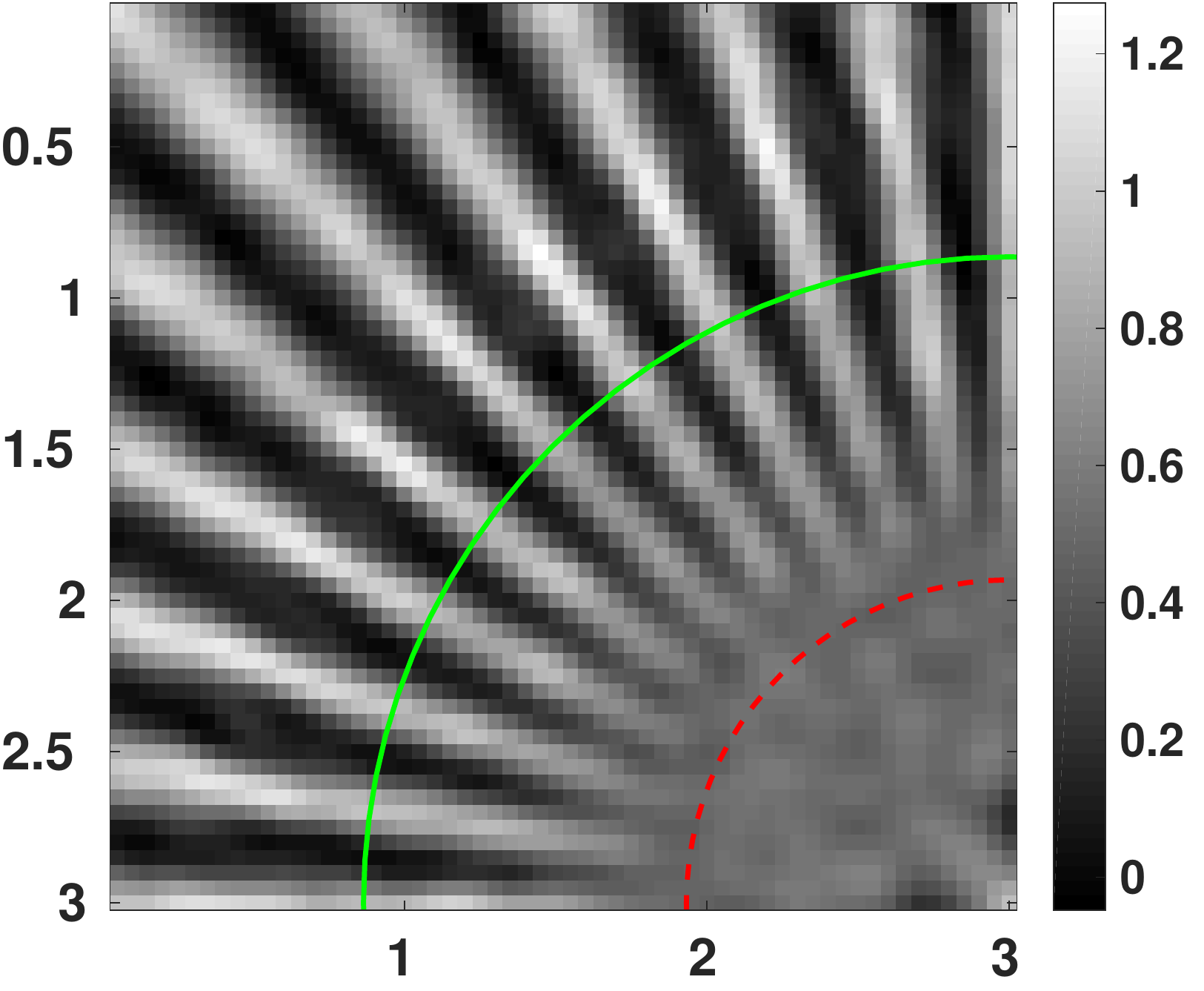}\\
\small e.) $\ell_1+ \ell_2$ regularizer \cite{labouesse2017joint} &  f.) Marginal estimator \cite{idier2017super} \\

	\end{tabular}  
	\caption{\textbf{Reconstructed object with 300 speckle patterns and 40dB Gaussian noise by minimizing $\ell_{p,q}$ norm of unconstrained form.} The green solid lines (resp. red dashed lines) correspond to spatial frequencies transmitted by OTF support (resp. 2 times OTF support) and the graduation in the images represents the wavelength $\lambda$.}
	
	\label{fig.lpqnorm} 
\end{figure}

\subsection{Reconstruction with different $\{p,q\}$ pairs }
\begin{figure}[t]
	\centering
	\begin{tabular}{cc}	
		\includegraphics[width=\wdth]{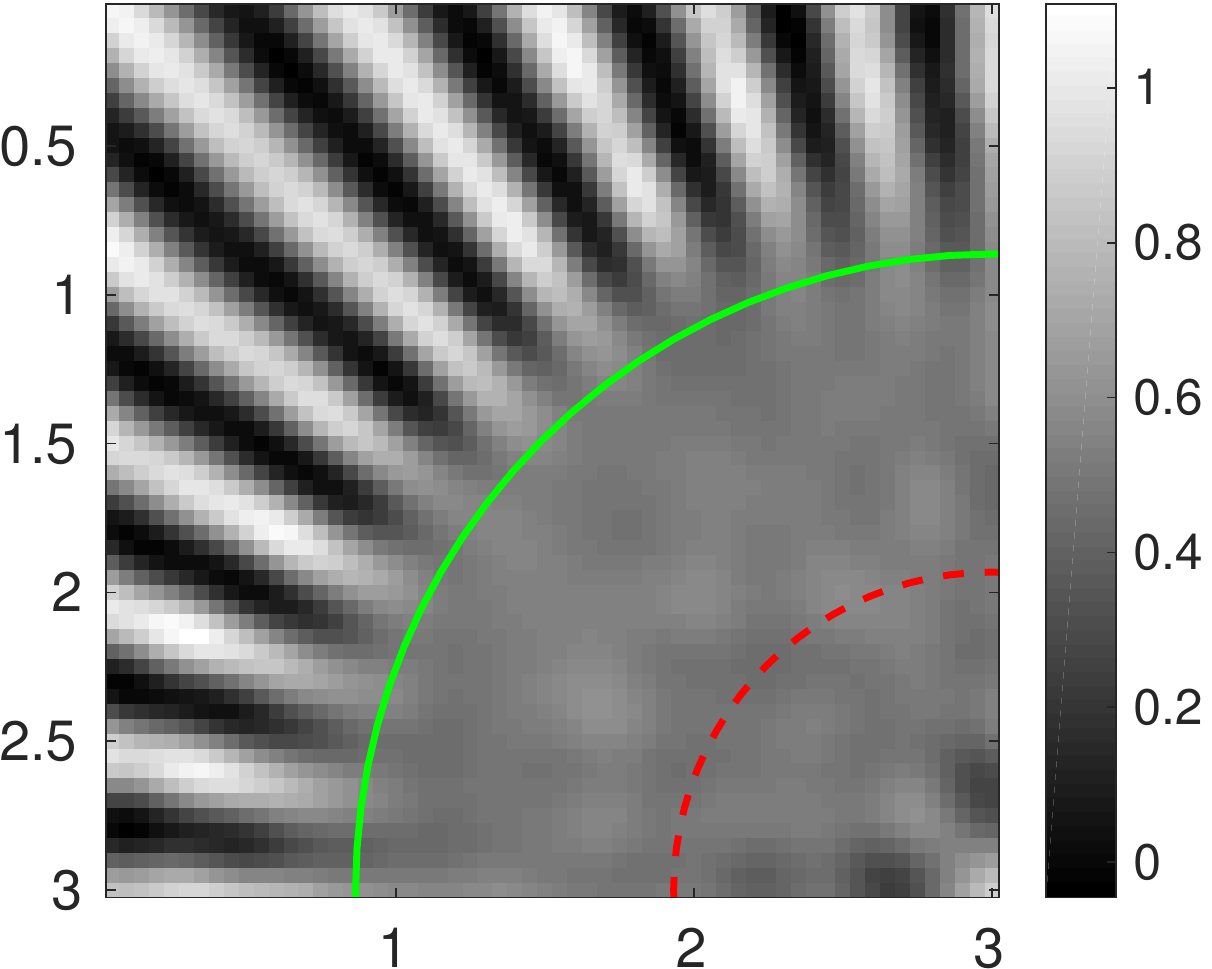}
		& \includegraphics[width=\wdth]{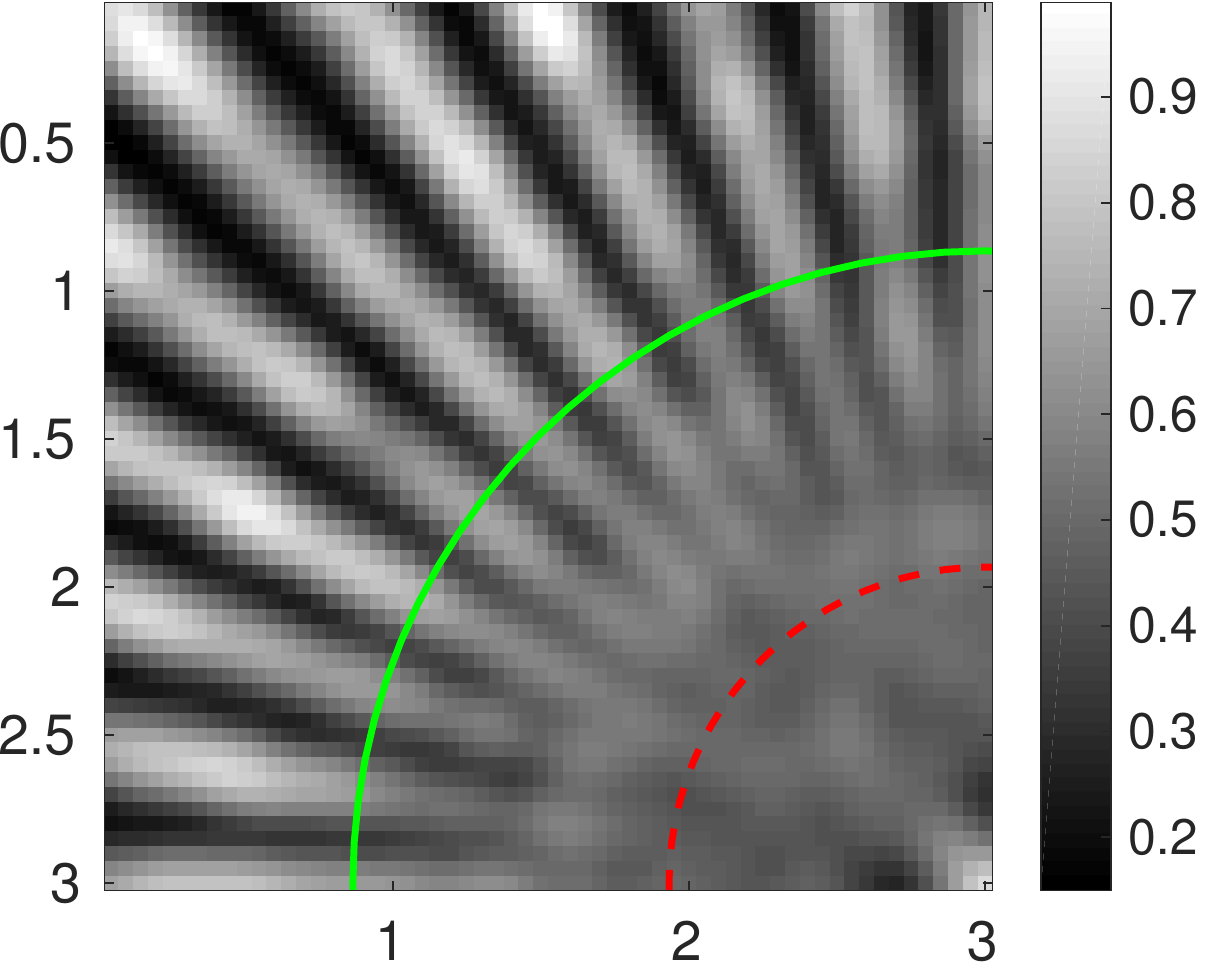}\\

		\small   a.) Mean of $\hat{\qb}_{mWiener}$ &  b.) Std of $\hat{\qb}_{mWiener}$ \\
			
		\includegraphics[width=\wdth]{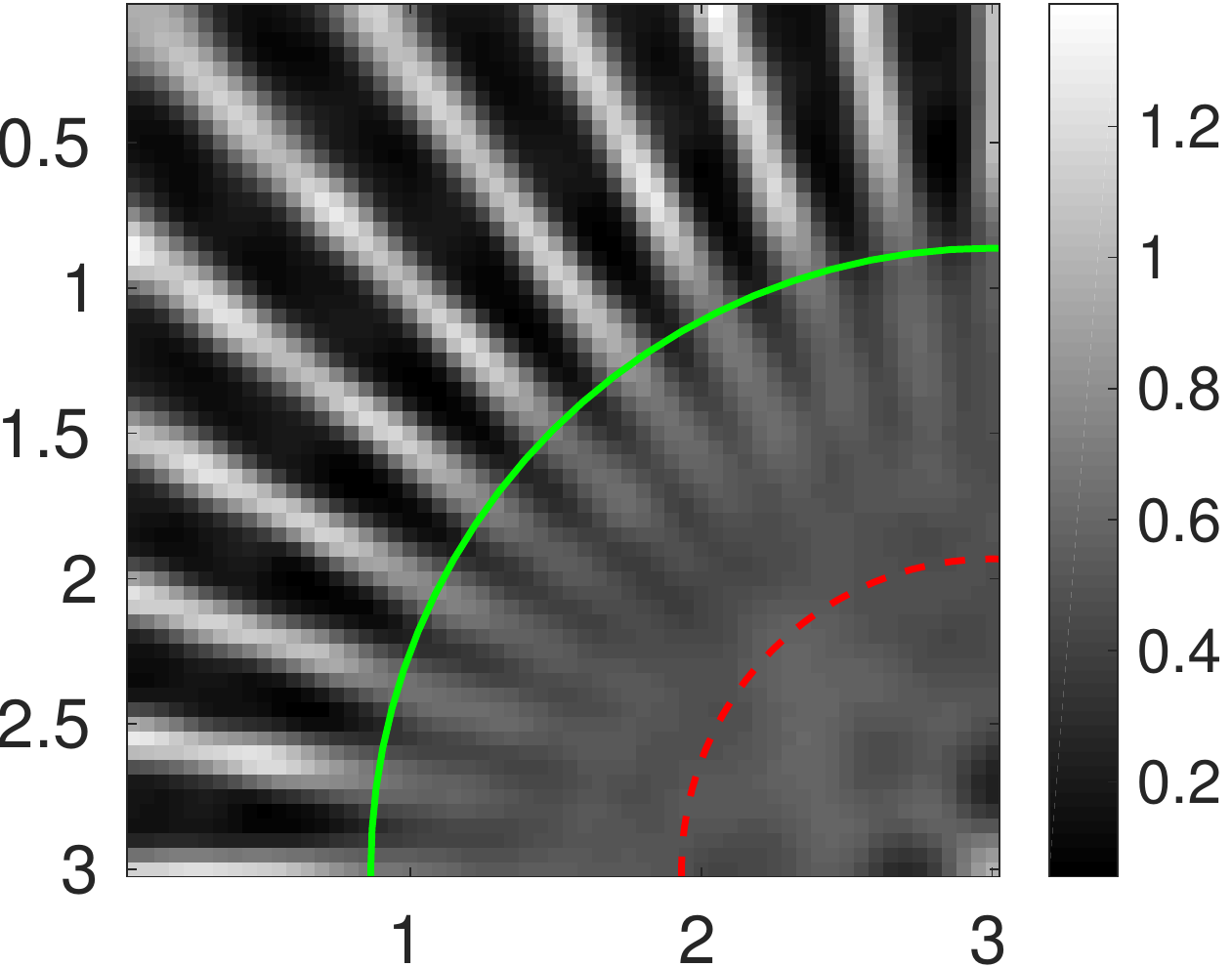}
		& \includegraphics[width=\wdth]{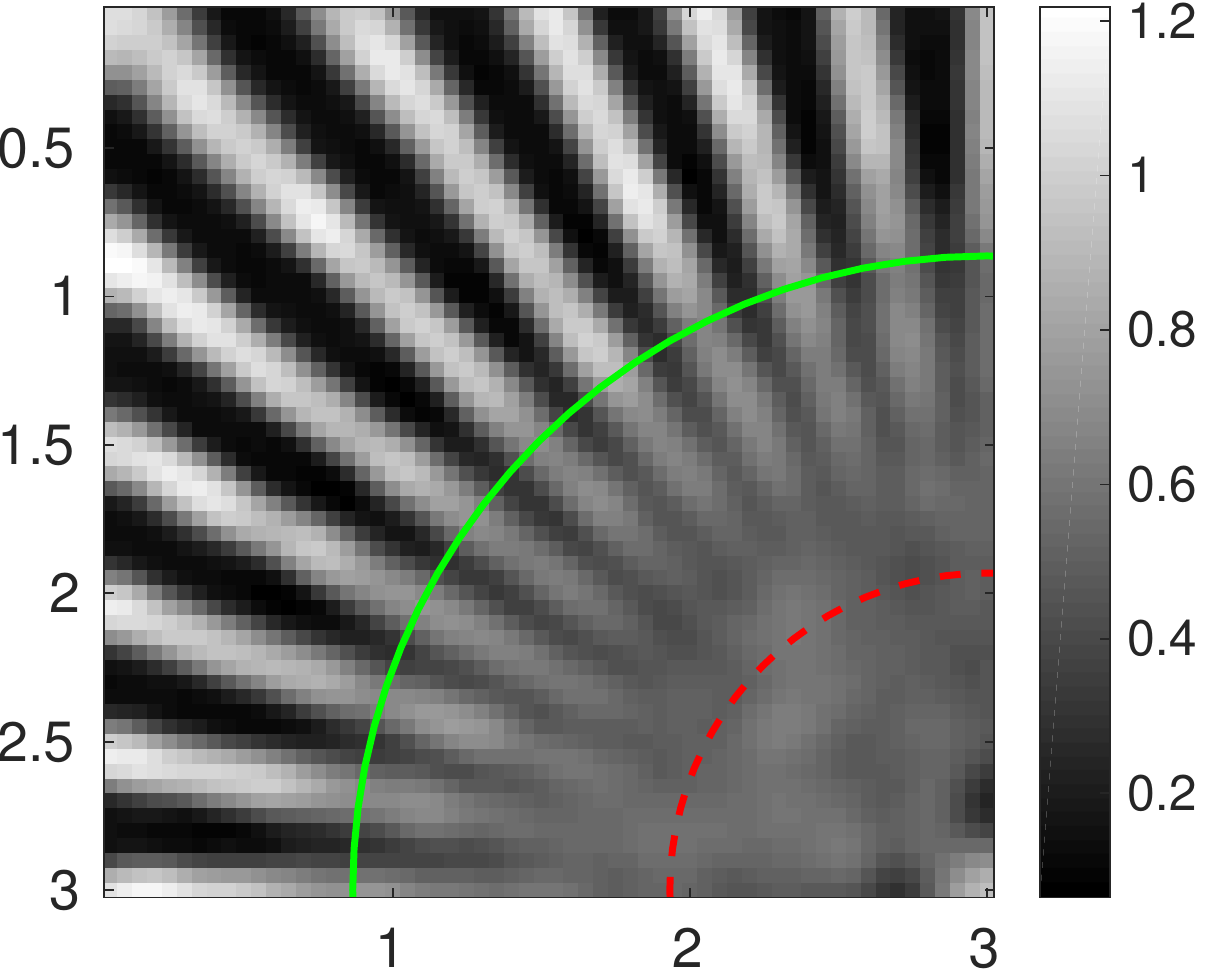}\\
		\small c.) $\ell_{1,1}\; \hat{\qb}_m$ mean &  d.) $\ell_{1,1}\; \hat{\qb}_m$ std  \\

		\includegraphics[width=\wdth]{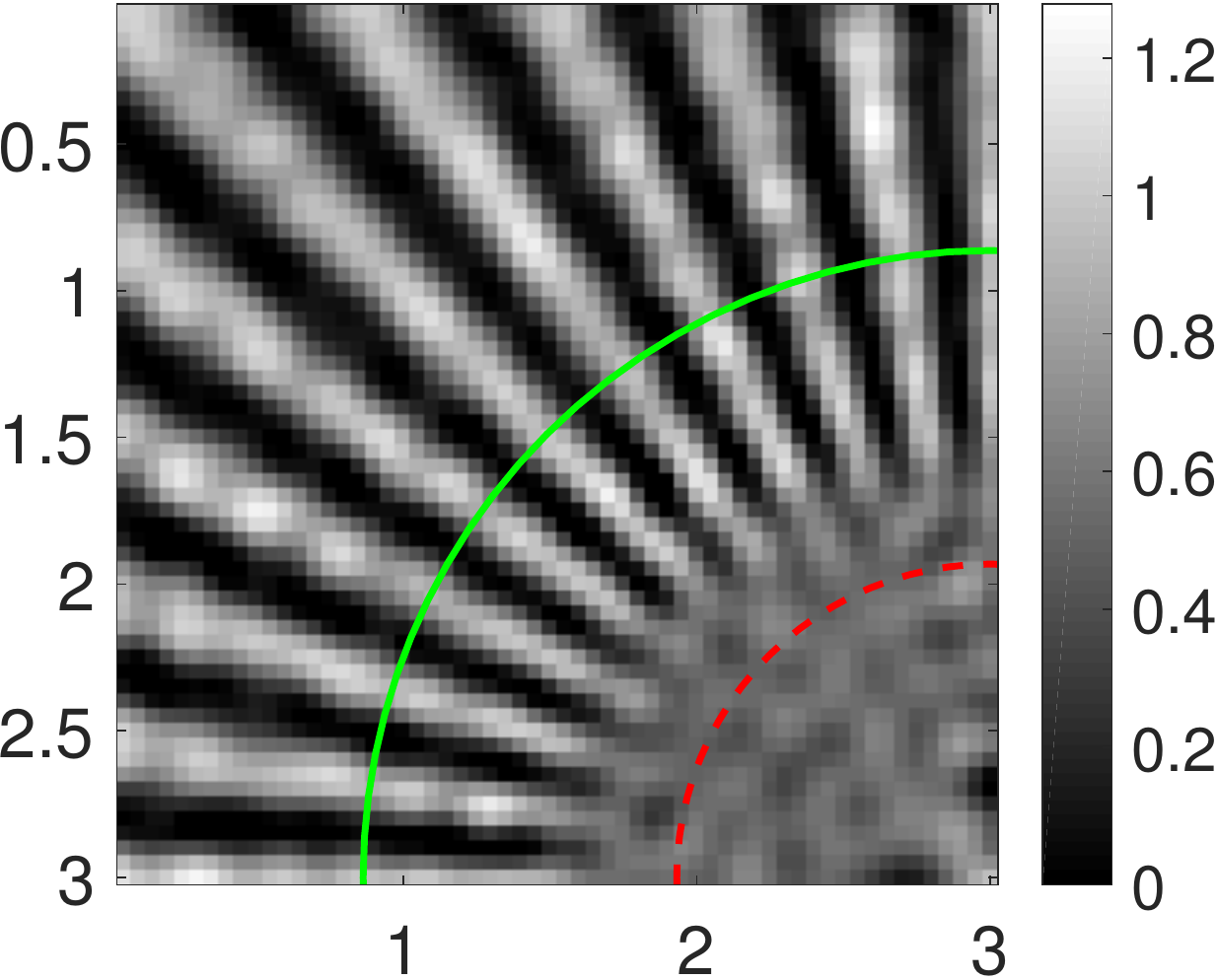}
		& \includegraphics[width=\wdth]{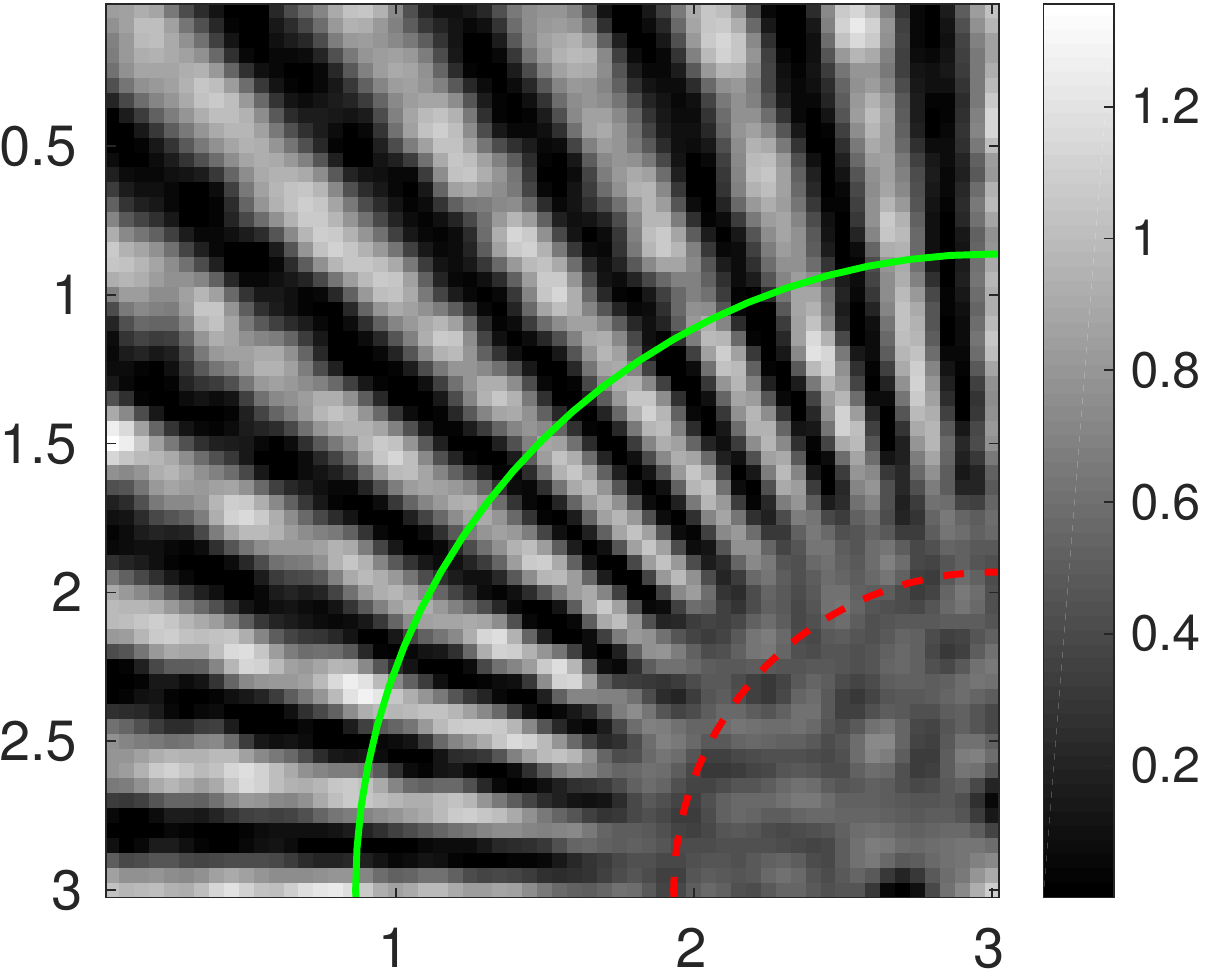}\\

		\small   e.) $\ell_{2,1}\; \hat{\qb}_m$ mean &  f.) $\ell_{2,1}\; \hat{\qb}_m$ std \\
		
			\includegraphics[width=\wdth]{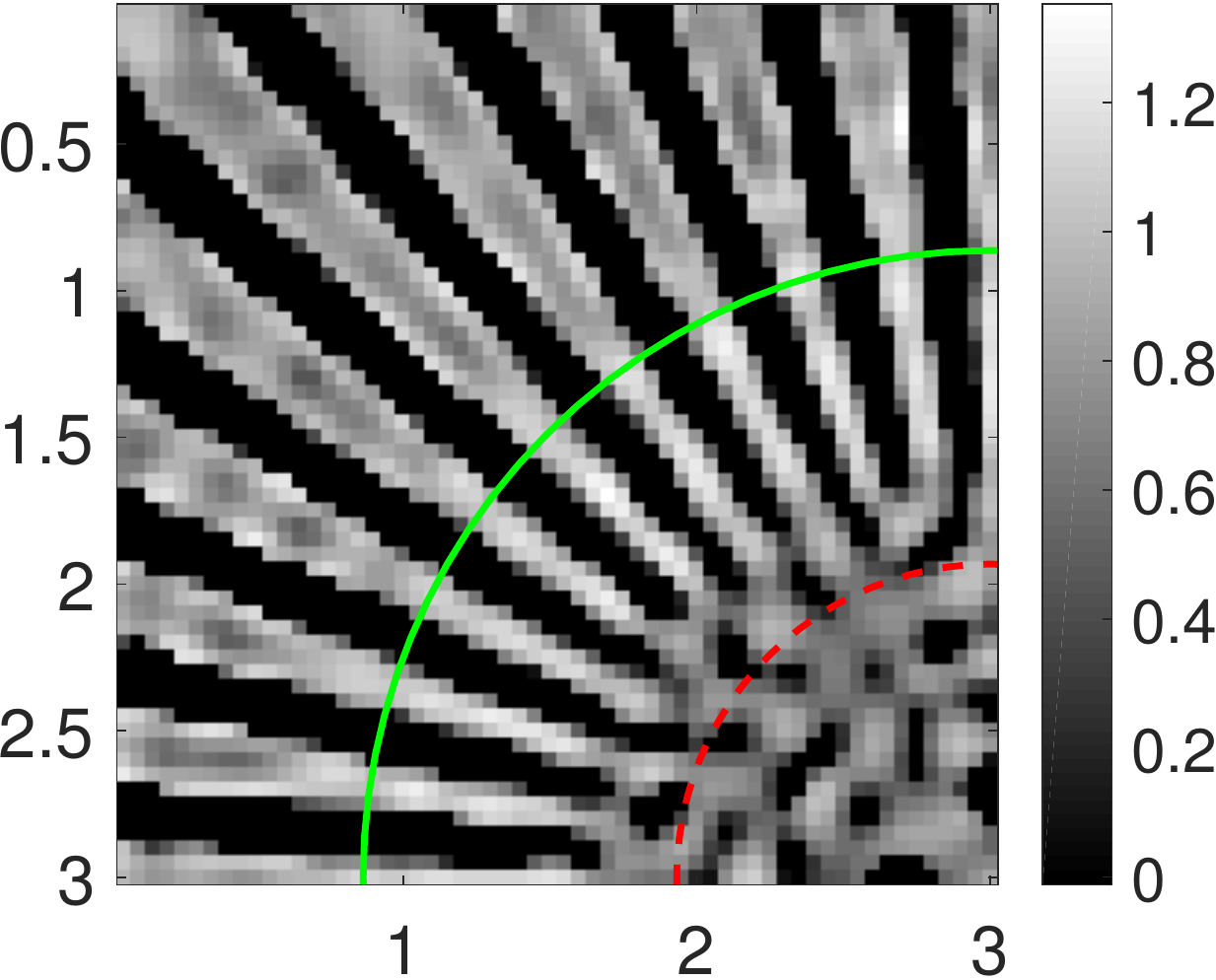}
			& \includegraphics[width=\wdth]{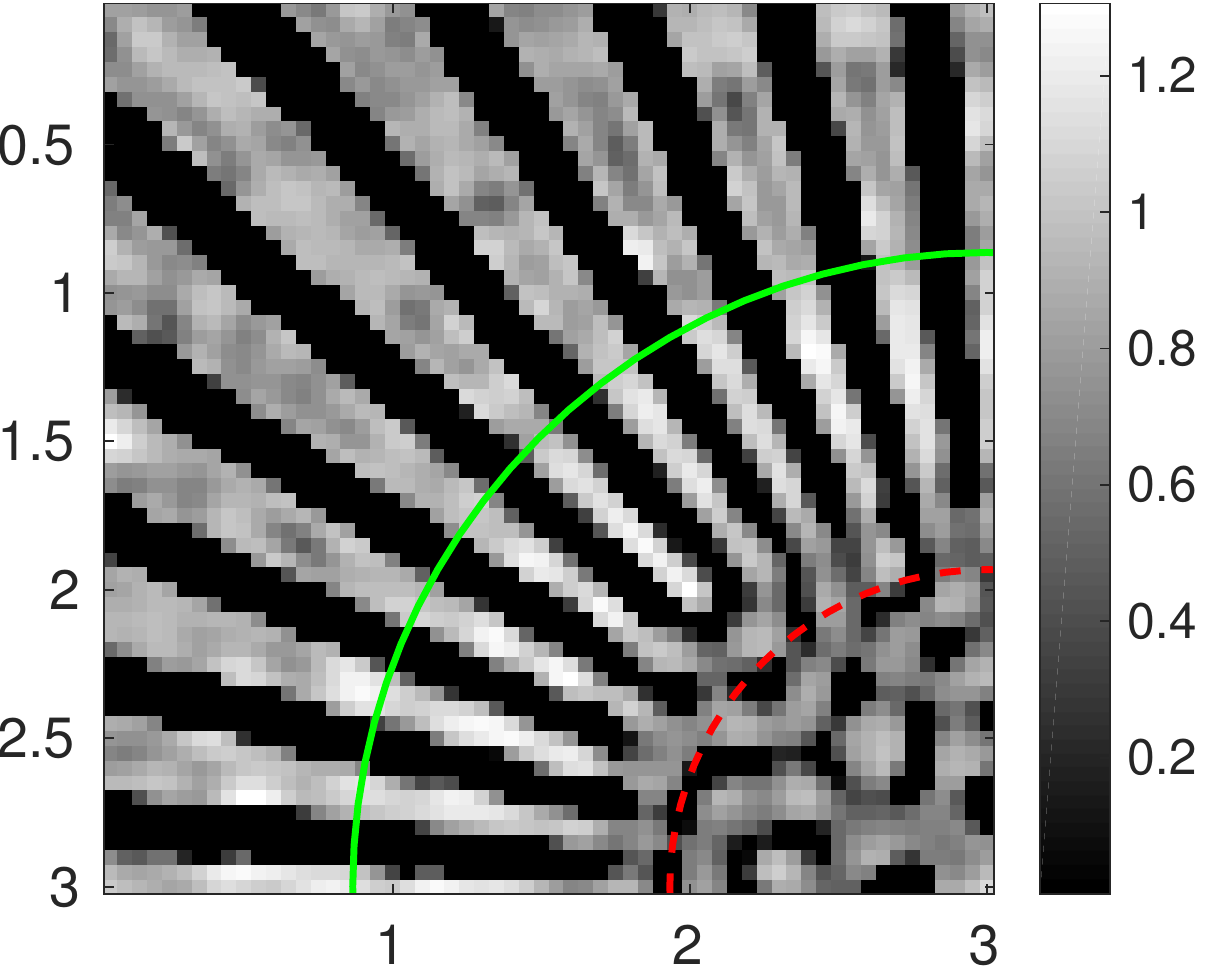}\\
			\small g.) $\ell_{2,1/2}\; \hat{\qb}_m$ mean &  h.) $\ell_{2,1/2}\; \hat{\qb}_m$ std  \\

			\includegraphics[width=\wdth]{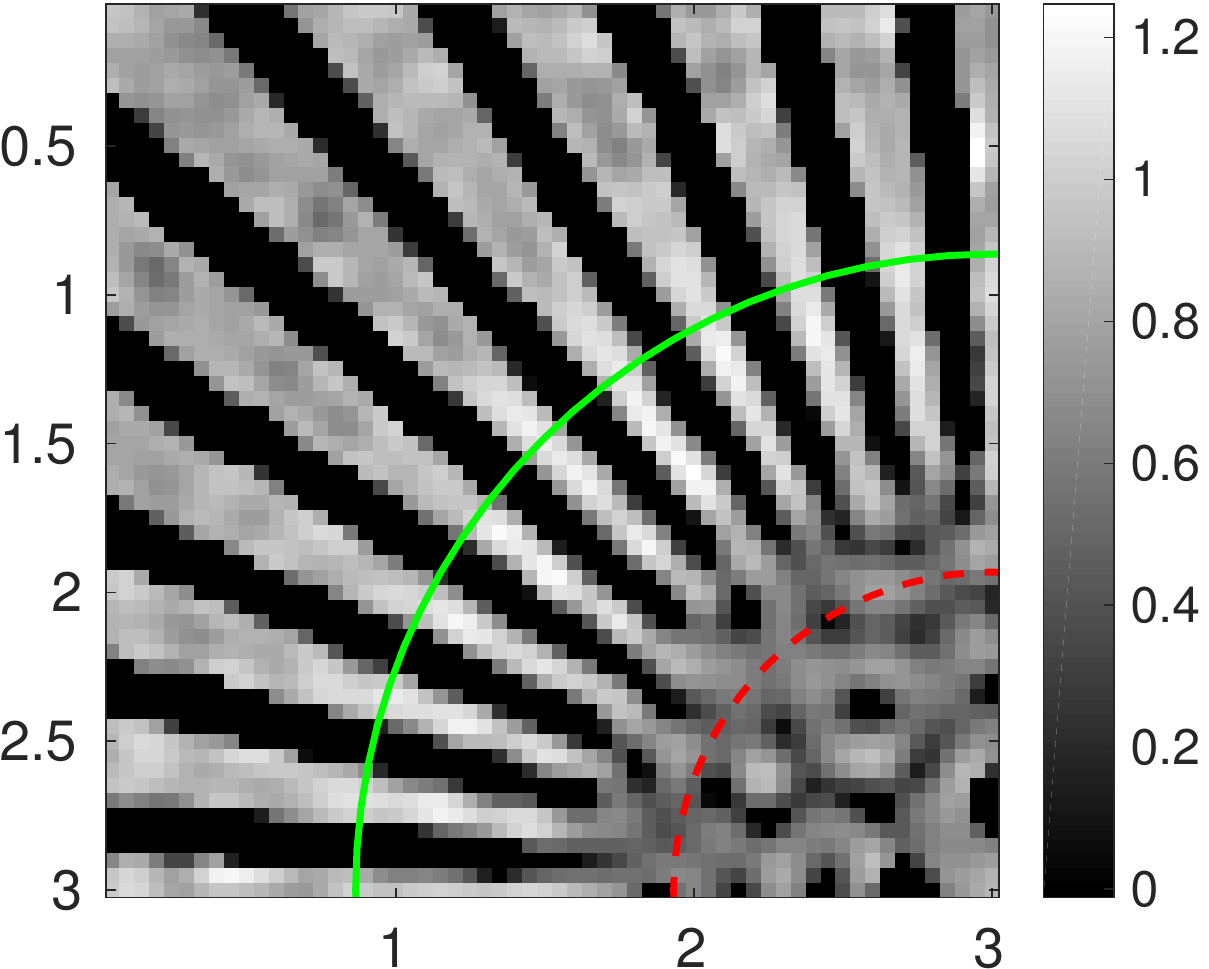}
			& \includegraphics[width=\wdth]{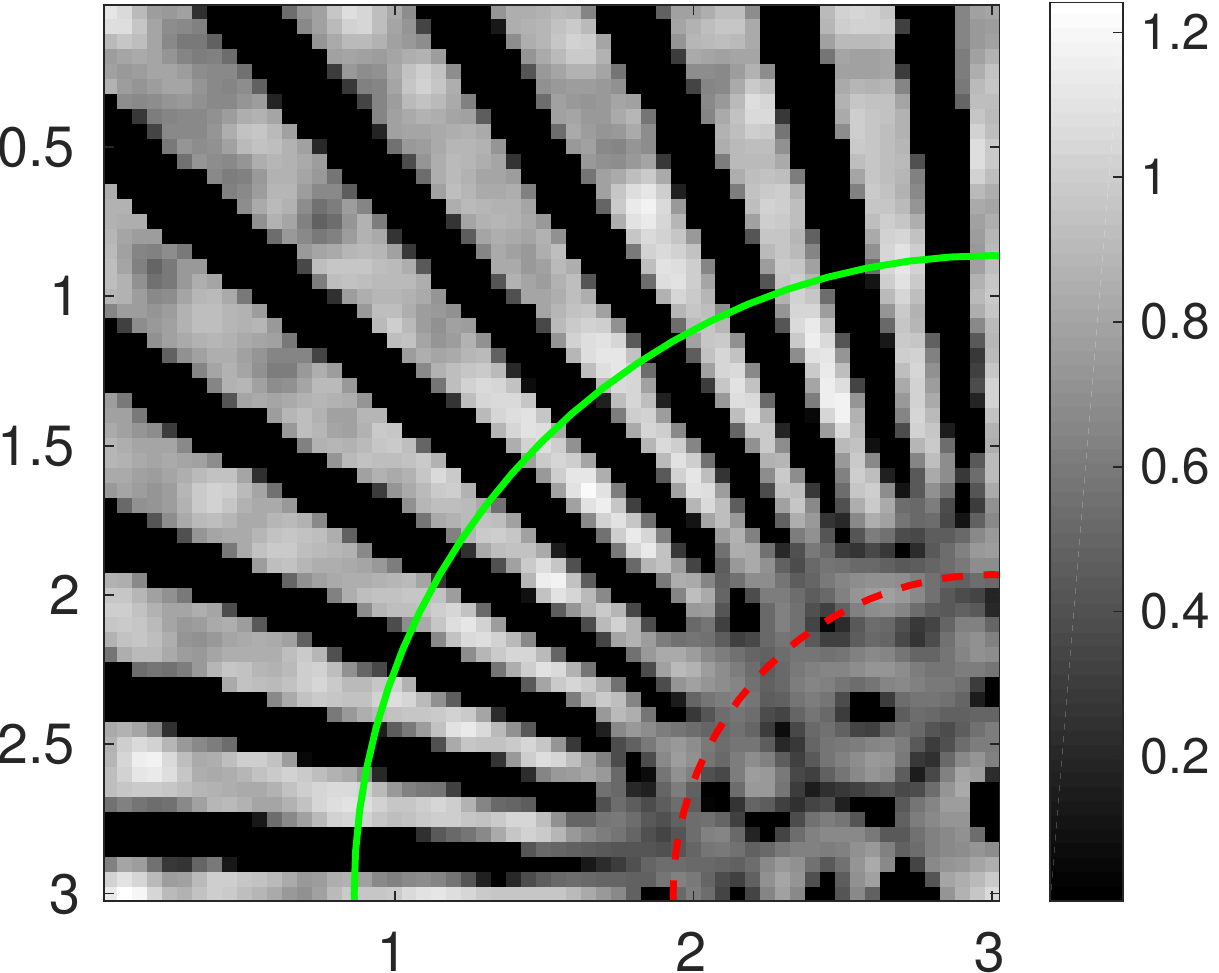}\\

			\small   i.) $\ell_{2,2/3}\; \hat{\qb}_m$ mean &  j.) $\ell_{2,2/3}\; \hat{\qb}_m$ std \\

	\end{tabular}  
	\caption{\textbf{Reconstructed object with 300 speckle patterns and 40dB Gaussian noise by minimizing $\ell_{p,q}$ norm of unconstrained form.} The green solid lines (resp. red dashed lines) correspond to spatial frequencies transmitted by OTF support (resp. 2 times OTF support) and the graduation in the images represents the wavelength $\lambda$.}
	
	\label{fig.lpqnorm2} 
\end{figure}

Figure \ref{fig.lpqnorm2} show the reconstruction results by minimizing the constrained $\ell_{p,q}$ model with different $(p,q)$ pairs. The first column are obtained by averaging $\qb_m$ while the second column by Eq. \eqref{eq.variance_qm}. 

The mean and the standard deviation of $\qb_m$ obtained from Wiener deconvolution are shown in Fig. \ref{fig.lpqnorm2}(a,b). The mean of Wiener deconvolution gives no super-resolution information as expected \cite{labouesse2017joint} , however, their standard deviation gives partial super-resolution.  

\begin{figure}[htbp]
	\centering
	\includegraphics[width=0.45\textwidth]{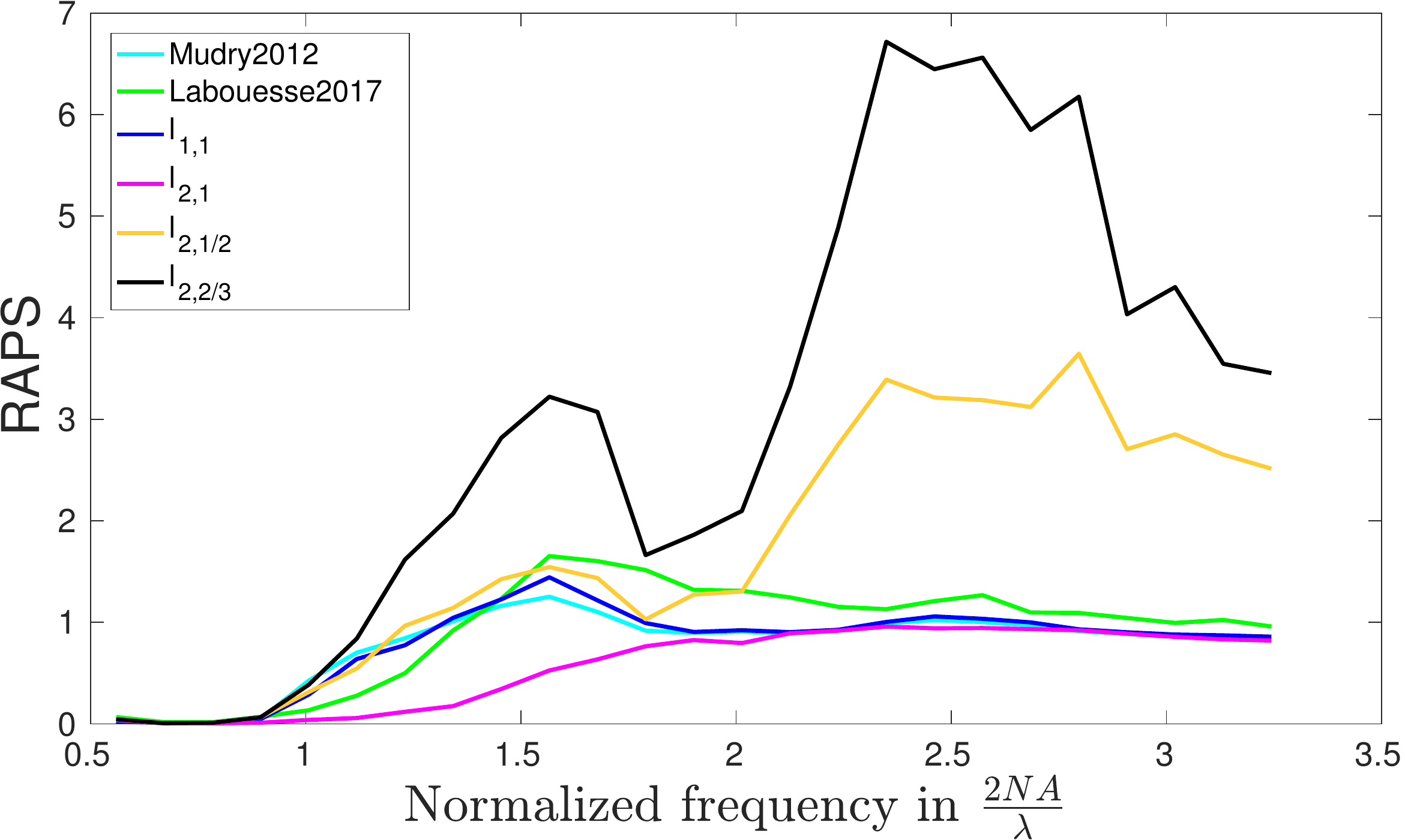}
	\caption{\textbf{The normalized RAPS of the reconstructed objects by averaging $\qb_m$ in blind-speckleSIM with different regularizers.} }
	\label{fig:rapsANDmcfOFjointEstimaters}
\end{figure}

We almost retrieve a super-resolution factor of two as marked by the red dashed lines with $\ell_{p,q}$ regularizer term, which is also the resolution limit can be reached by the standard SIM in the epi-illumination geometry. To measure the quality of reconstructed images, the normalized radially averaged power spectrum (RAPS) of the error images are plotted, which is defined as:

\begin{equation}
\begin{aligned}
\label{eq.normalizedRAPSDefine}
f(r) = \frac{\int_{-\pi}^{\pi} \big\lvert\widetilde{\hat{\rho}}(\ub)-\widetilde{\rho^*}(\ub) \big\rvert^2 d\theta}{\int_{-\pi}^{\pi} \lvert\widetilde{\rho^*}(\ub) \rvert^2 d\theta}\\
\qquad \text{with} \quad \ub =\left[ \begin{array}{c} r\cos \theta\\ r\sin\theta \end{array}\right],\; r>0,\;\theta \in (0,2\pi)
\end{aligned}
\end{equation}
where $\hat{\rho}$ and $\rho^*$ denote the estimated and true object, respectively. The normalized RAPS of errors by different regularizers are displayed in Figure \ref{fig:rapsANDmcfOFjointEstimaters}. The $\ell_{2,1}$ regularizer has the lowest error power in almost all the spectrum. The strong error power in high frequency part with $\ell_{2,1/2}$ and $\ell_{2,2/3}$ regularizers are probably caused by the binary effect in the reconstructed images. The capacity of standard deviation estimator \eqref{eq.variance_qm} to remove out-of-focus background signal is presented in section \ref{secref.bgRemoval}.

\subsection{Two estimators and background removal}\label{secref.bgRemoval}
The real images recorded by the microscopy are always blurred by out-of-focus background. A more accurate model than \eqref{eq.obs} to describe the imaging process is:
\begin{equation}
\yb_m = \Hv\qb_m + \epsilonb_m + \bb
\end{equation}
with $\bb \in \mathbb{R}^L$ denoting the background noise. So the reconstructed $m$-th column of $\Qv$  is in fact $\hat{\qb}_m =\qb_m + \Hv^+(\epsilonb_m + \bb)$, with $\Hv^+$ the pseudo inverse of $\Hv$. If we continue estimating $\rhob$ by averaging $\hat{\qb}_m$, then the estimated object will be blurred by $\Hv^+\bb$:
\begin{equation}
\hat{\rhob} = \frac{1}{MI_0} \sum_m \hat{\qb}_m =  \rhob + \frac{1}{I_0}\Hv^+\bb
\end{equation}

Unlike the ensemble mean, the empirical variance of $\hat{\qb}_m$ is not blurred with the background, so  \eqref{eq.variance_qm} holds even when strong background signal is presented. 

\begin{figure}[t]
	\centering
	\begin{tabular}{cc}
		
		\includegraphics[width=\wdth]{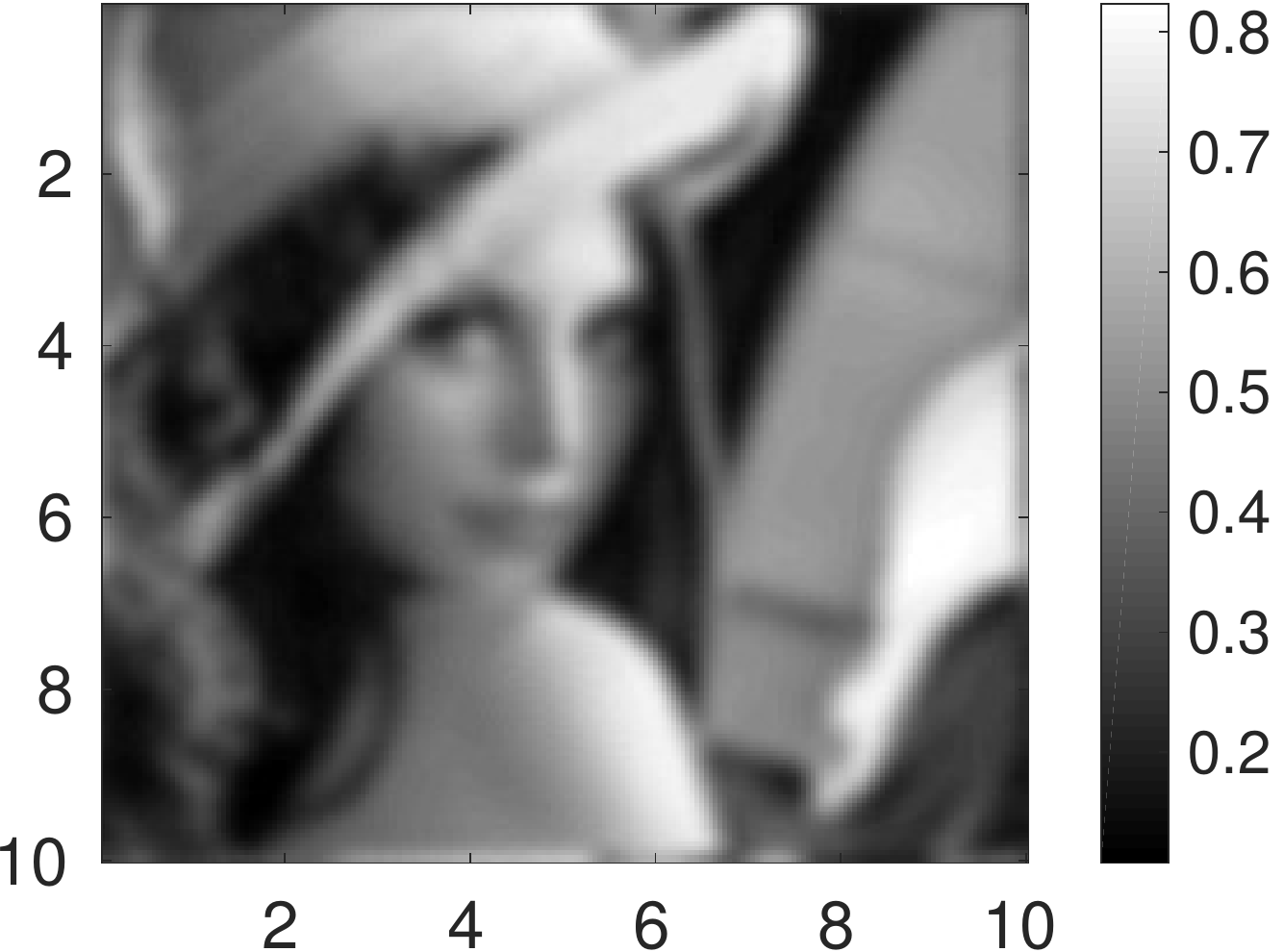}
		& \includegraphics[width=\wdth]{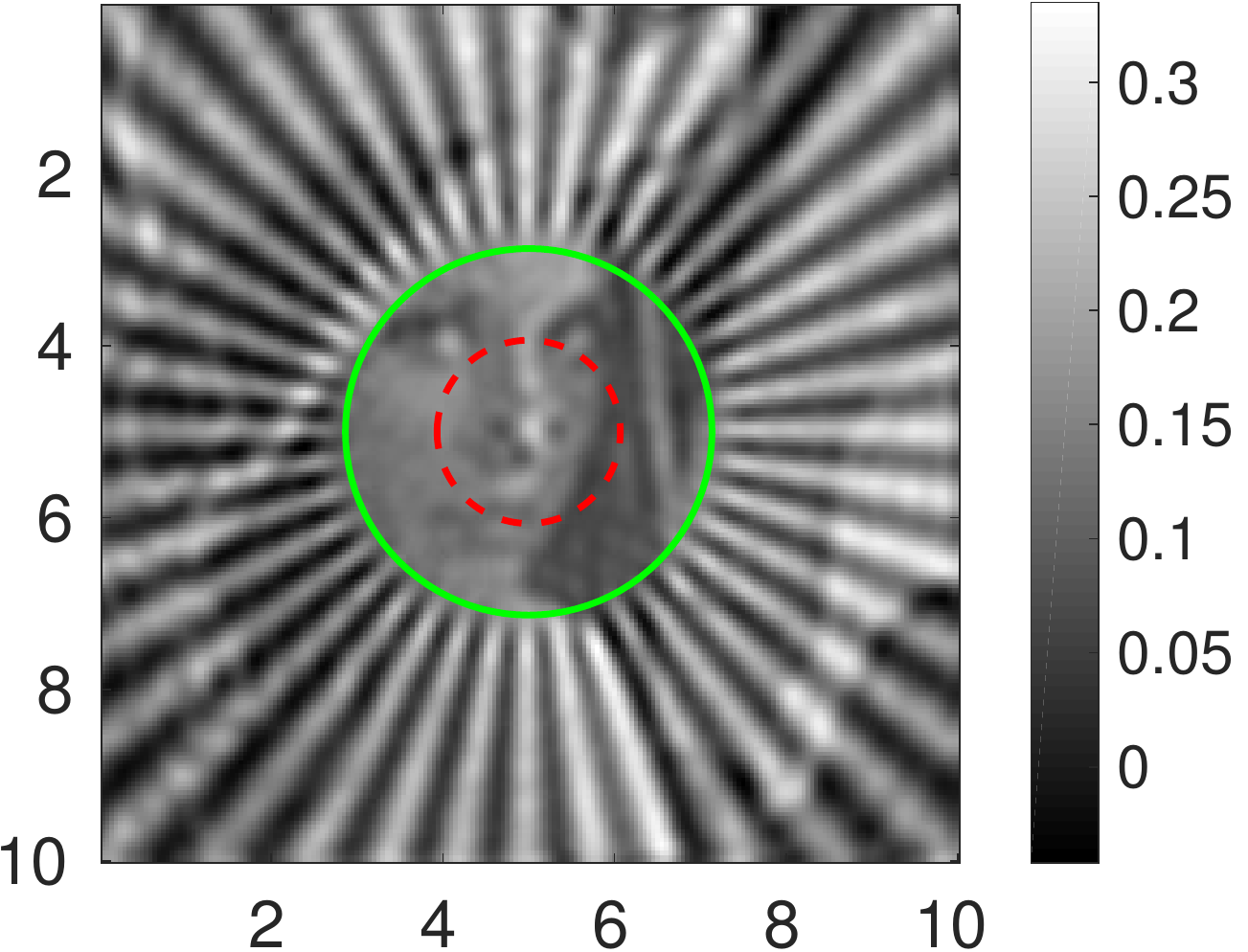}\\
		\small a.) Background  &  b.) Wiener deconvolution of $\bar{\yb}$ \\
		
		\includegraphics[width=\wdth]{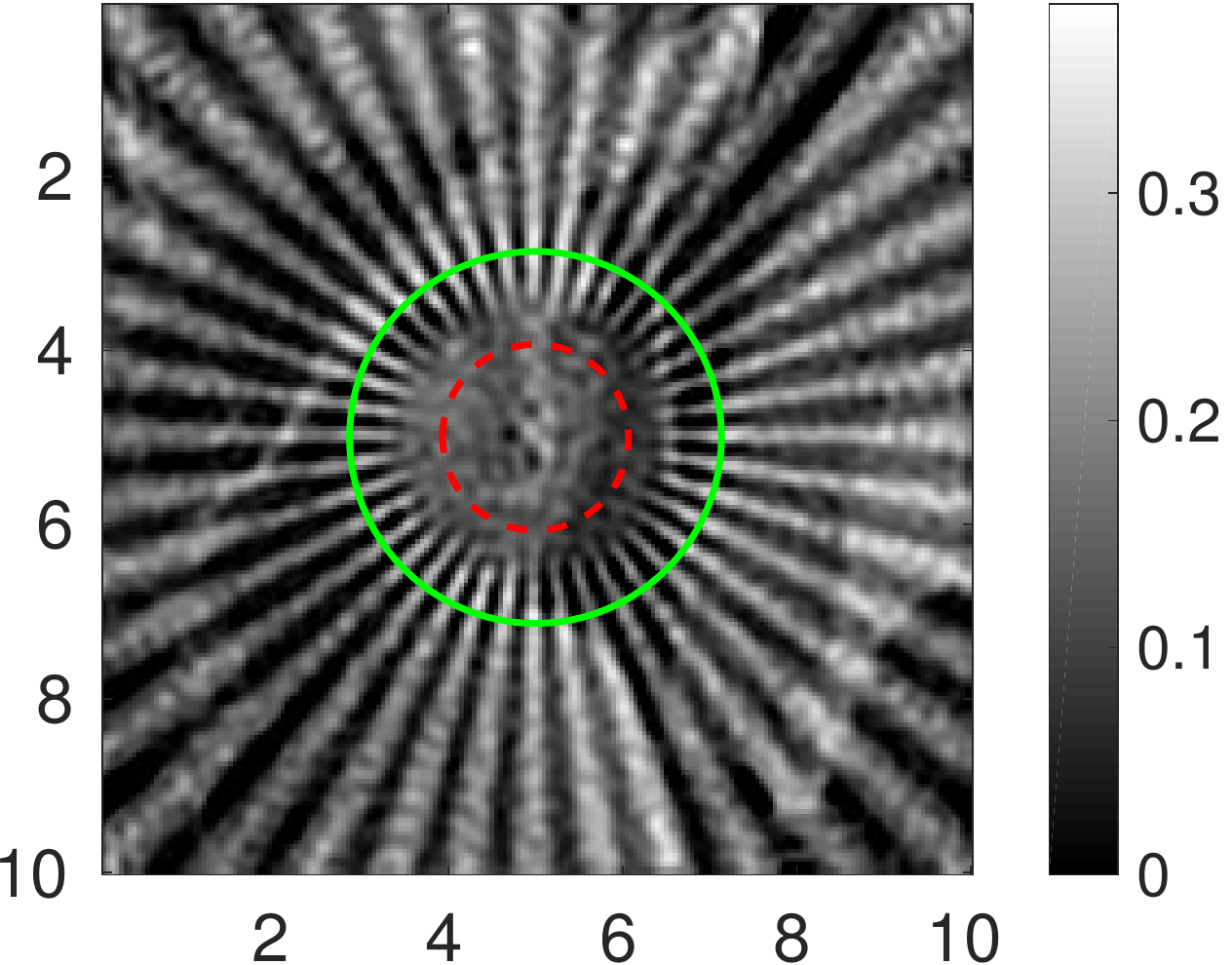}
		& \includegraphics[width=\wdth]{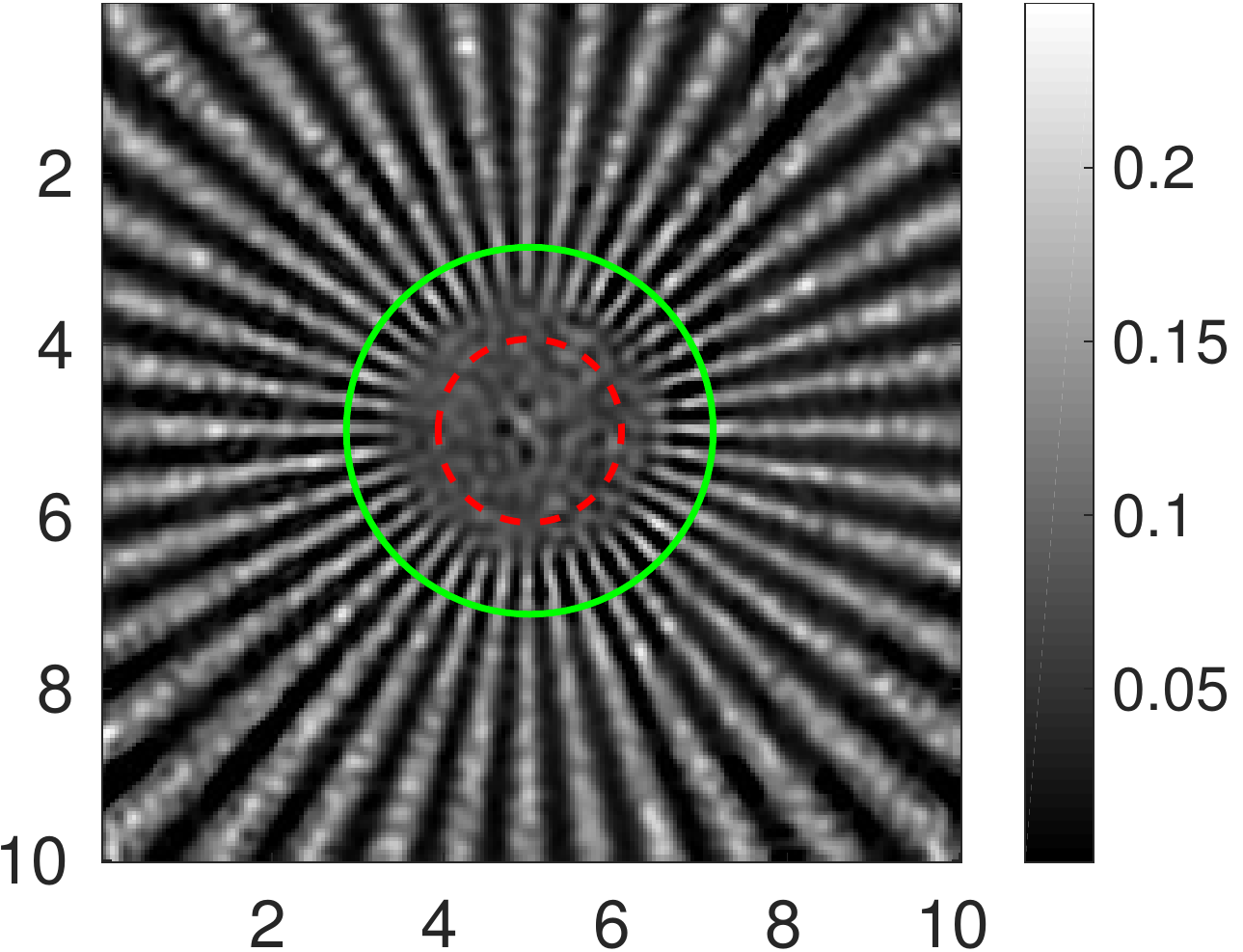}\\

		\small   c.) Mean of $\hat{\qb}_m$ &  d.) Standard deviation of $\hat{\qb}_m$ \\

	\end{tabular}  
	\caption{\textbf{Reconstructed object with 300 speckle patterns and 40dB Gaussian noise with a fixed (lena) background.}}
	
	\label{fig.background} 
\end{figure}

To verify this point, simulation results using 300 speckle patterns and 40dB Gaussian noise with a fixed background (lena) are shown in figure \ref{fig.background}. The images shown in second line are obtained by minimizing the constrained $\ell_{21}$ regularizer. As expected, both the Wiener deconvolution of wide-field image and the mean of $\hat{\qb}_m$ are blurred by the background while the standard deviation of $\hat{\qb}_m$ (Fig. \ref{fig.background}d) are rather clear. 

\subsection{Influence of the hyperparameter}
\begin{figure}[t]
	\centering
	\begin{tabular}{cc}

		\includegraphics[width=\wdth]{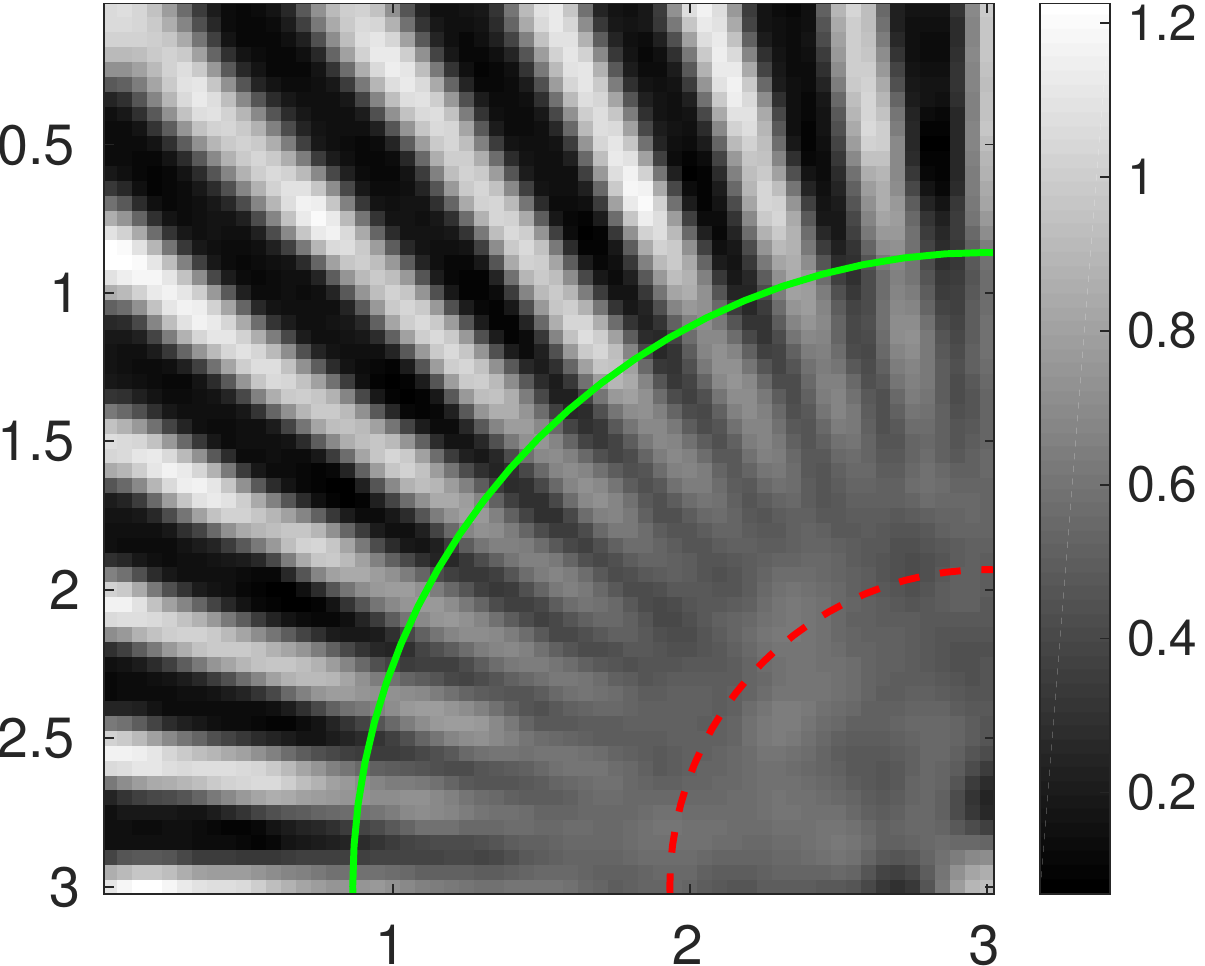}
		& \includegraphics[width=\wdthnar]{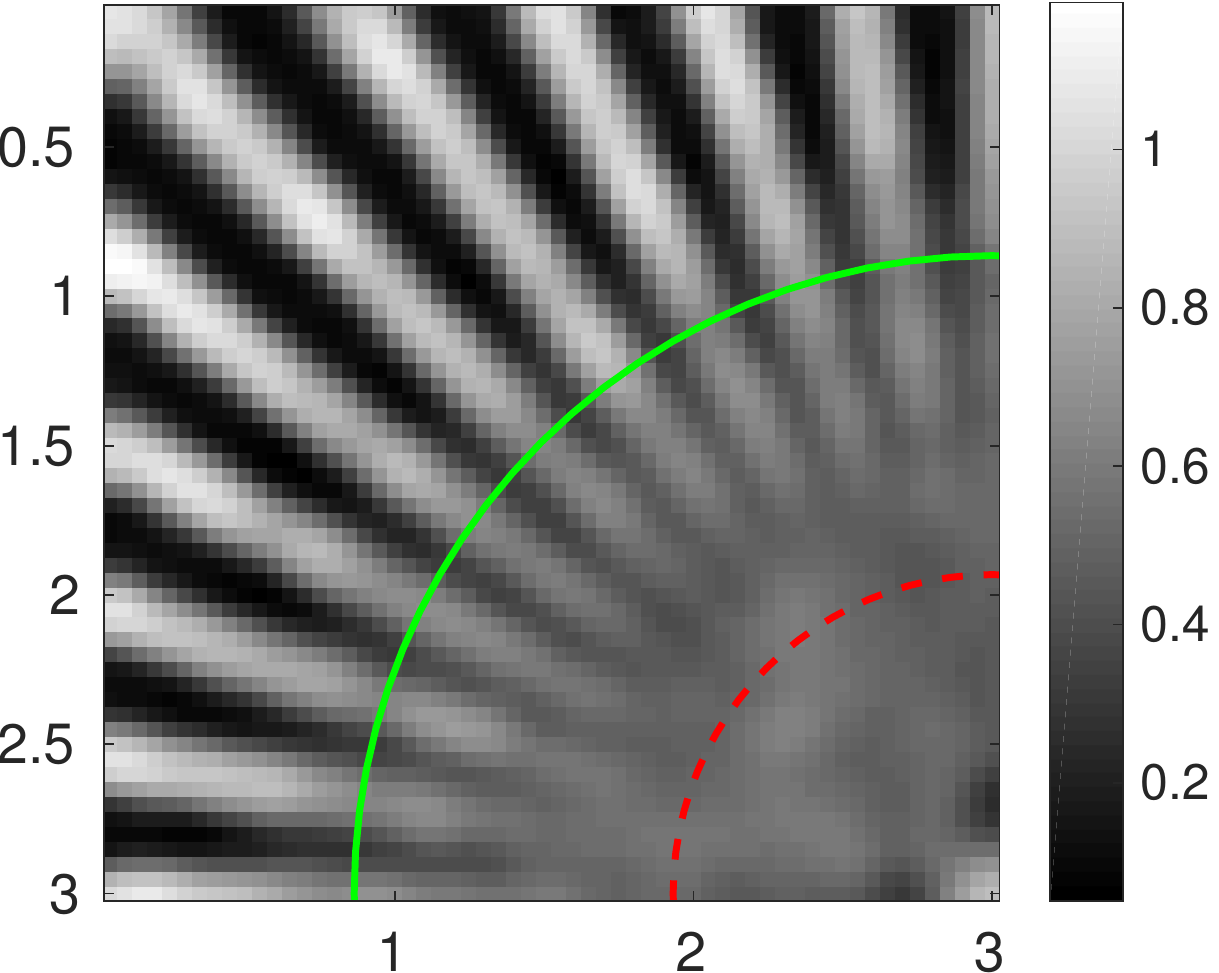}\\
		\small $\ell_{1,1}\; \hat{\qb}_m$ Std  &  $\ell_{1,1}\; \hat{\qb}_m$ Std  \\

		\includegraphics[width=\wdth]{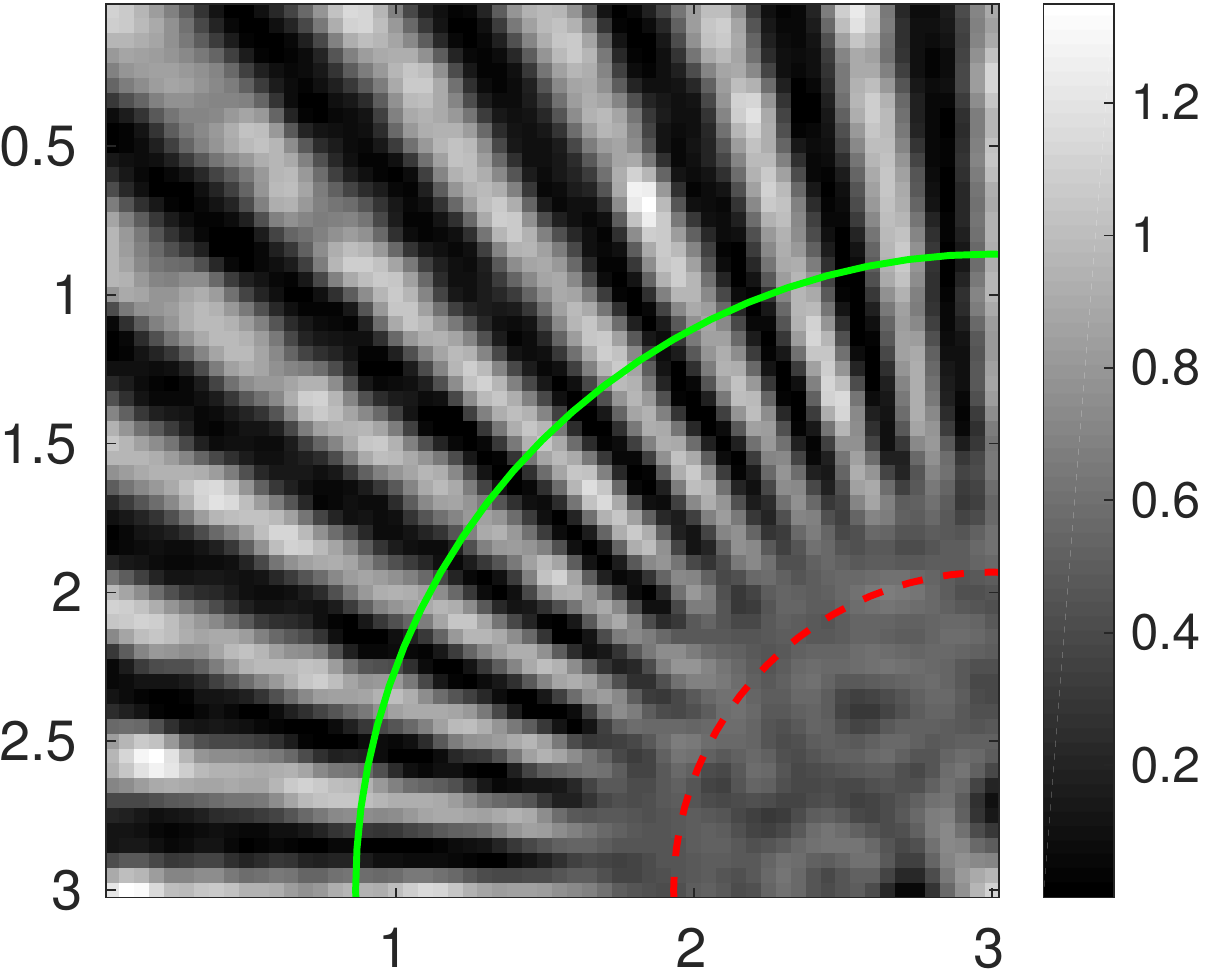}
		& \includegraphics[width=\wdthnar]{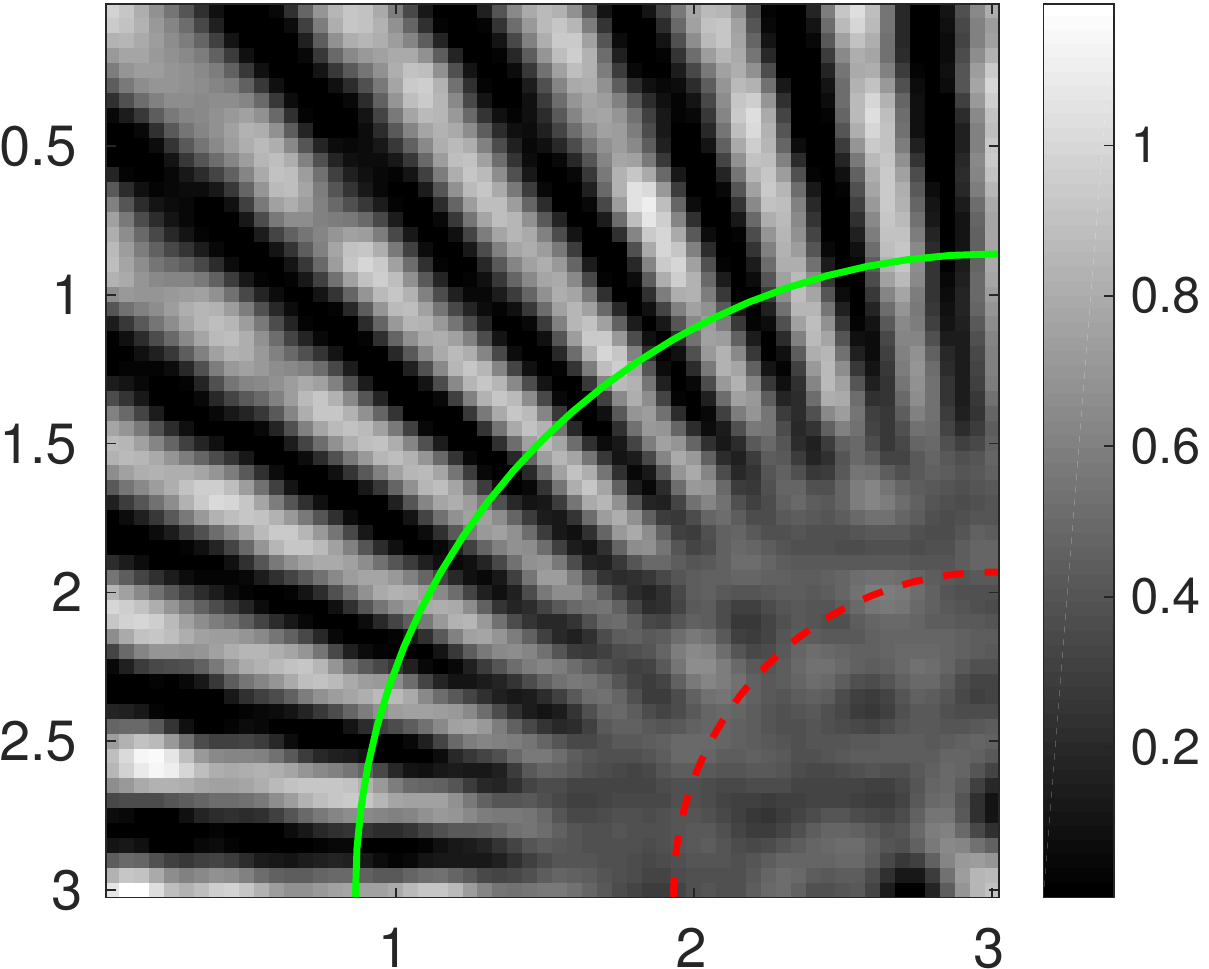}\\
		\small $\ell_{2,1}\; \hat{\qb}_m$ Std &  $\ell_{2,1}\; \hat{\qb}_m$ Std  \\
		\\
		\small col.1) $\xi = 0.2 \xi_{real}$ & col.2)  $\xi = 5 \xi_{real}$
		
	\end{tabular}  
	\caption{\textbf{Reconstructed object with 300 speckle patterns and 40dB Gaussian noise with different hyperparameter $\xi$.}}
	
	\label{fig.speckleSIMHyperparameter} 
\end{figure}

When the TV norm is not considered, the only hyperparameter  $\xi$ in the model, denoting the variance of the additive noise,  is assumed to be known in previous simulations. In this section, its influence on the estimator is explored when it is not correctly set. The reconstruction results with 300 speckle patterns and 40dB white noise using different $\xi$ value are shown in Fig. \ref{fig.speckleSIMHyperparameter}. When $\xi$ is much lower than its true value as shown in the first column of  Fig. \ref{fig.speckleSIMHyperparameter} ($\xi = \frac{1}{5} \xi_{\text{real}}$), no evident visual differences are observed in comparison to the reconstructions in Fig. \ref{fig.lpqnorm2}(d,f). When $\xi$ equal 5 times its true value, we lost partial super-resolution in comparison to the situation when it is correctly set. 

\subsection{SR under different frequency support of speckle patterns}
\begin{figure}[htb]
	\centering
	\begin{tabular}{cc}
\includegraphics[width=\wdth]{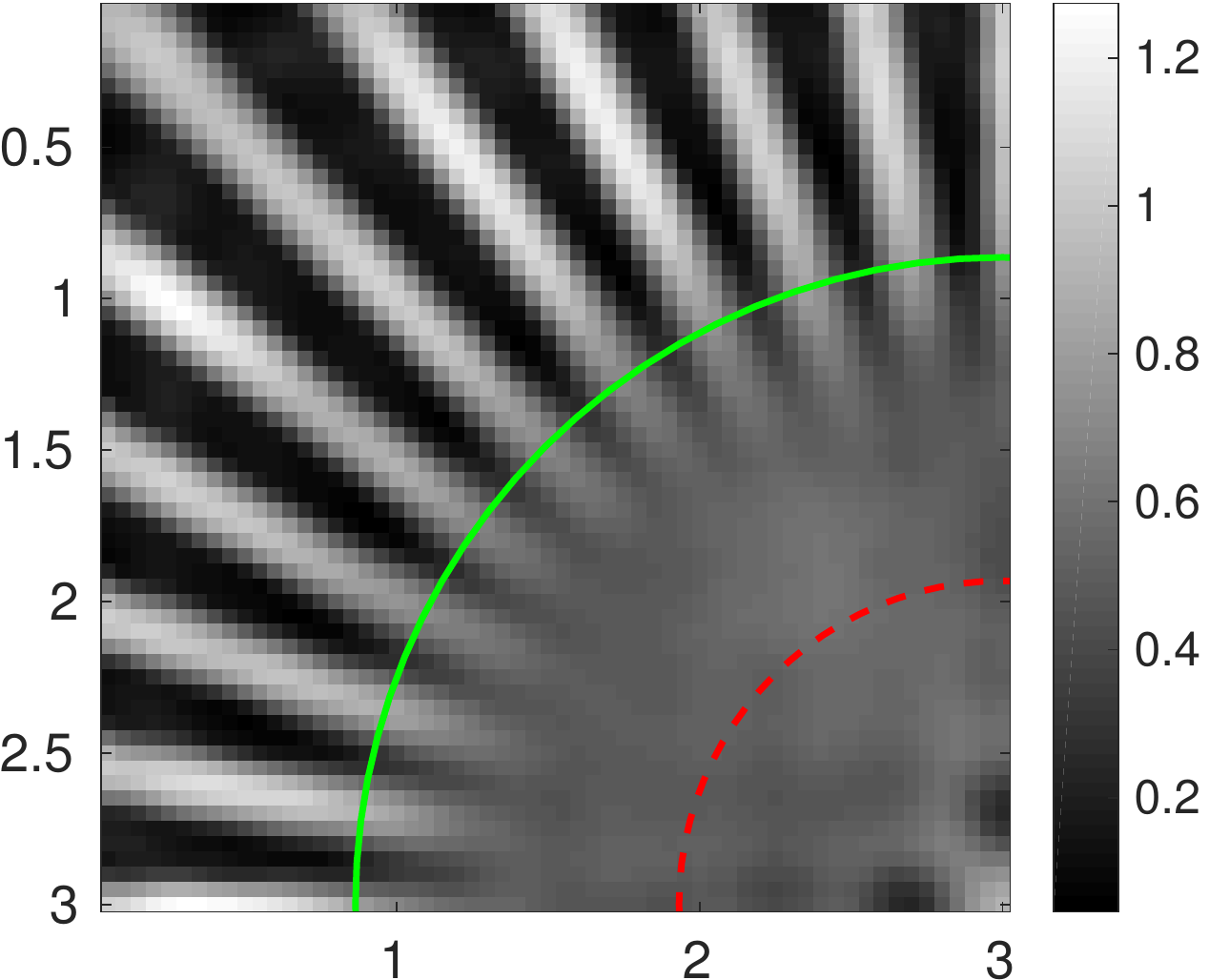}
& \includegraphics[width=\wdth]{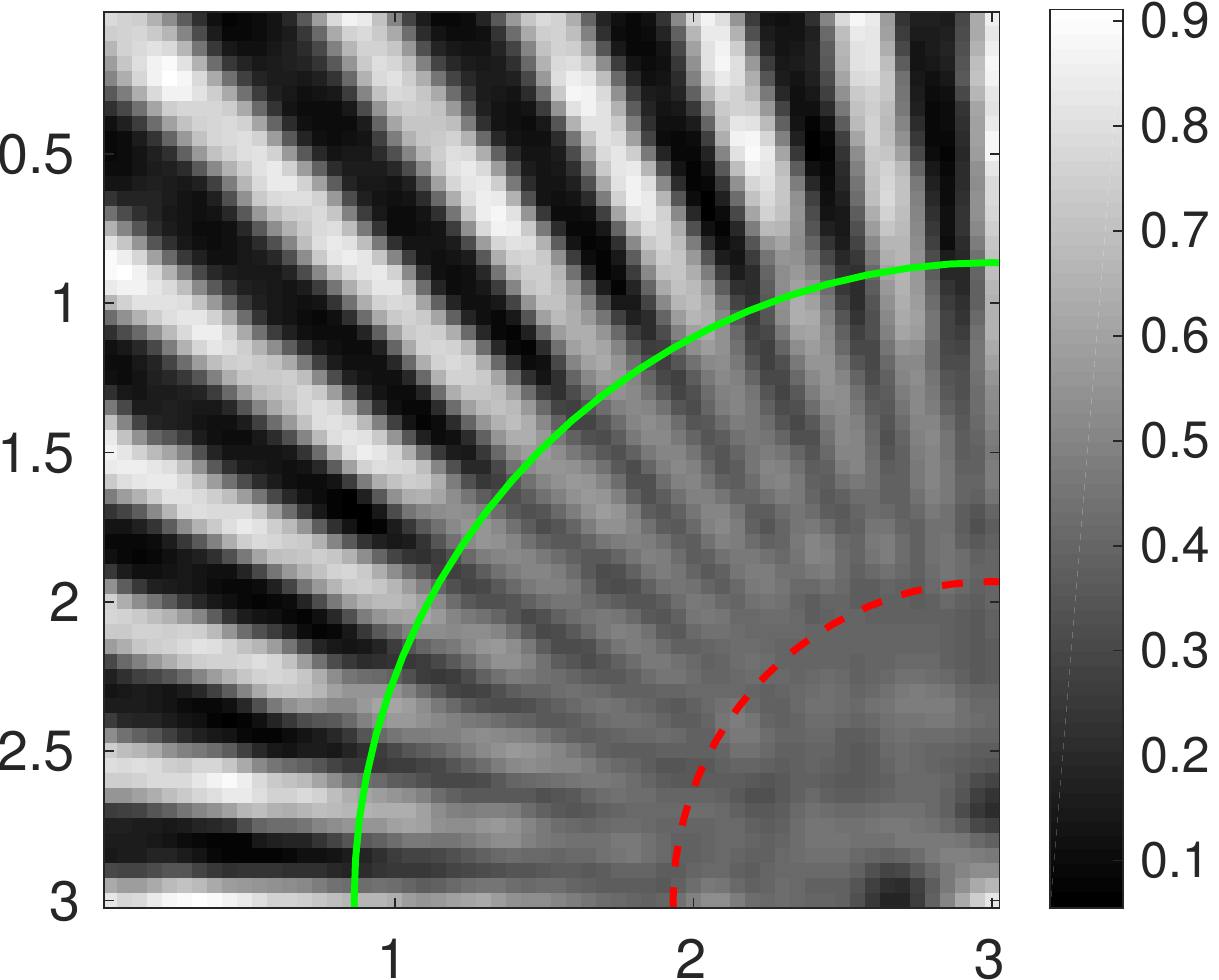}\\
\small $\ell_{1,1}\; \hat{\qb}_m$ Std  &  $\ell_{1,1}\; \hat{\qb}_m$ Std  \\

          \includegraphics[width=\wdth]{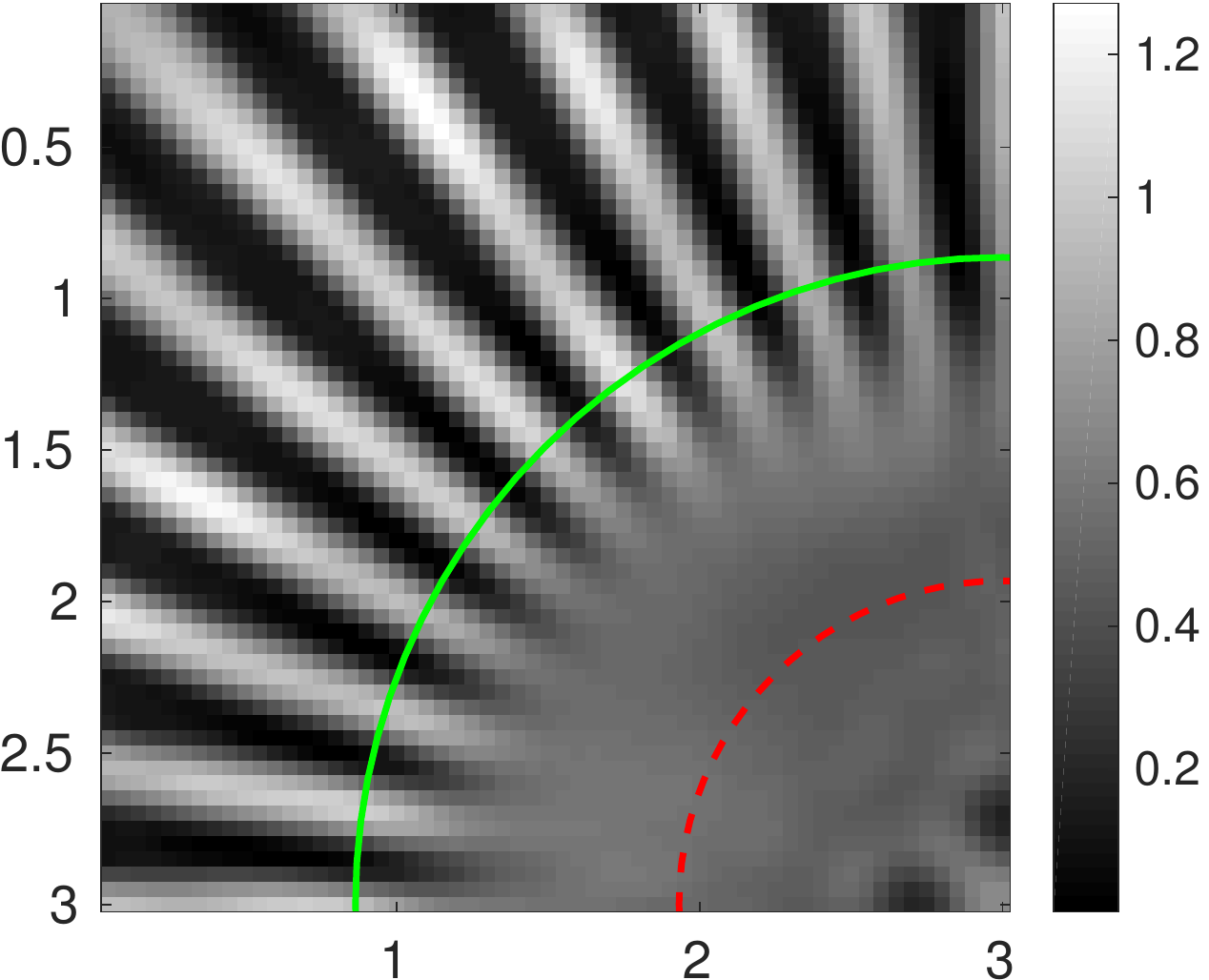}
           & \includegraphics[width=\wdth]{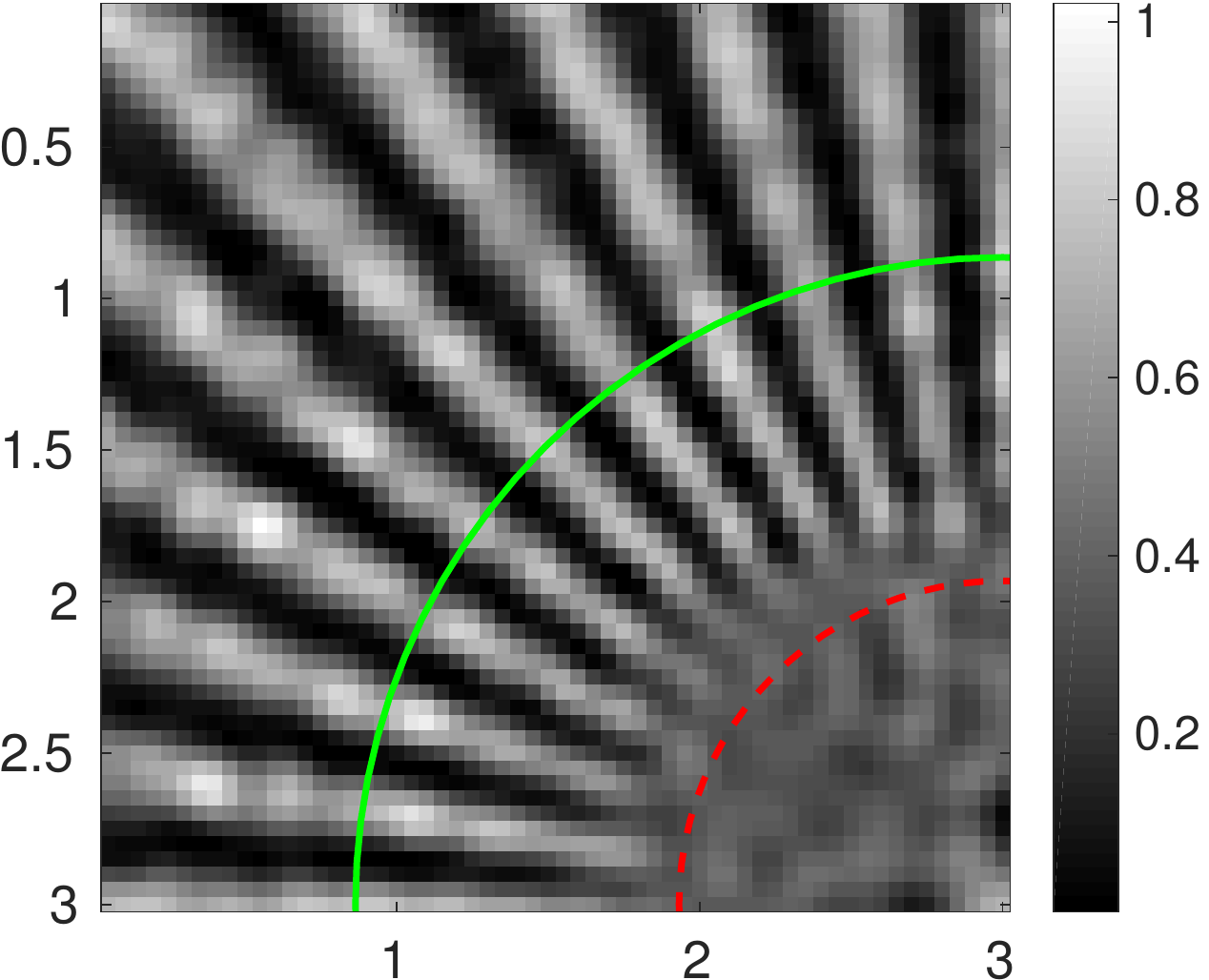}\\
\small $\ell_{2,1}\; \hat{\qb}_m$ Std  &  $\ell_{2,1}\; \hat{\qb}_m$ Std  \\

\\
\small col.1) $\text{NA}_{\text{ill}} = 0.5\text{NA}$ & col.2)  $\text{NA}_{\text{ill}} = 2\text{NA}$

	\end{tabular}  
	\caption{\textbf{Reconstructed object with 300 speckle patterns and 40dB Gaussian noise with different frequency support of speckle patterns.}}
	
	\label{fig.speckleSIMNonlinear} 
\end{figure}

In this section the influence by Fourier support of speckle on the super-resolution capacity is explored. The reconstruction results of the proposed method with speckle generated with different numerical aperture ($\text{NA}_{\text{ill}}$) under 300 illumination and 40dB SNR are shown in Fig. \ref{fig.speckleSIMNonlinear}. When the support of power spectral density of the speckle patterns becomes smaller, we lost partial super-resolution, as shown in the first column in Fig. \ref{fig.speckleSIMNonlinear}. With the enlarged support of speckle spectral density, the $\ell_{1,1}$ regularizer retrieves better super-resolution information in comparison to Fig. \ref{fig.lpqnorm2}d.

\begin{figure}[htb]
	\centering
	\begin{tabular}{cc}		
		\includegraphics[width=\wdth]{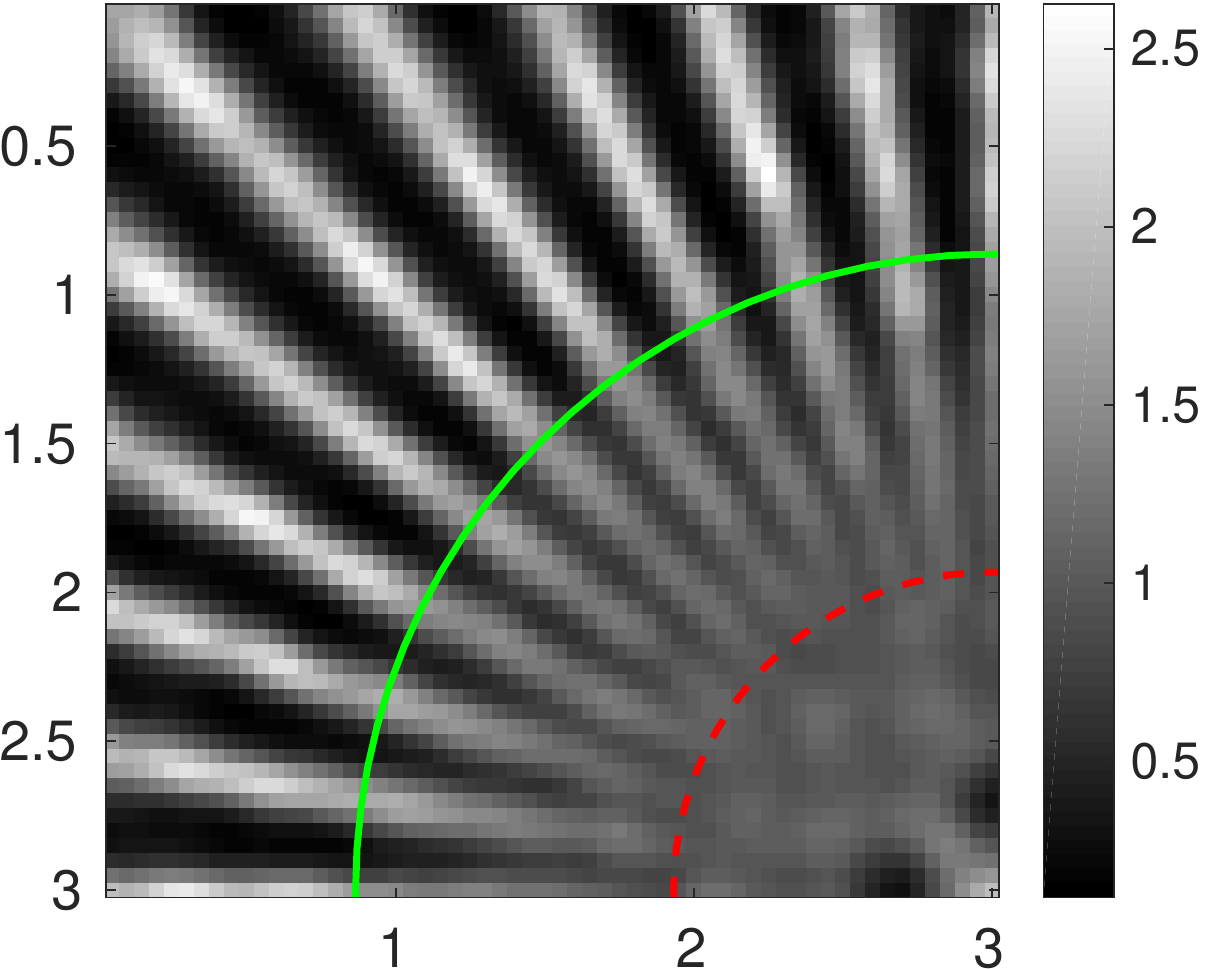}
		& \includegraphics[width=\wdthnar]{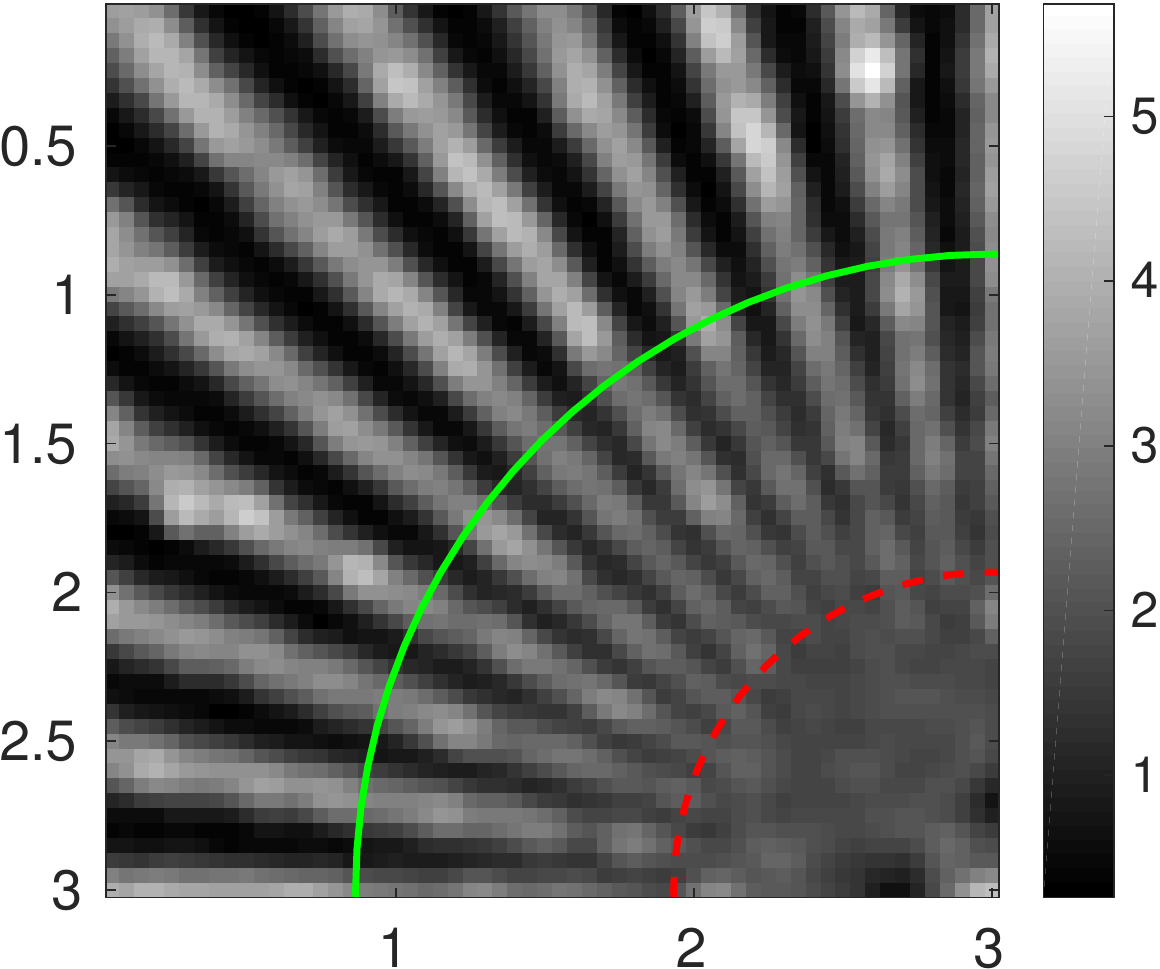}\\
		\small a.) $\ell_{1,1}\; \hat{\qb}_m$ mean &  b.) $\ell_{1,1}\; \hat{\qb}_m$ std  \\

		\includegraphics[width=\wdth]{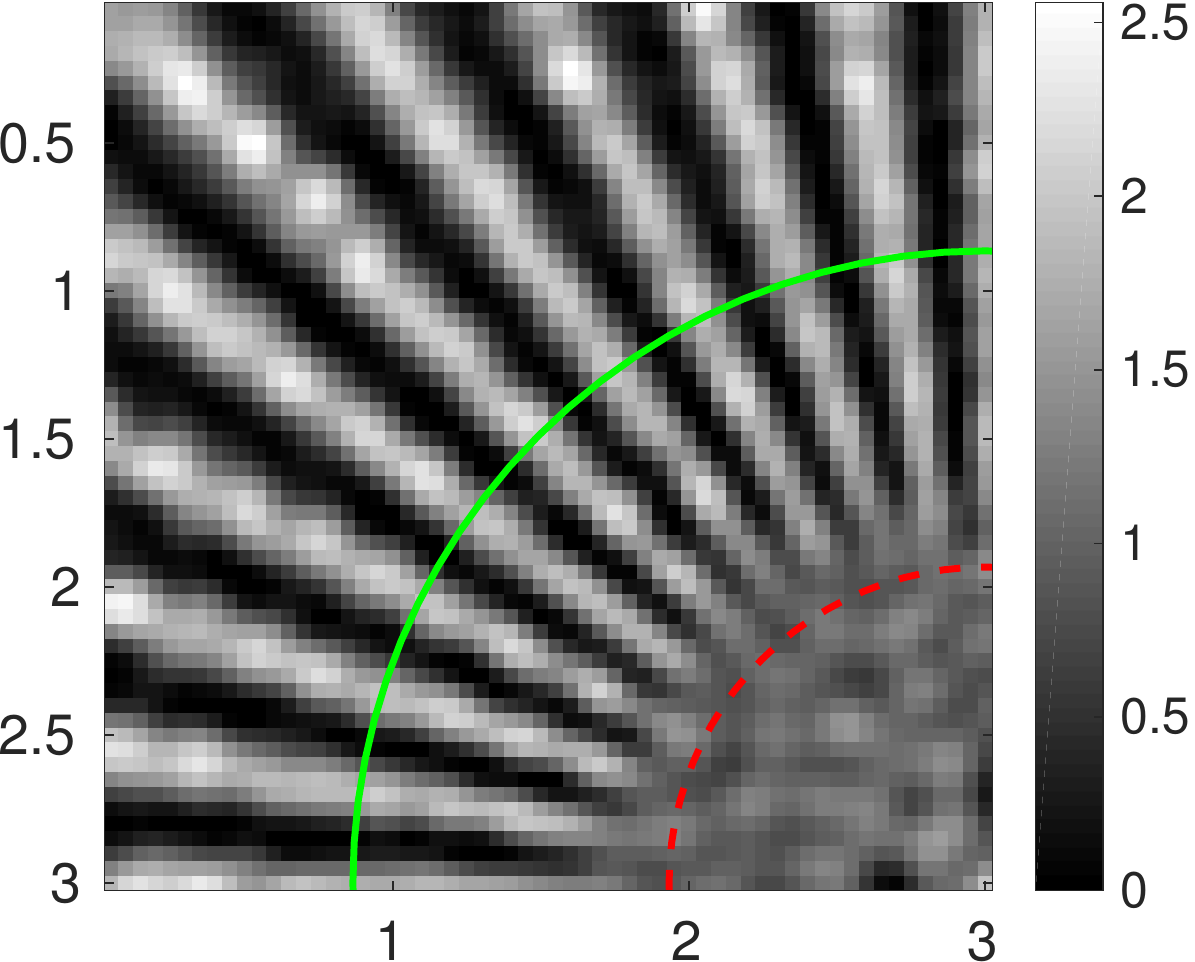}
		& \includegraphics[width=\wdth]{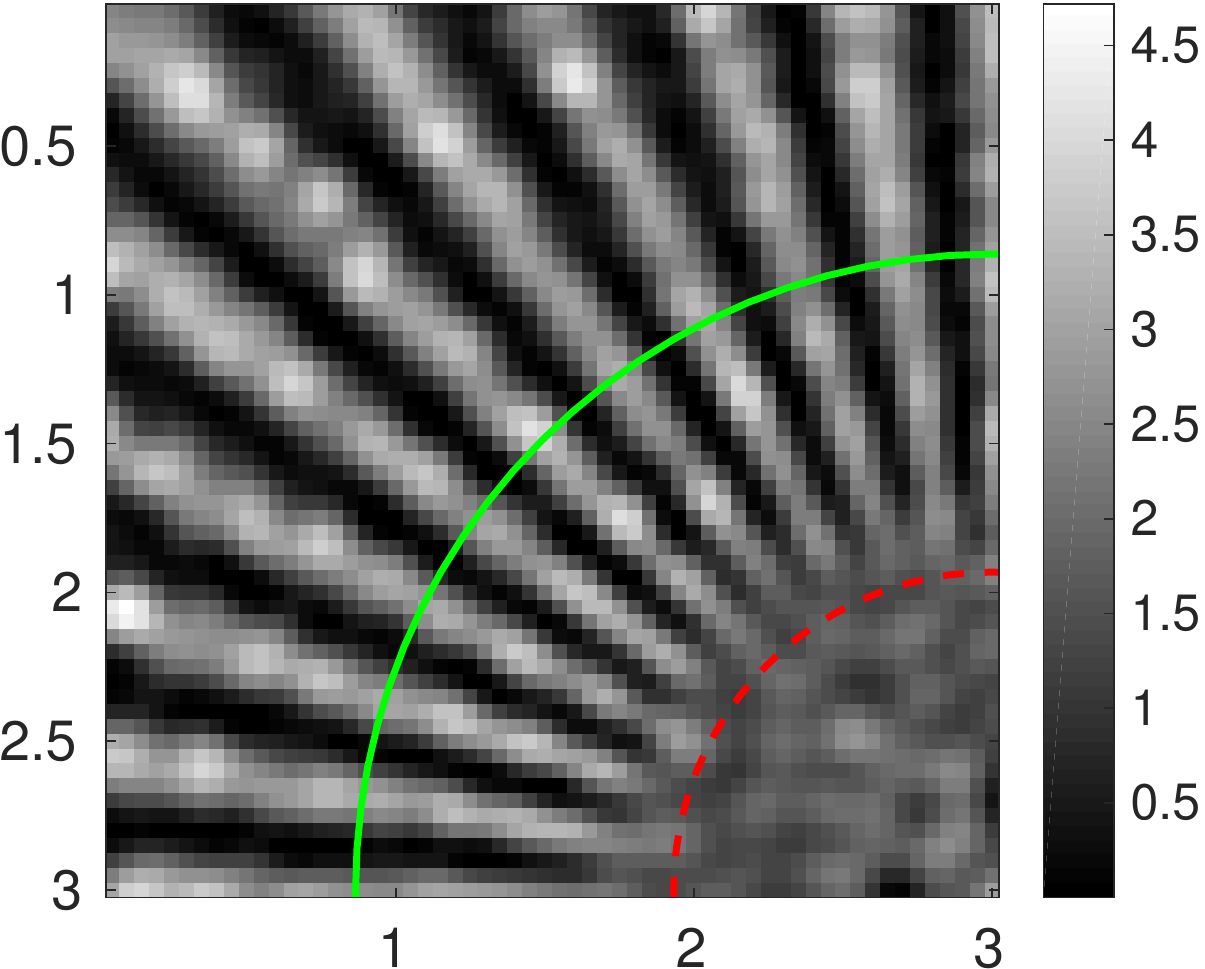}\\

		\small   c.) $\ell_{2,1}\; \hat{\qb}_m$ mean &  d.) $\ell_{2,1}\; \hat{\qb}_m$ std \\
		
	\end{tabular}  
	\caption{\textbf{Reconstructed images with  1000 squared speckle patterns where $\text{NA}_\text{ill} =2\text{NA}$ under 40dB Gaussian noise .}}
	
	\label{fig.2NA2Photon} 
\end{figure}

It has been reported in reference \cite{labouesse2017joint} that the sparsity of illumination play a pivotal role in blind-SIM technique. Simulations with ``squared'' speckle patterns, which are sparser than the ``standard'' speckle \cite{Negash:18}, are shown in Fig. \ref{fig.2NA2Photon}. The contrast in super-resolution part becomes better using $\ell_{11}$ regularizer term under ``squared'' speckle cases in comparison to ``standard'' speckle, while super-resolution information beyond a factor of two is still inaccessible.

\subsection{Resolution under Poisson noise}
In the previous image formation model in Eq. \eqref{eq.SIM}, the shot noise of CCD caused by the random arrival of photons is neglected. For a given photon, the probability of its arrival within a given time period is governed by Poisson distribution.

\begin{figure}[t]
	\centering
	\begin{tabular}{cc}
		\includegraphics[width=\wdth]{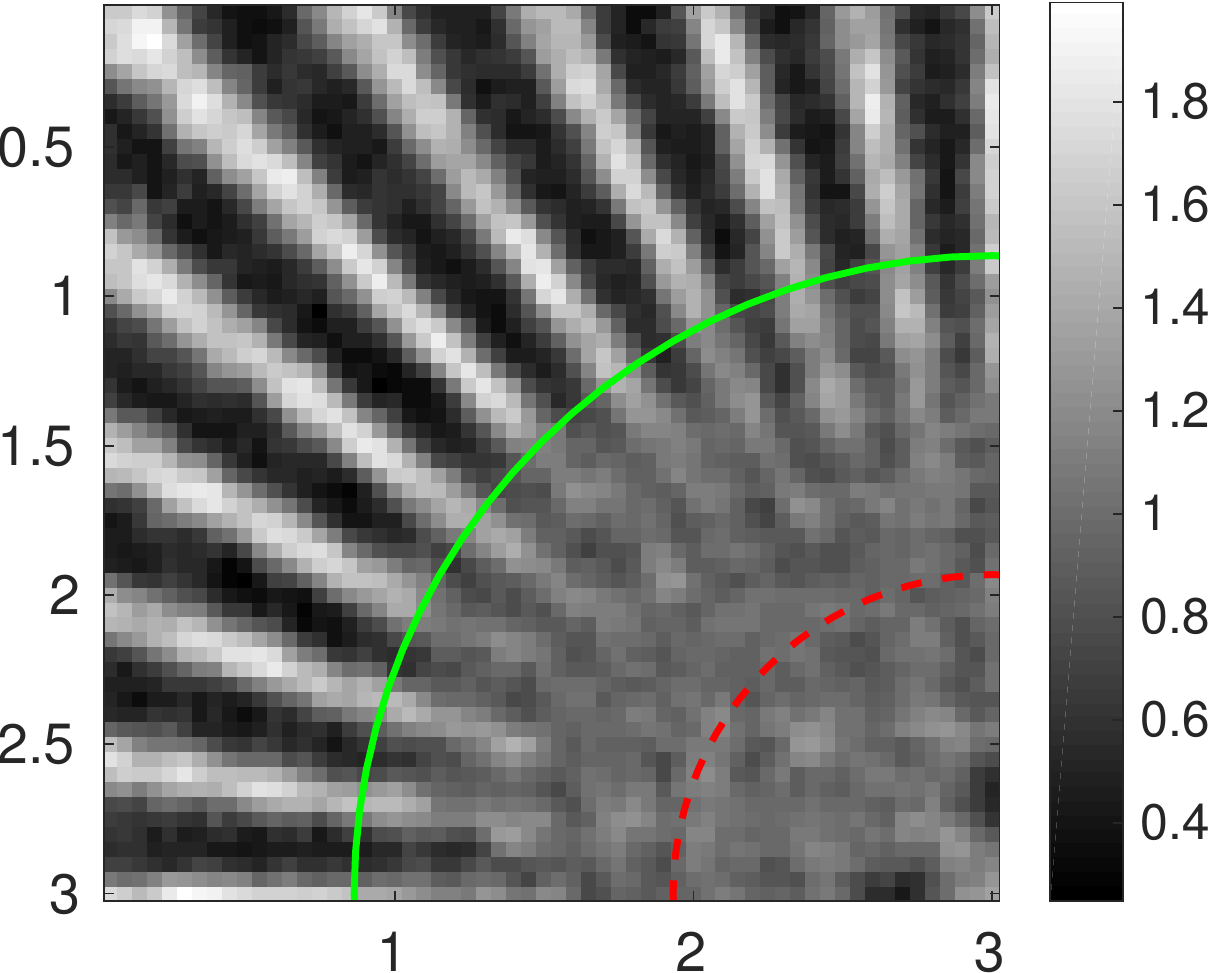}
                &\includegraphics[width=\wdth]{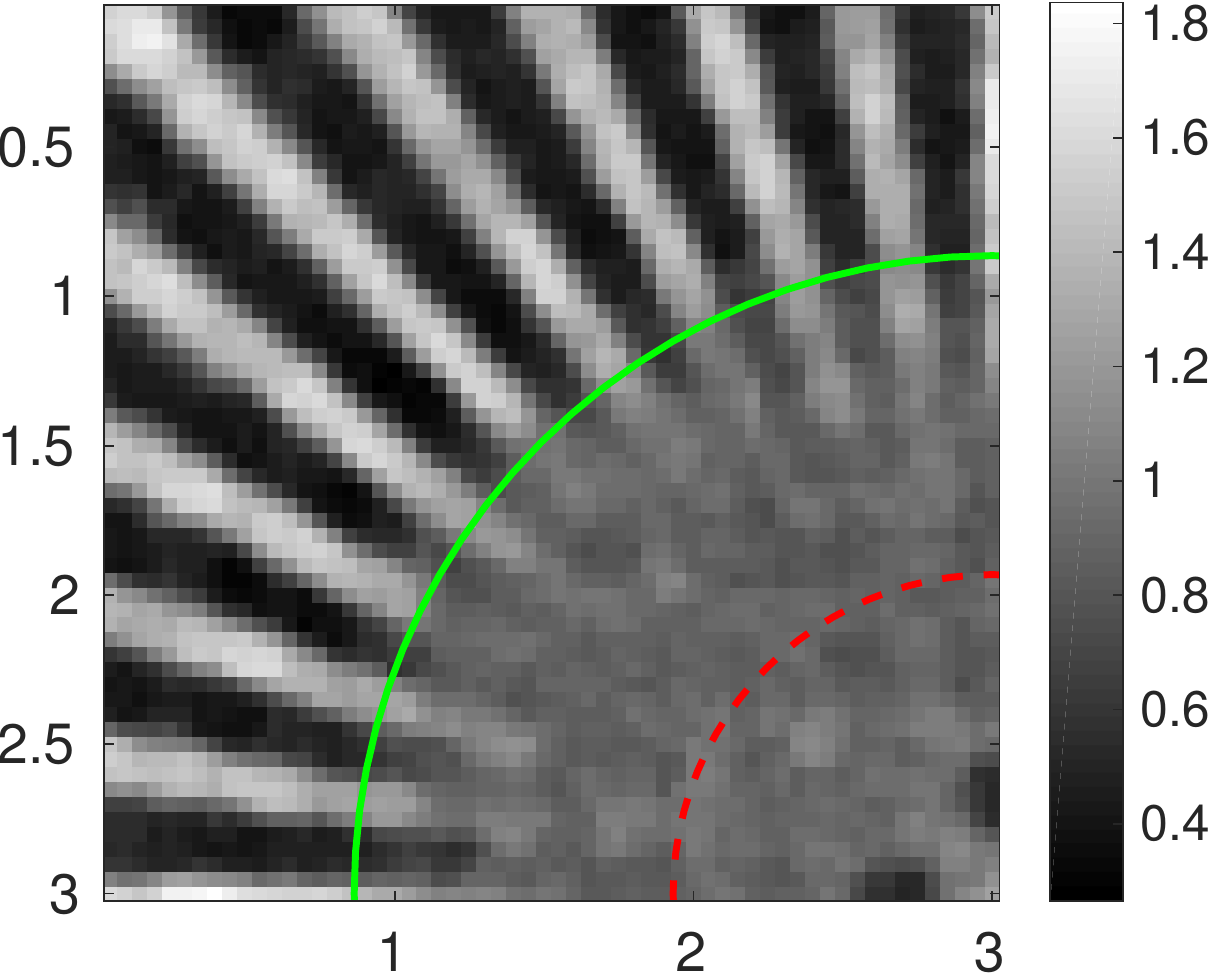}\\
\small $\ell_{1,1}\; \hat{\qb}_m$ Std  &  $\ell_{1,1}\; \hat{\qb}_m$ Std  \\
\includegraphics[width=\wdth]{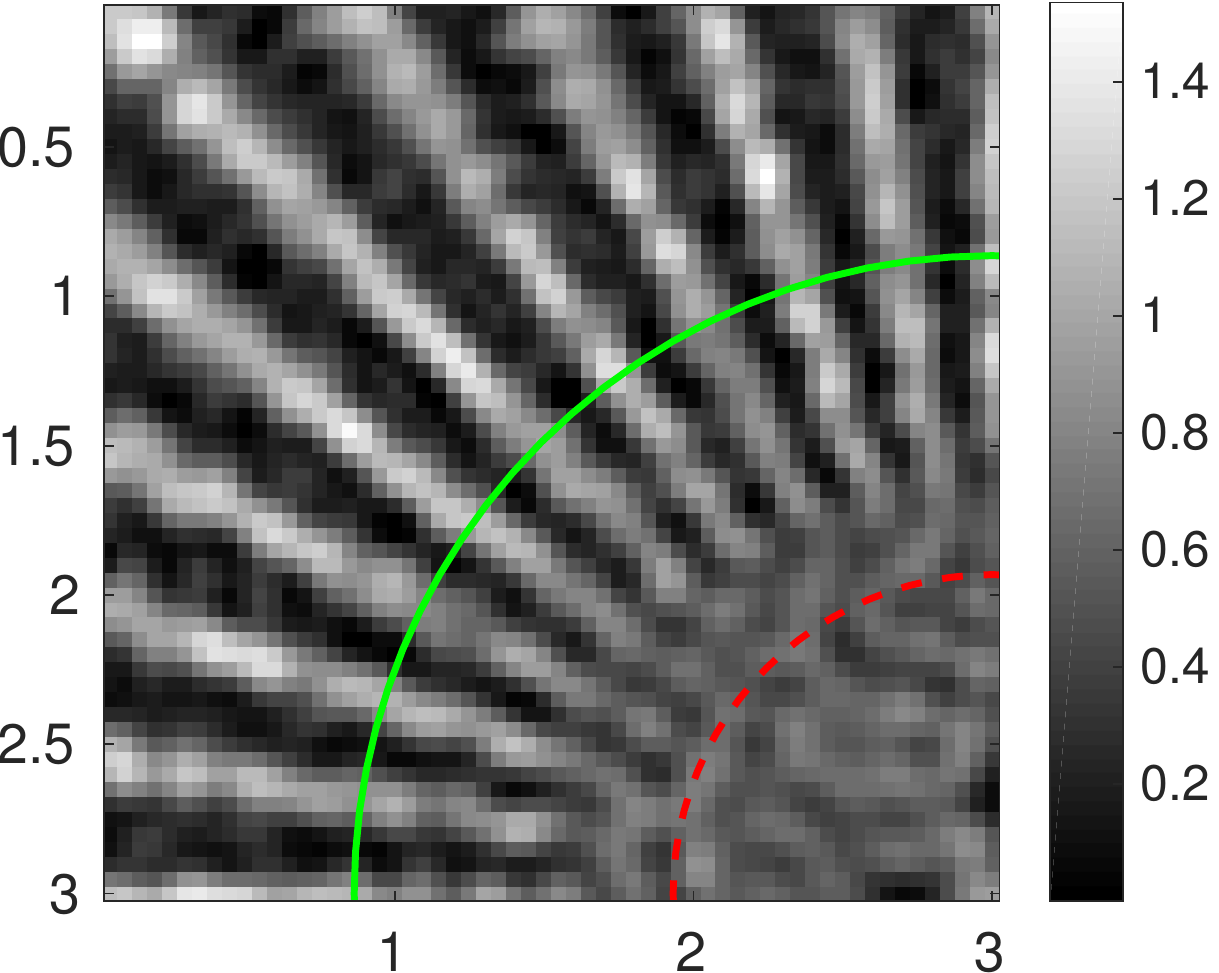}
&\includegraphics[width=\wdth]{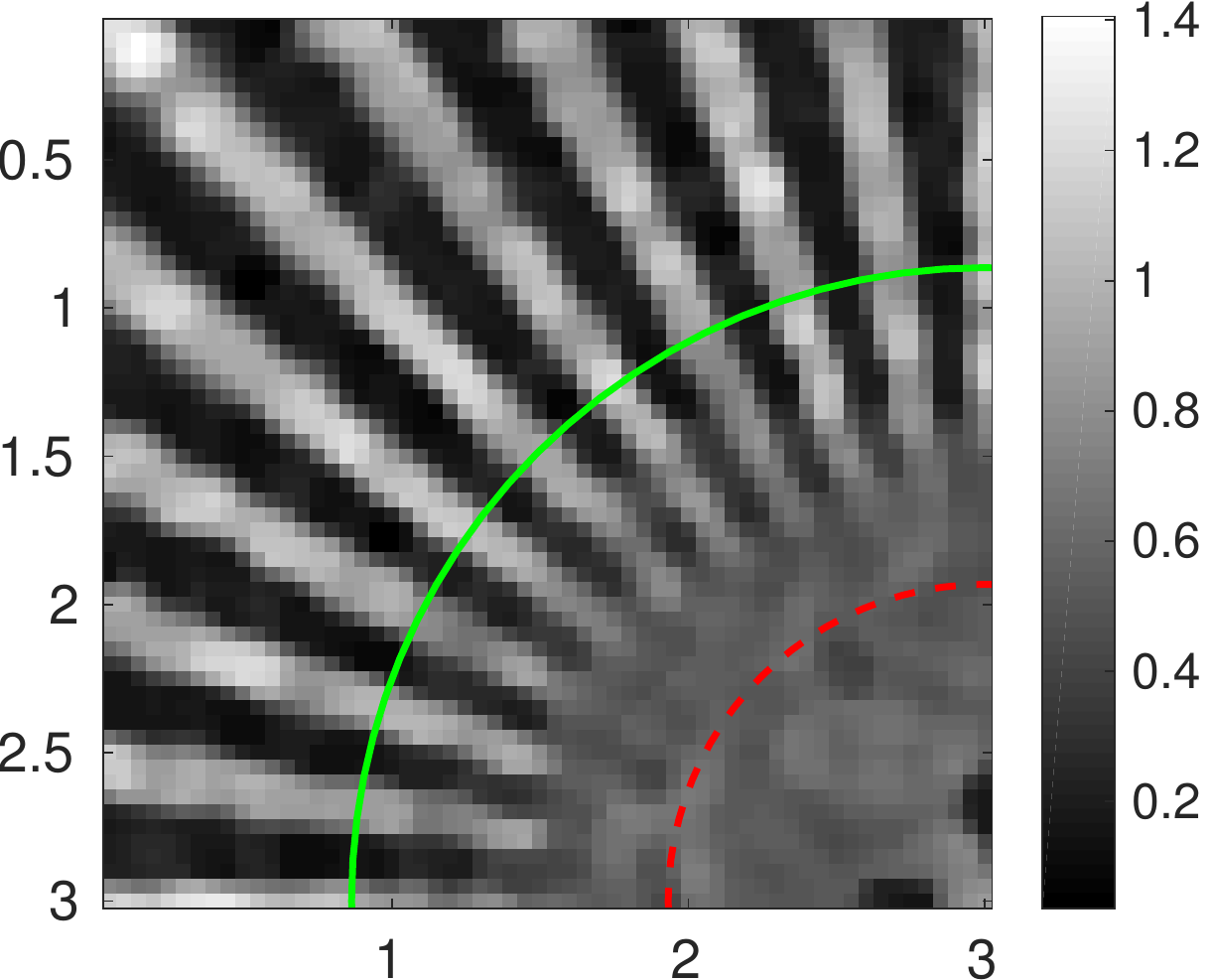}\\
\small $\ell_{2,1}\; \hat{\qb}_m$ Std  &  $\ell_{2,1}\; \hat{\qb}_m$ Std  \\
	 
\\
		
		\small (col 1.)  $\mu=0$  &  (col 2.) $\mu = 0.3$  \\

\end{tabular}  
\caption{\textbf{Reconstructed object with 300 speckle patterns under mixture of Poisson and Gaussian noise.} The SNR of Gaussian noise is 15dB and the number of photons per pixel per measurement is set to 100.}

\label{fig.lpqnormTVMixtureNoise} 
\end{figure}

The reconstruction results with mixture of Poisson and Gaussian noise using 300 speckle patterns are presented in Figure \ref{fig.lpqnormTVMixtureNoise}. The results shown in the first column are obtained by only considering $\ell_{p,q}$ norm regularizer while the images shown in the second column are obtained using $\ell_{p,q}$ norm regularizer plus the TV regularizer, with the hyperparameters $\mu=0\ldotp 3$. Strong degradation in super-resolution part are viewed by $\ell_{1,1}$ regularizer term 
and after introducing TV norm regularizer, the reconstructed images become smoother, especially in the low resolution part. 

\subsection{Simulations with more complex object}
\begin{figure*}[t]
	\centering
	\begin{tabular}{c@{\kern1pt}c@{\kern1pt}c}
		
			\includegraphics[height=0.33\textwidth] {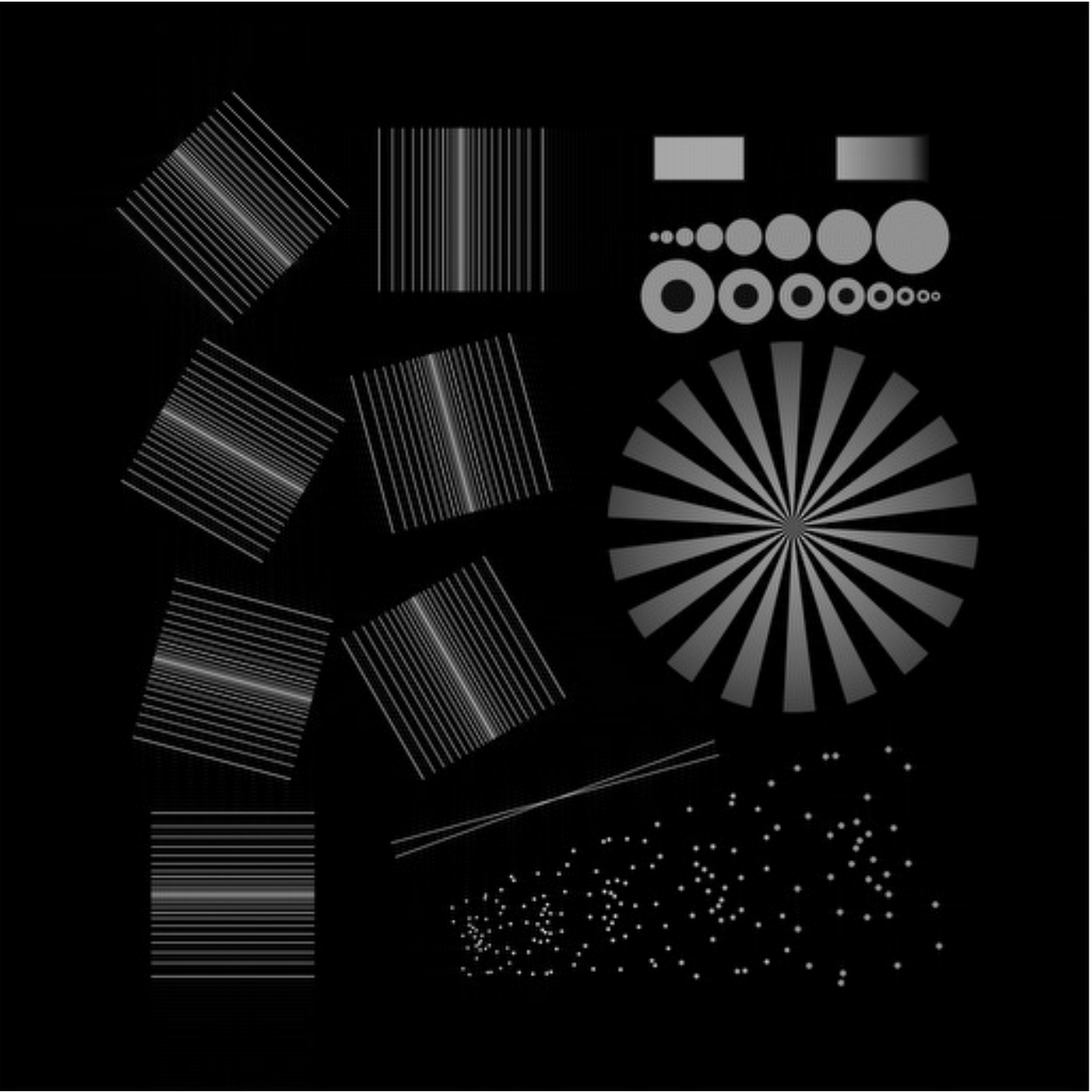}
		&
			\includegraphics[height=0.33\textwidth] {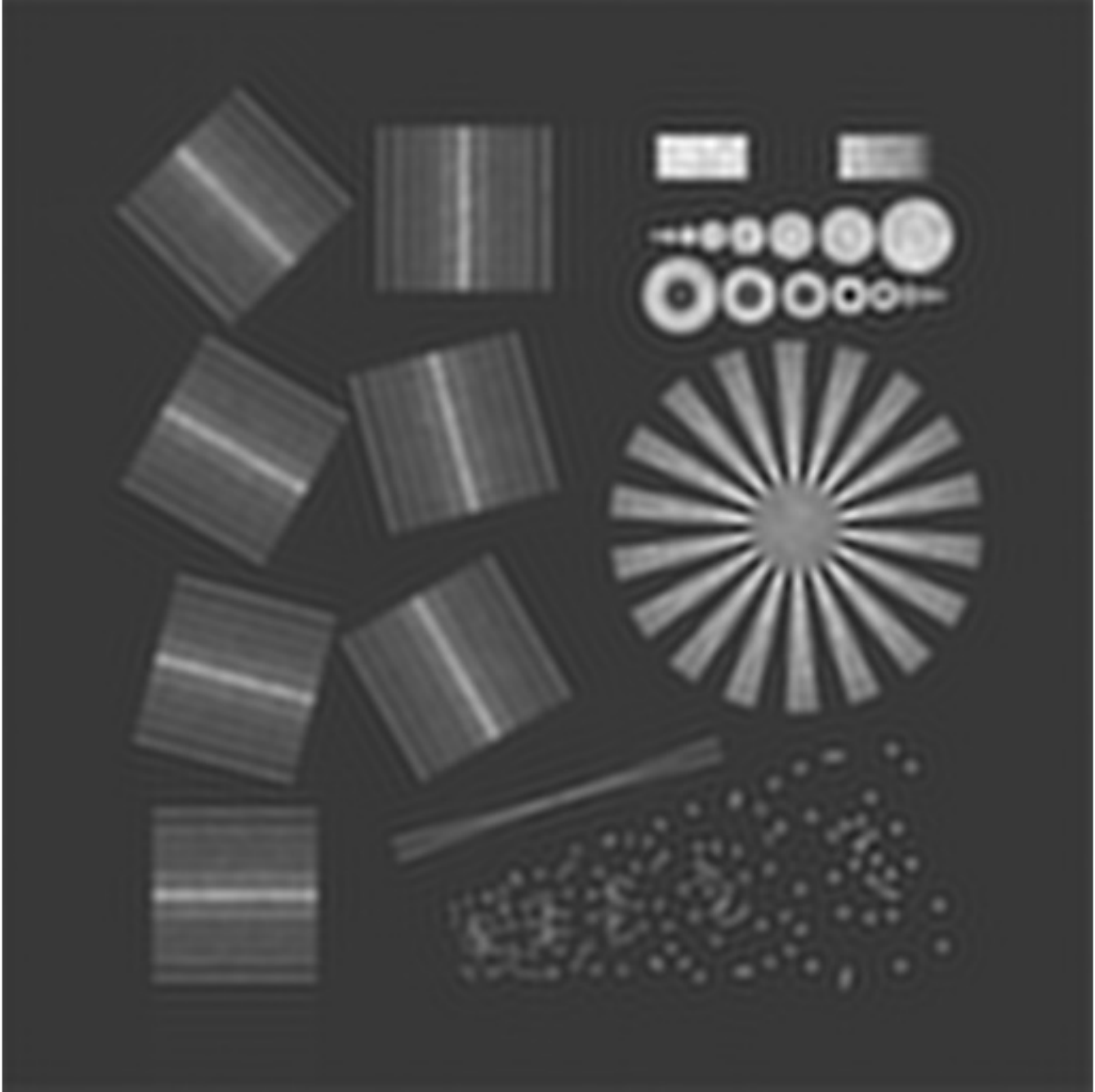}
		&	\includegraphics[height=0.33\textwidth]{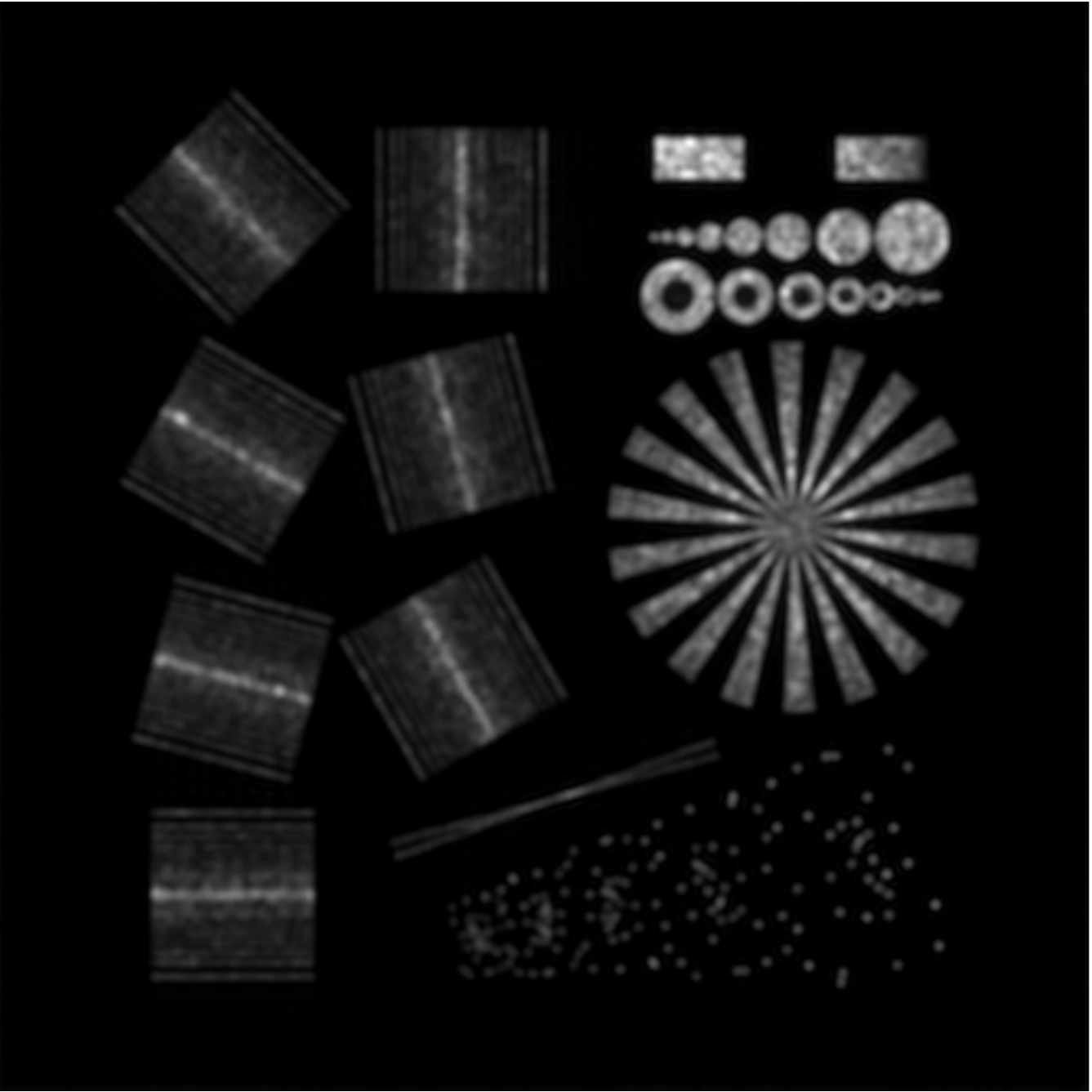}\\
		\small  a.) The true object  &   b.) Wiener deconvolution of $\bar{\yb}$ & c.) $\ell_{2,1}\; \hat{\qb}_m$ Std\\
\\		
		\includegraphics[height=0.33\textwidth]{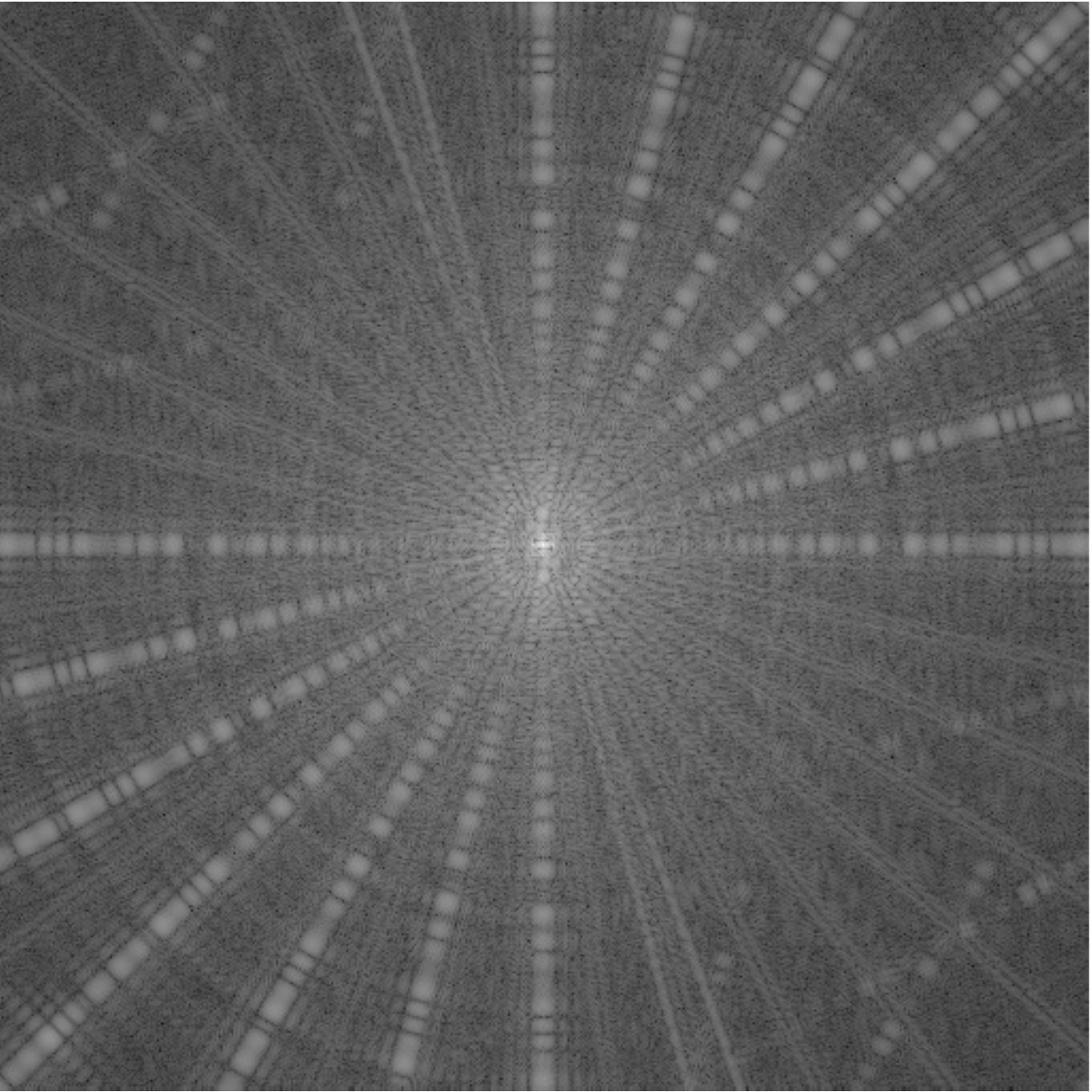}
		&\includegraphics[height=0.33\textwidth]{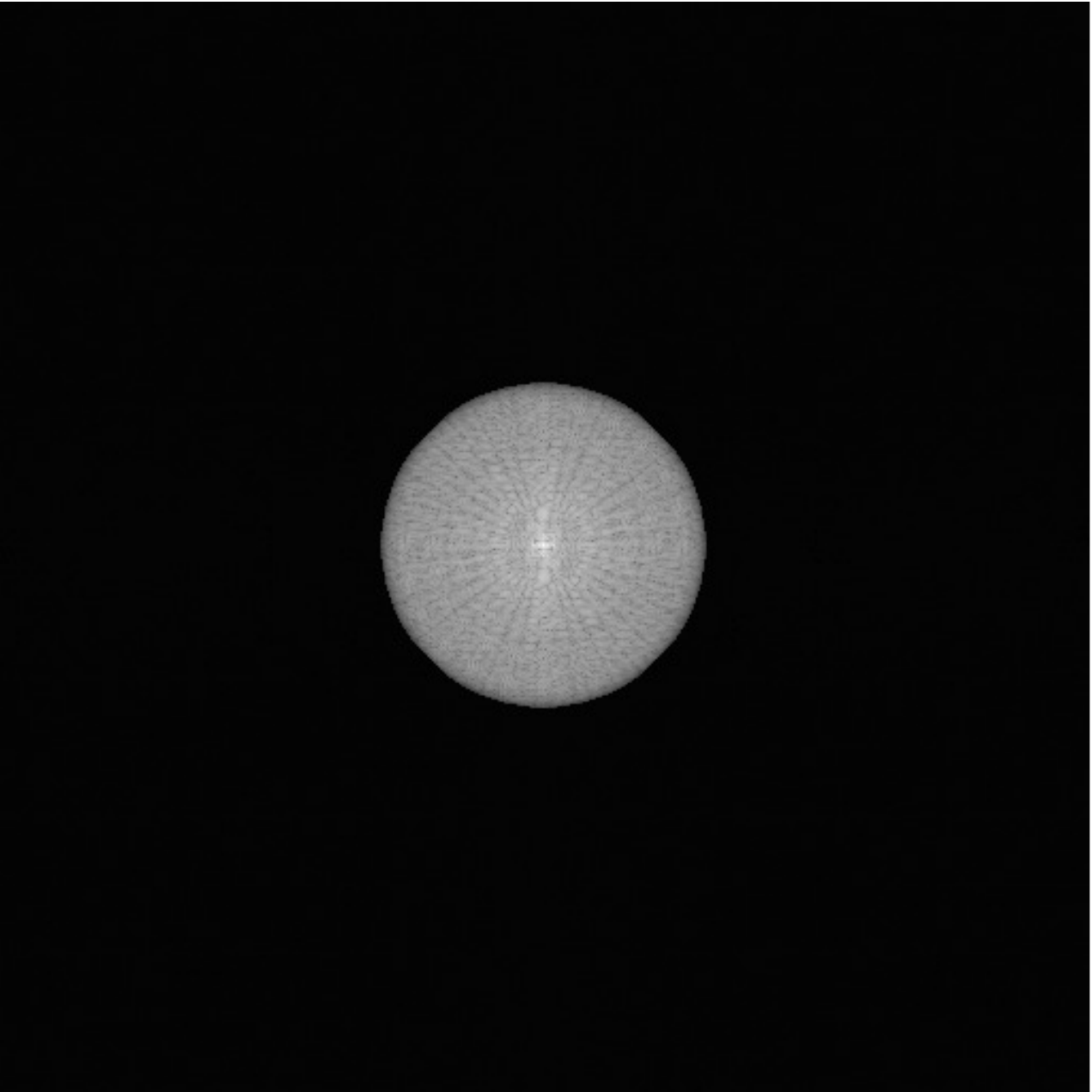}
		&\includegraphics[height=0.33\textwidth]{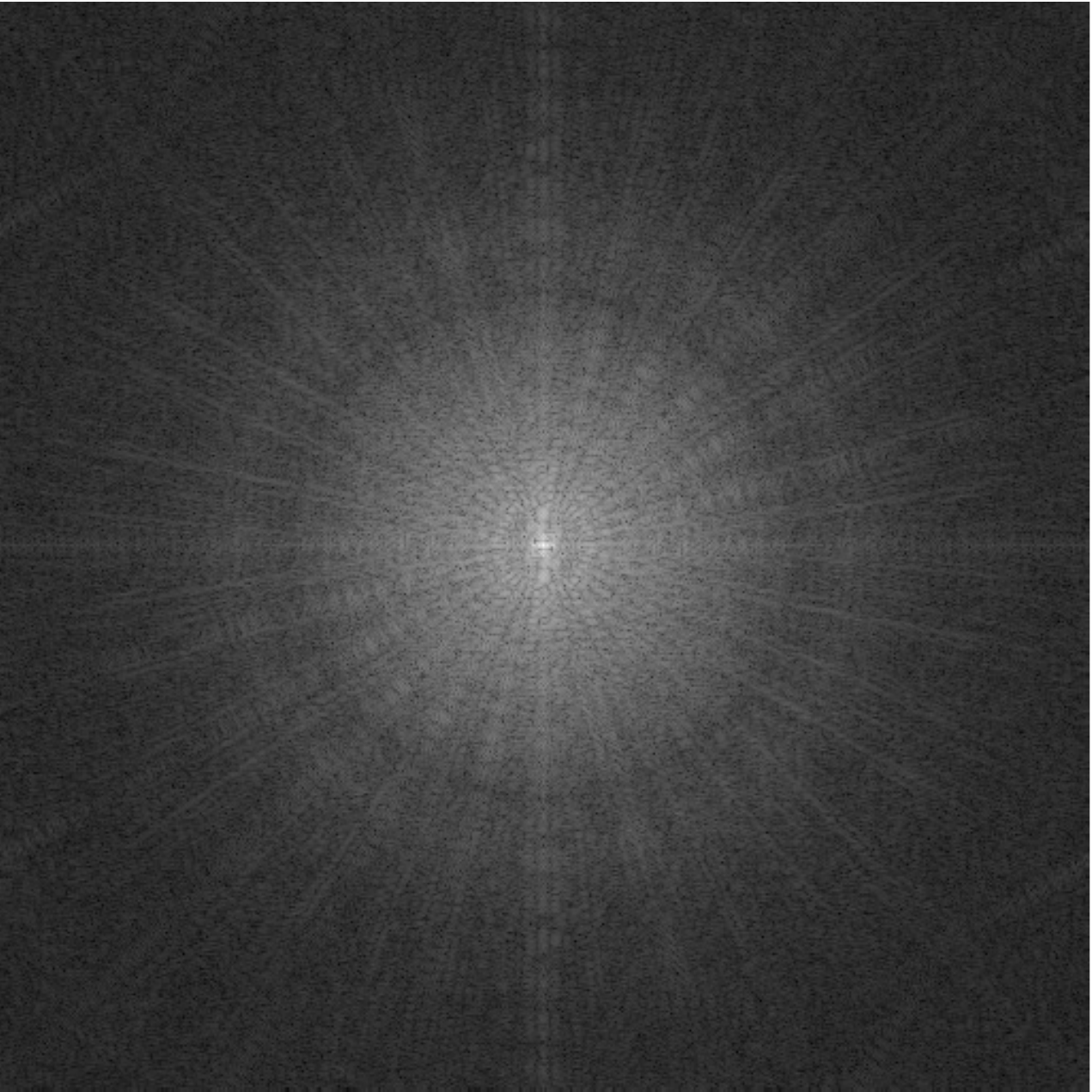}\\
		
		\small  d.) FFT of (a)  &   e.) FFT of (b) & f.) FFT of (c)\\
		\\
		
	\end{tabular}
	
	\caption{
		{\bf Simulation results of the second object using 200 speckle patterns under 40dB Gaussian noise with $\ell_{2,1}$ regularizer term.}  }
	\label{fig:Heltzmann}
\end{figure*}

To demonstrate the versatility of the proposed methods, simulations using synthetic image with more complex structures are shown in Fig. \ref{fig:Heltzmann}. Reconstructed image by $\ell_{2,1}$ regularizer  using 200 speckle patterns under 40dB Gaussian noise are shown in Fig. \ref{fig:Heltzmann}(c). In comparision to the Wiener deconvolution of the widefield image (Fig. \ref{fig:Heltzmann}b), we see the resolution improvement in both spatial and Fourier domain.

\subsection{Reconstructions from experimental data}
To provide a more convincing illustration of the super-resolution capacity of blind-speckleSIM, the processing of experimental datasets is presented in this section. The raw images are obtained with an objective of $NA=1.49$ and $100\times$ magnification. 
The PSF used is simulated using a ICY plug-in called PSF Generator with Gibson \& Lanni 3D Optical Model. The spatial sampling rate is set to be equal or slightly above the Nyquist rate $\frac{\lambda}{4NA}$.

\begin{figure*}[t]
	\centering
	\begin{tabular}{ccc}
		\includegraphics[height=0.32\textwidth] {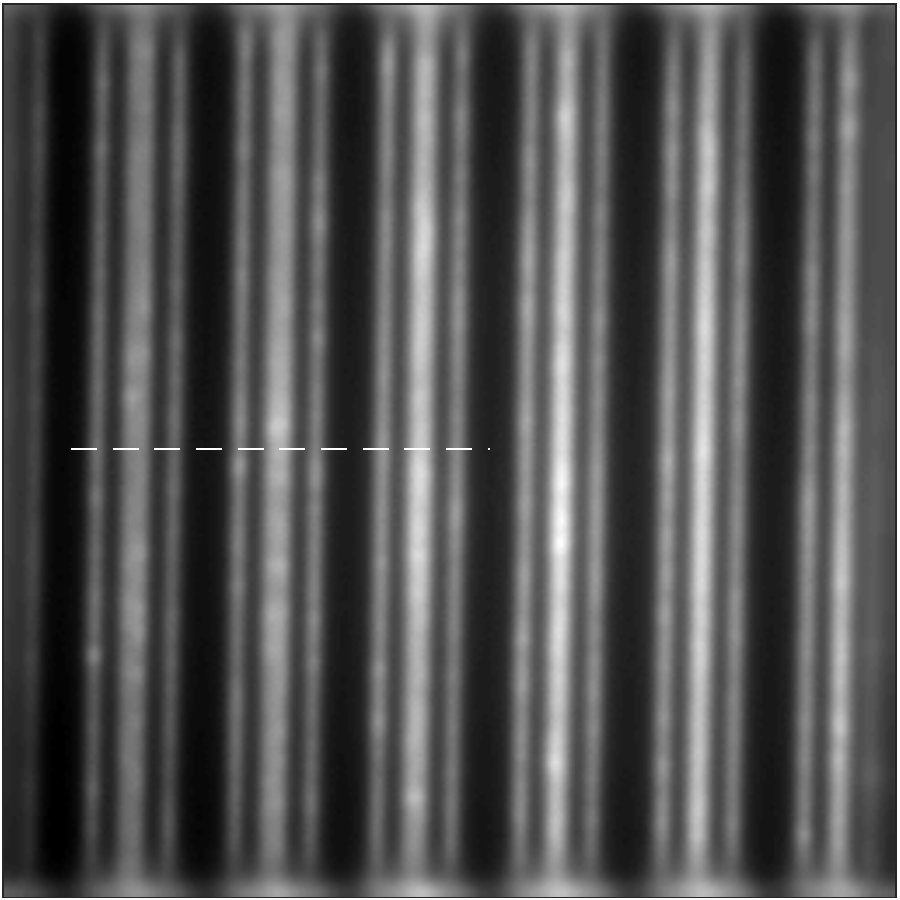}		
		&\includegraphics[height=0.32\textwidth] {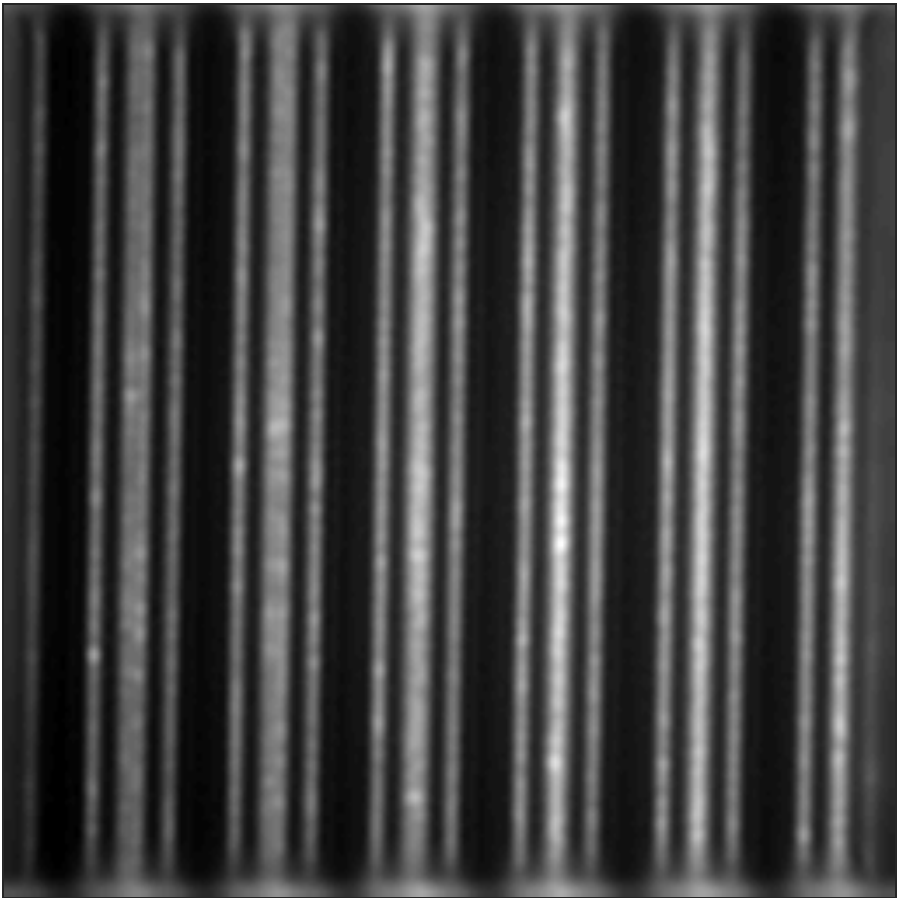}		
		&\includegraphics[height=0.32\textwidth]{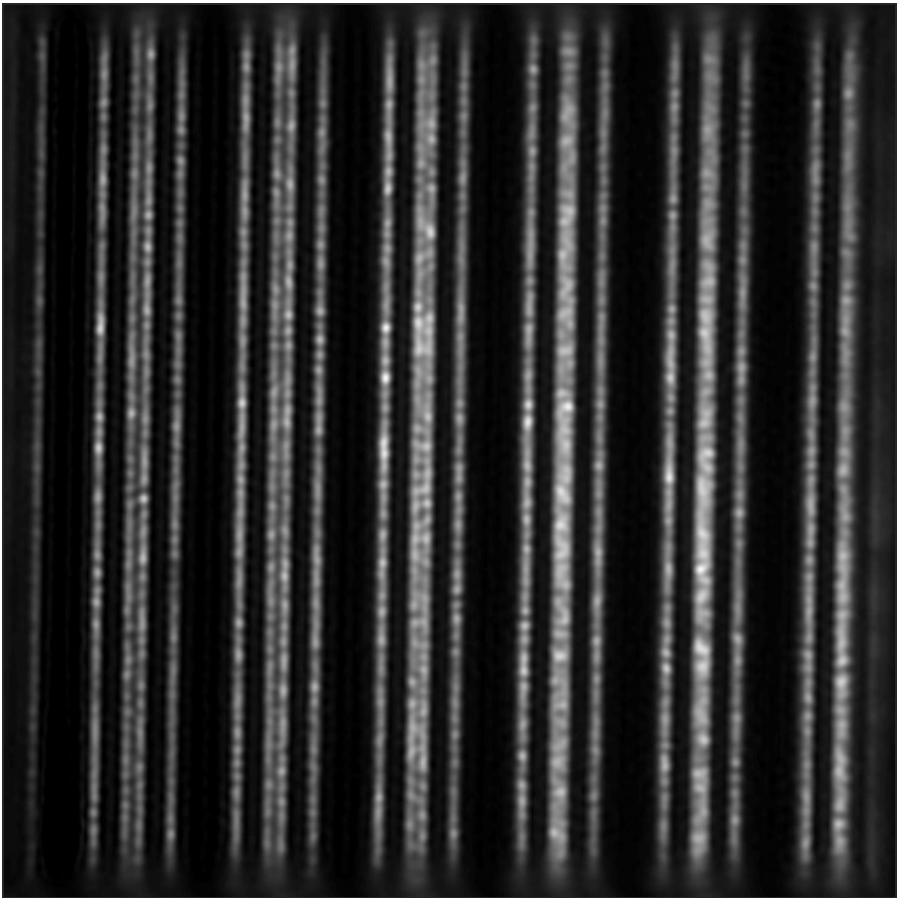}\\
		
		\includegraphics[height=0.32\textwidth] {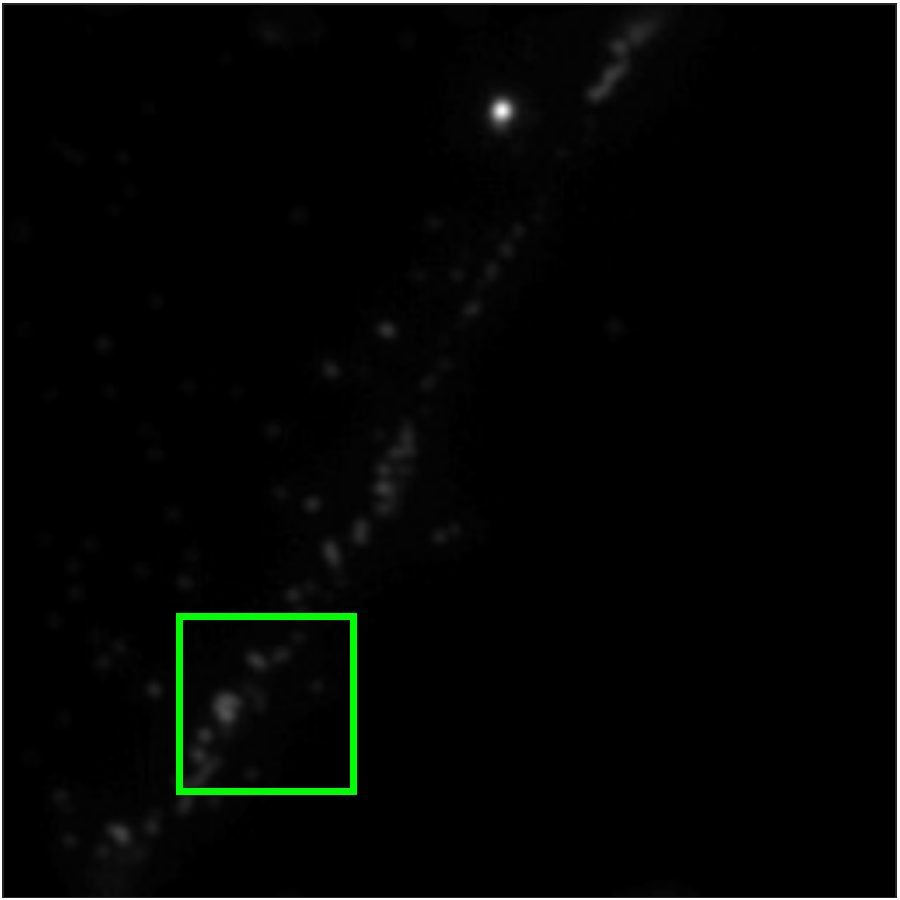}		
		&\includegraphics[height=0.32\textwidth] {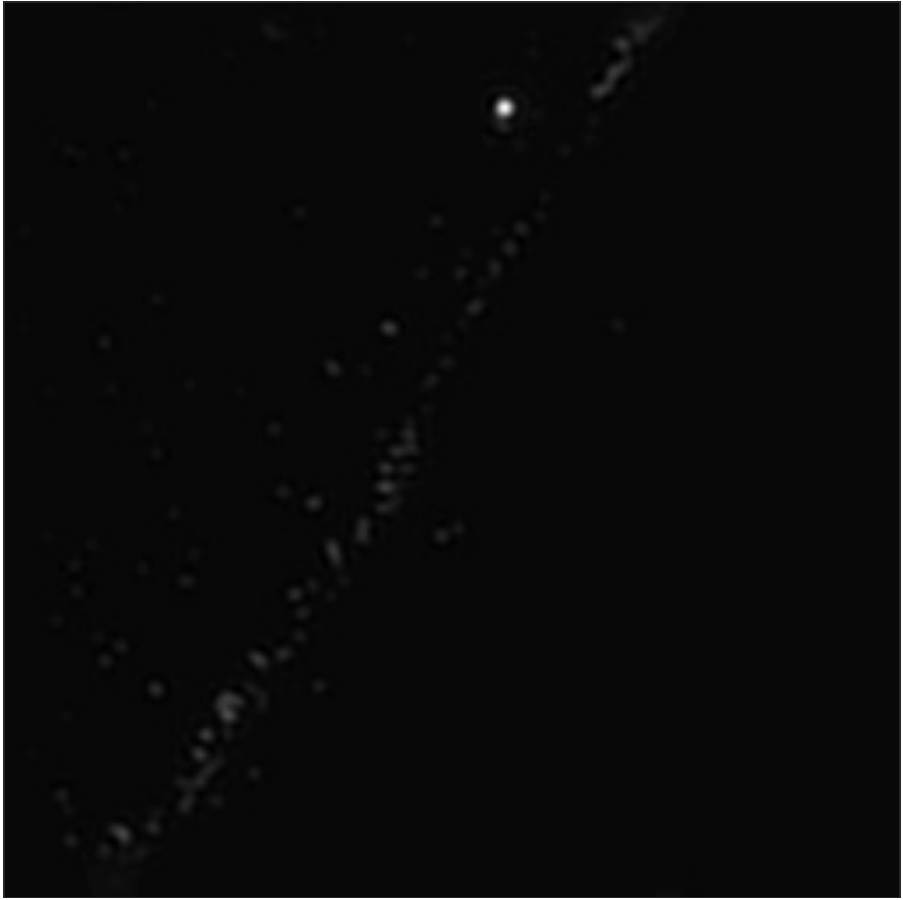}		
		&\includegraphics[height=0.32\textwidth]{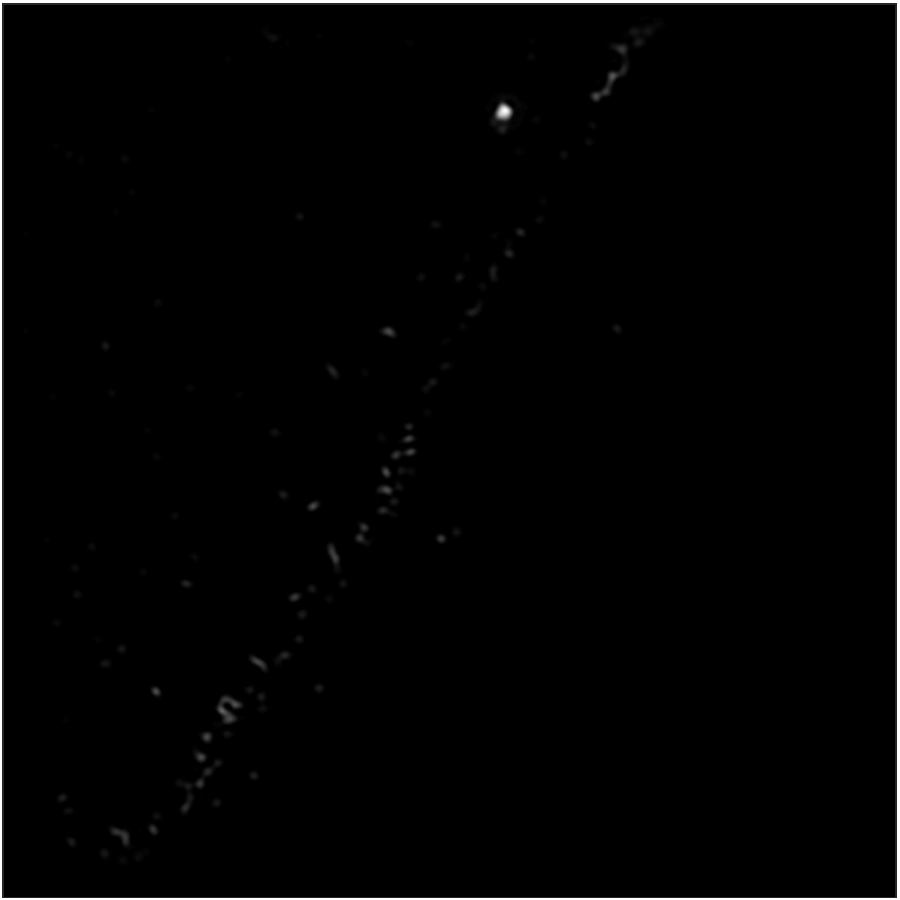}\\

		\includegraphics[height=0.32\textwidth] {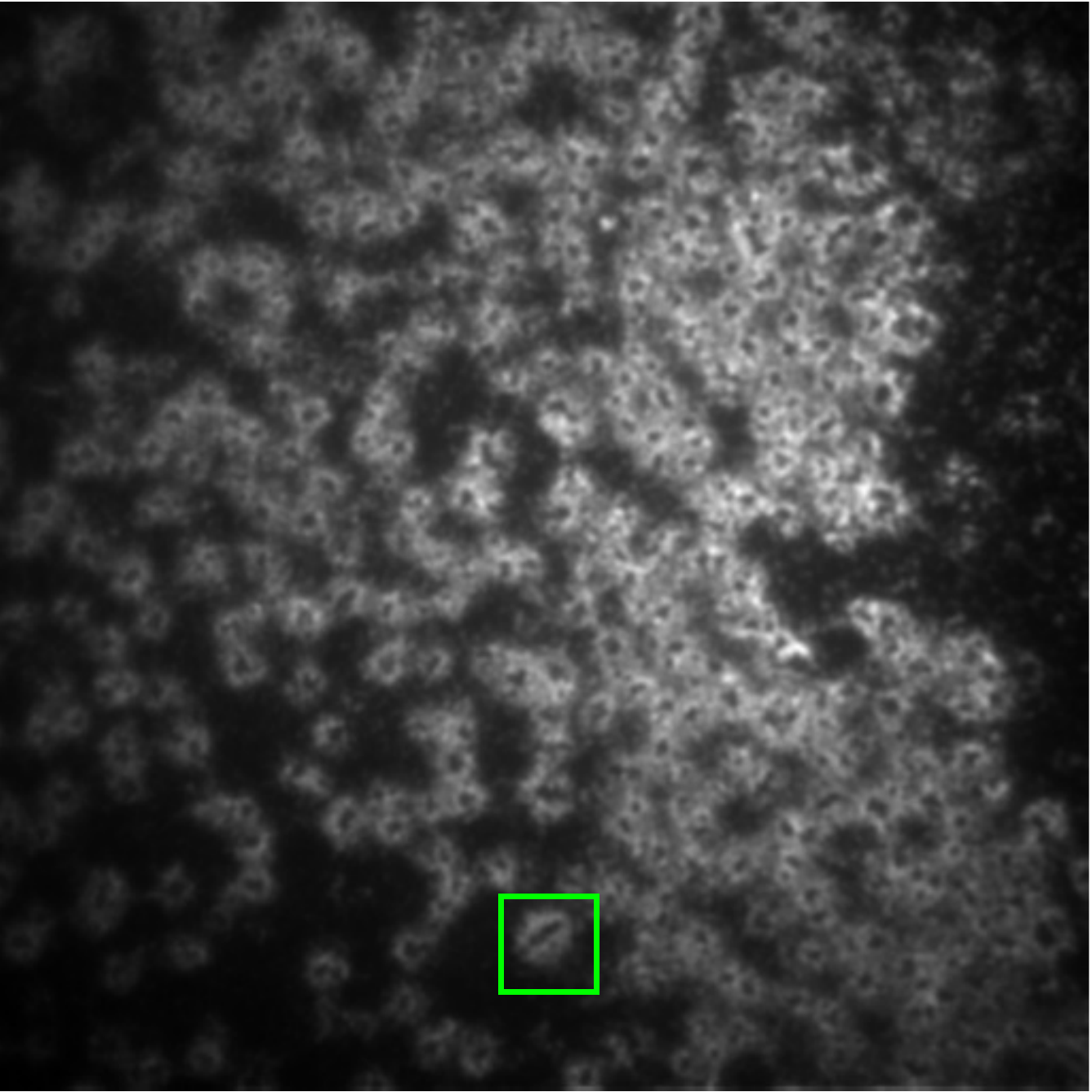}
		
		&\includegraphics[height=0.32\textwidth] {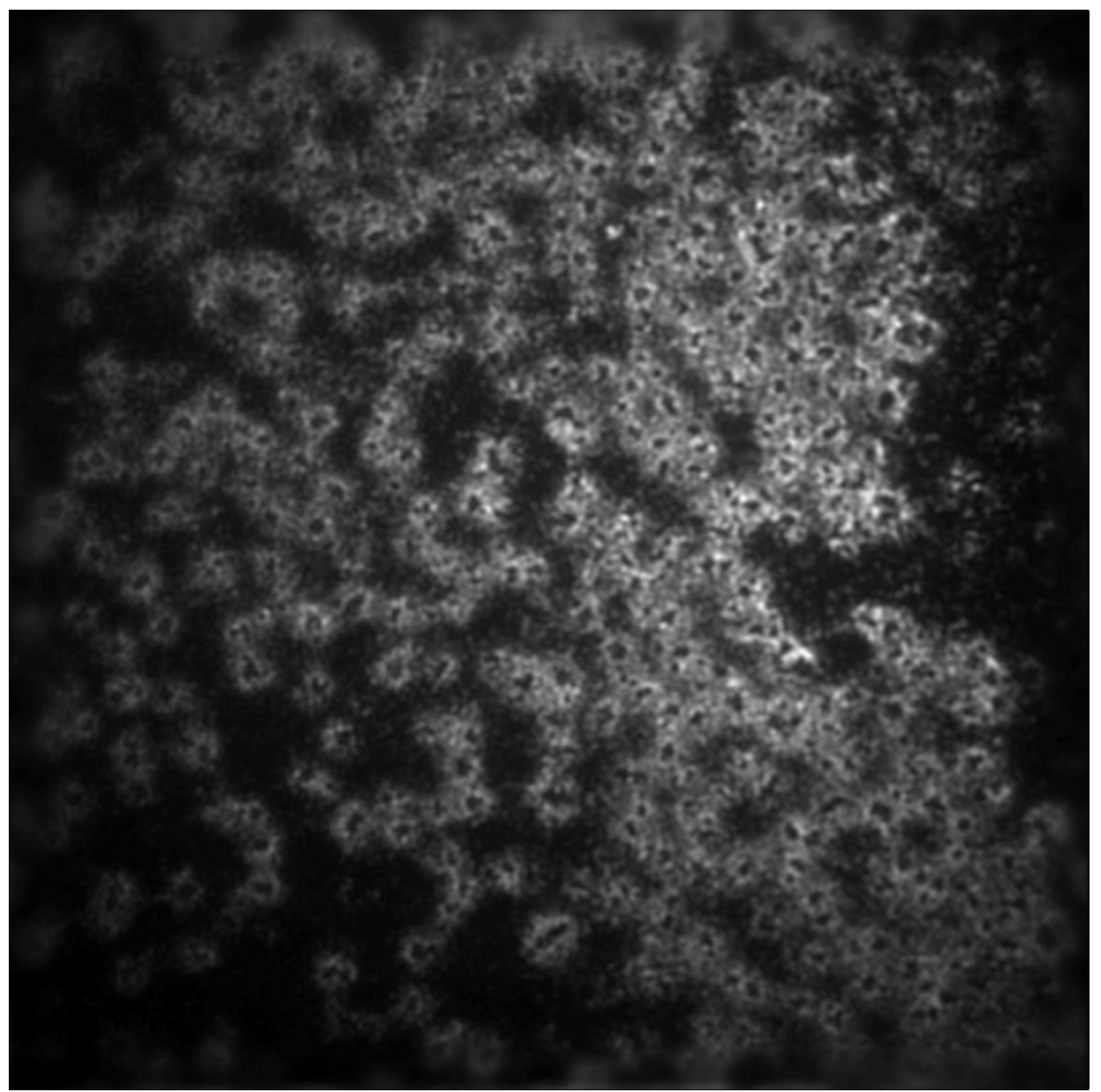}
		
		&\includegraphics[height=0.32\textwidth]{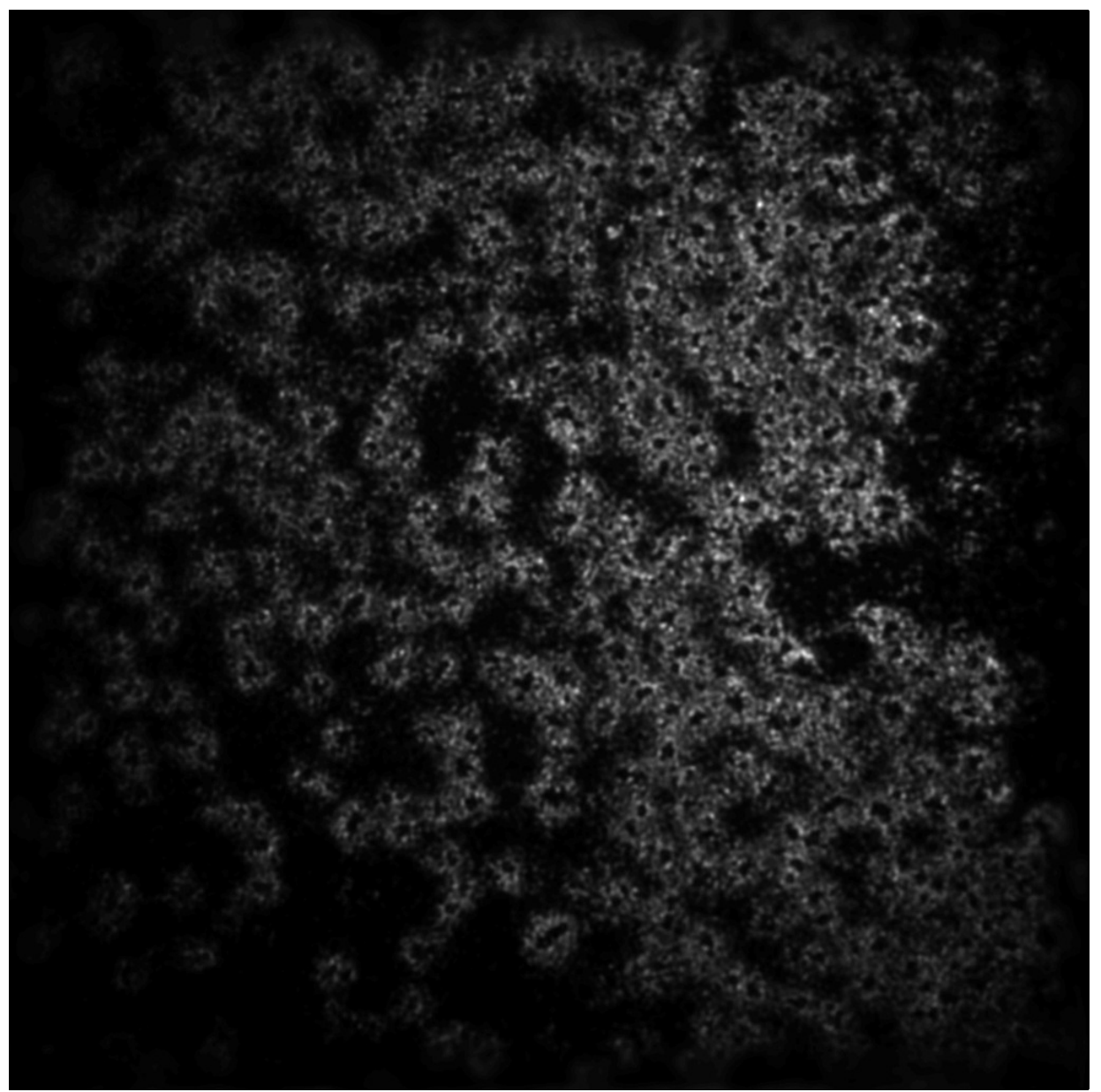}\\

	\small col. 1.) Measurements average $\bar{\yb}_m$ &\small col. 2.) Wiener deconvolution of $\bar{\yb}_m$	& \small col. 3.) $\ell_{2,1}$ regularized blind-SIM\\	
	\end{tabular}

	\caption{
		{\bf Processing of real data obtained from Argolight (Row first), beads (Row second) and Podosome (Row third) samples .} The raw images of Argolight and beads contains 100 speckle patterns while reconstructions of Podosome uses 80 speckle patterns.}
	\label{fig:realdata}
\end{figure*}

Reconstructed images by minimizing the constrained $\ell_{2,1}$ regularizer are displayed in Fig. \ref{fig:realdata}. The Argolight sample shown in the first row is a designed slide, in which from left to right, the spacing between two middle lines becomes narrower. The data shown in second row is composed of fluorescent beads with diameters of $\SI{100}{\nano\meter}$ and the images in the third row are obtained from Podosome sample.

Line section plot of Argolight reconstructions in Fig. \ref{fig.argolight45LineSection} reveals that blind-speckleSIM is superior in resolution. The zoom of a small part of the images from beads and Podosome samples (marked by green square in Fig. \ref{fig:realdata}) are shown in Fig. \ref{fig.lpqPodosomePart}. The reconstructions of the corresponding subimages by marginal approach are shown in the fourth column in Fig. \ref{fig.lpqPodosomePart}. To remove the out-of-focus information in raw images, I slightly adapt the objective function introduced in Ref. \cite{idier2017super} and reconstruct $\rhob$ with only the second-order statistics (that is, the mean values are abandoned) by minimizing:
\begin{equation}
\begin{aligned}
\label{eq.KL_covonly}
D_R(\rhob) &= D_{KL}\big(\Nc (0,\hat{\Gammab}_y)\|\Nc (0,\Gammab_y)\big) \\
&=  \frac{1}{2}\log \lvert \Gammab_y \rvert+ \frac{1}{2}\text{Tr} \big(\Gammab_y^{-1}\hat{\Gammab}_y\big) + K
\end{aligned}
\end{equation}
where $\Gammab_y$ indicates the theoretical covariance of $\yb_m$ (respectively $\hat{\Gammab}_y$ the empirical covariance) and $K$ is a constant number. The gradient of  $D_R(\rhob)$ is:
\begin{equation}
\nabla D_R(\rhob) = \Big(\Big(\Omegab^T\big(\Gammab_y-\hat{\Gammab}_y\big)\Omegab\Big)\circ \Gammab_{\text{s}}\Big)\rhob
\end{equation}
with $\circ$ denoting the Hadamard ({\it i.e.,} element-wise) product, $\Omegab = \Gammab_y^{-1}\Hv$ and $\Gammab_{\text{s}}$ the covariance of speckle patterns. The L-BFGS algorithm \cite{liu1989limited}\cite{minFunc} is chosen to optimize the marginal criterion \eqref{eq.KL_covonly}. Clearly we see better details after introducing the $\ell_{2,1}$ regularizer in comparison to the Wiener deconvolution of wide-field images. The high-frequency structures are verfied by the marginal approach which is built on elegant theoretical cornerstone.

\def\wdthh{3.3 in}
\begin{figure}[htb]
	\centering
	\includegraphics[width=\wdthh]{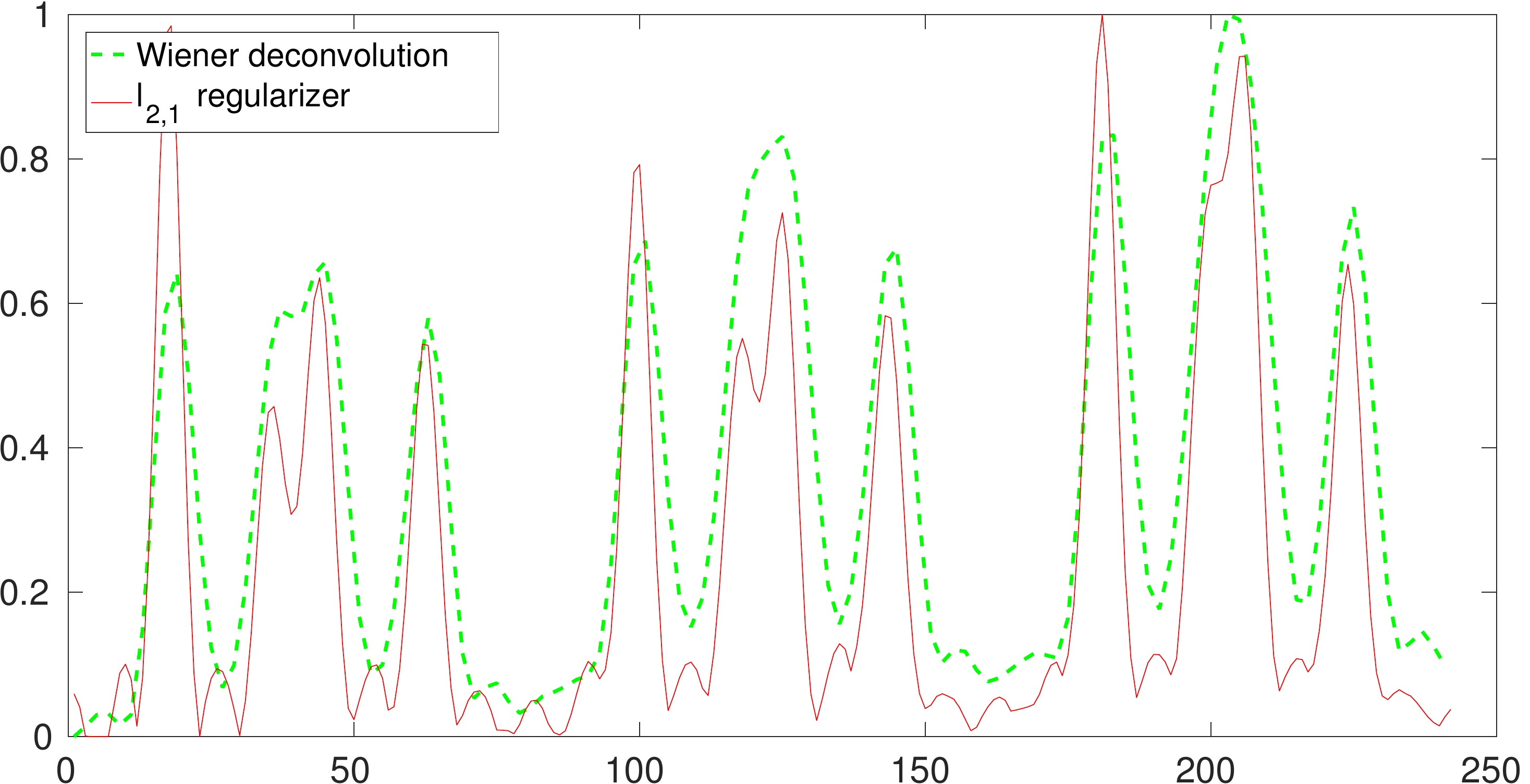}\\
	
	\caption{\textbf{Line section plot extracted from the argolight reconstructions shown in Fig. \ref{fig:realdata}.}}
	\label{fig.argolight45LineSection}
\end{figure}

\def\wdthh{1.57in}
\begin{figure*}[t]
	\centering
	\begin{tabular}{cccc}
			 
		\includegraphics[width=0.24\textwidth]{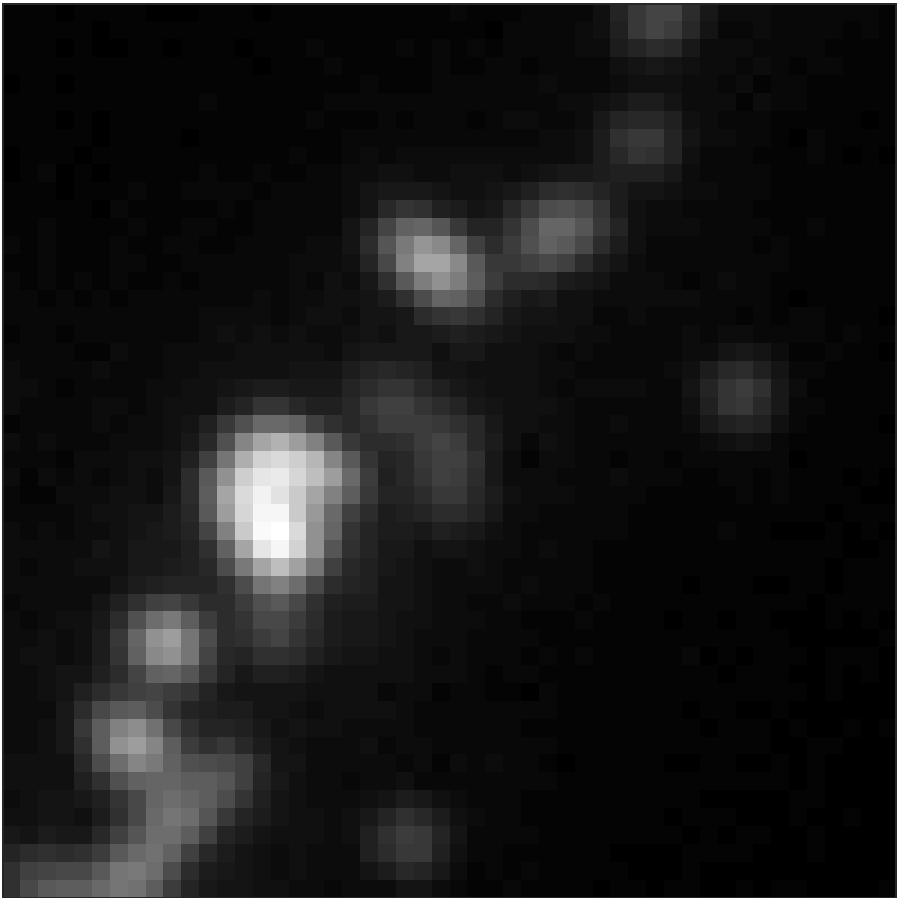}
		&\includegraphics[width=0.24\textwidth]{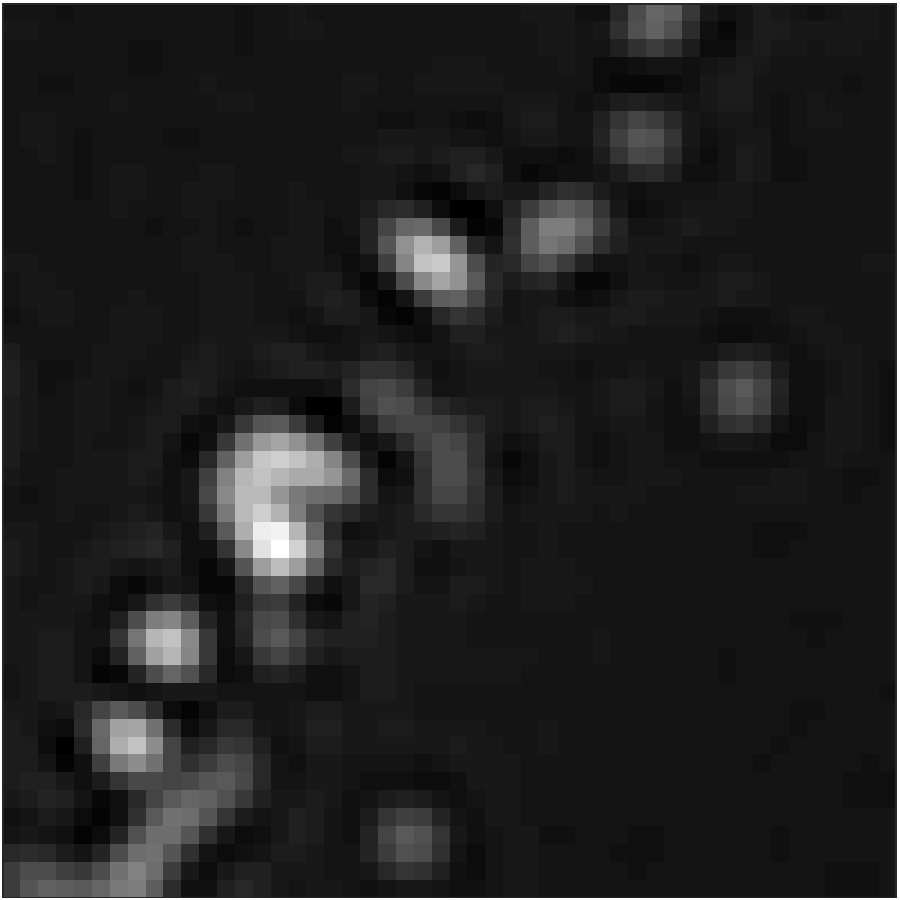}
		&\includegraphics[width=0.24\textwidth]{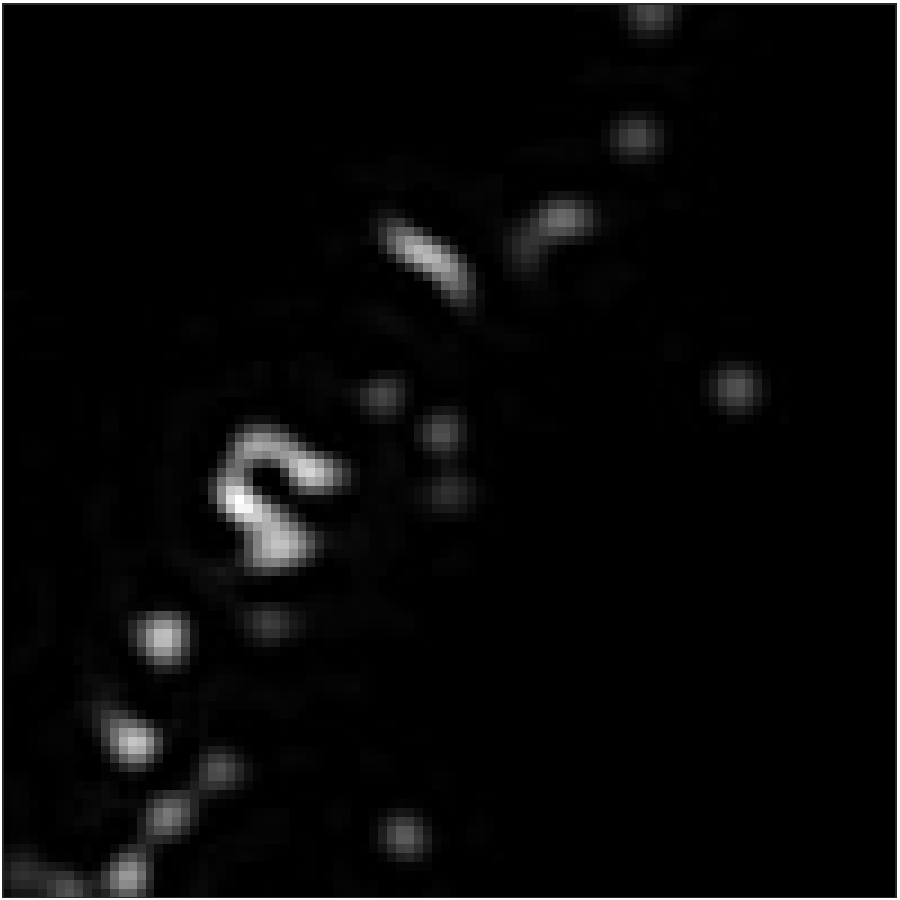}
		&\includegraphics[width=0.24\textwidth]{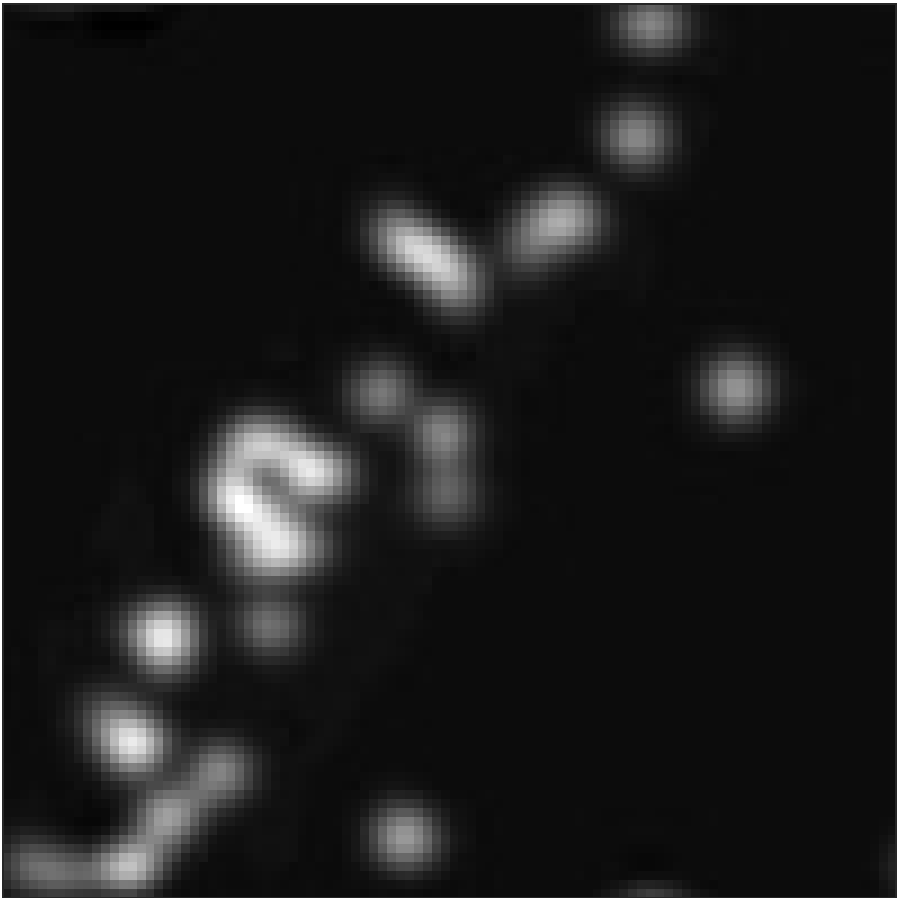}\\
	\\	
		 \includegraphics[width=0.24\textwidth]{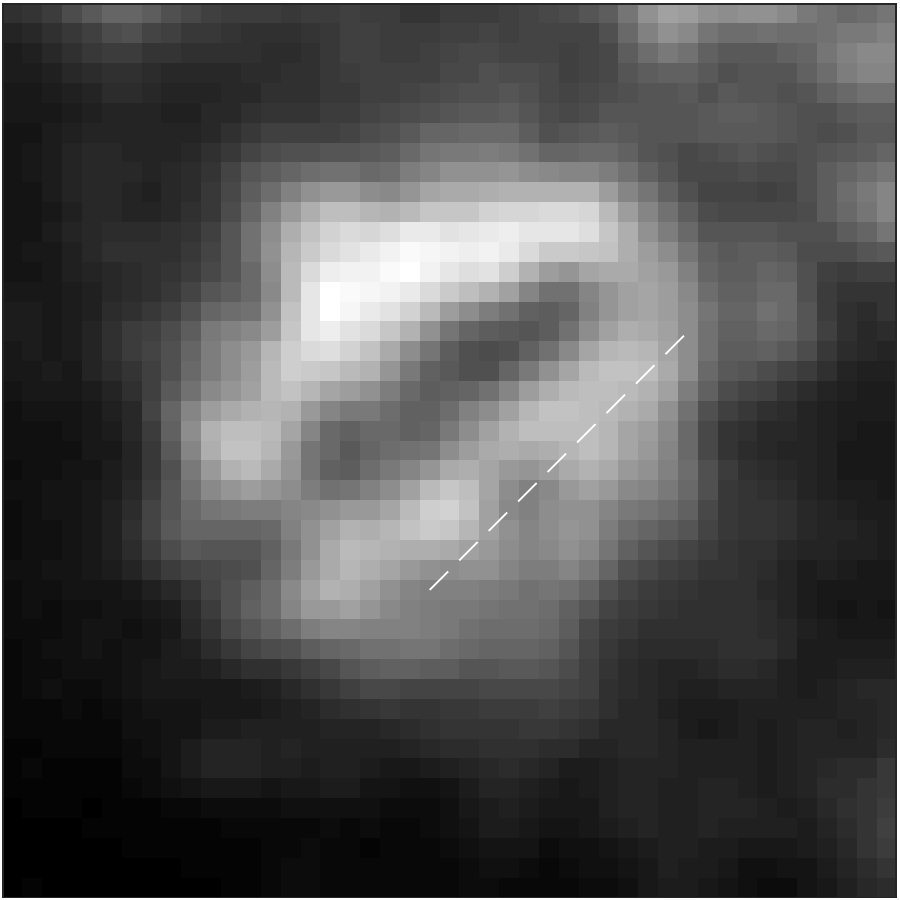}
		 &\includegraphics[width=0.24\textwidth]{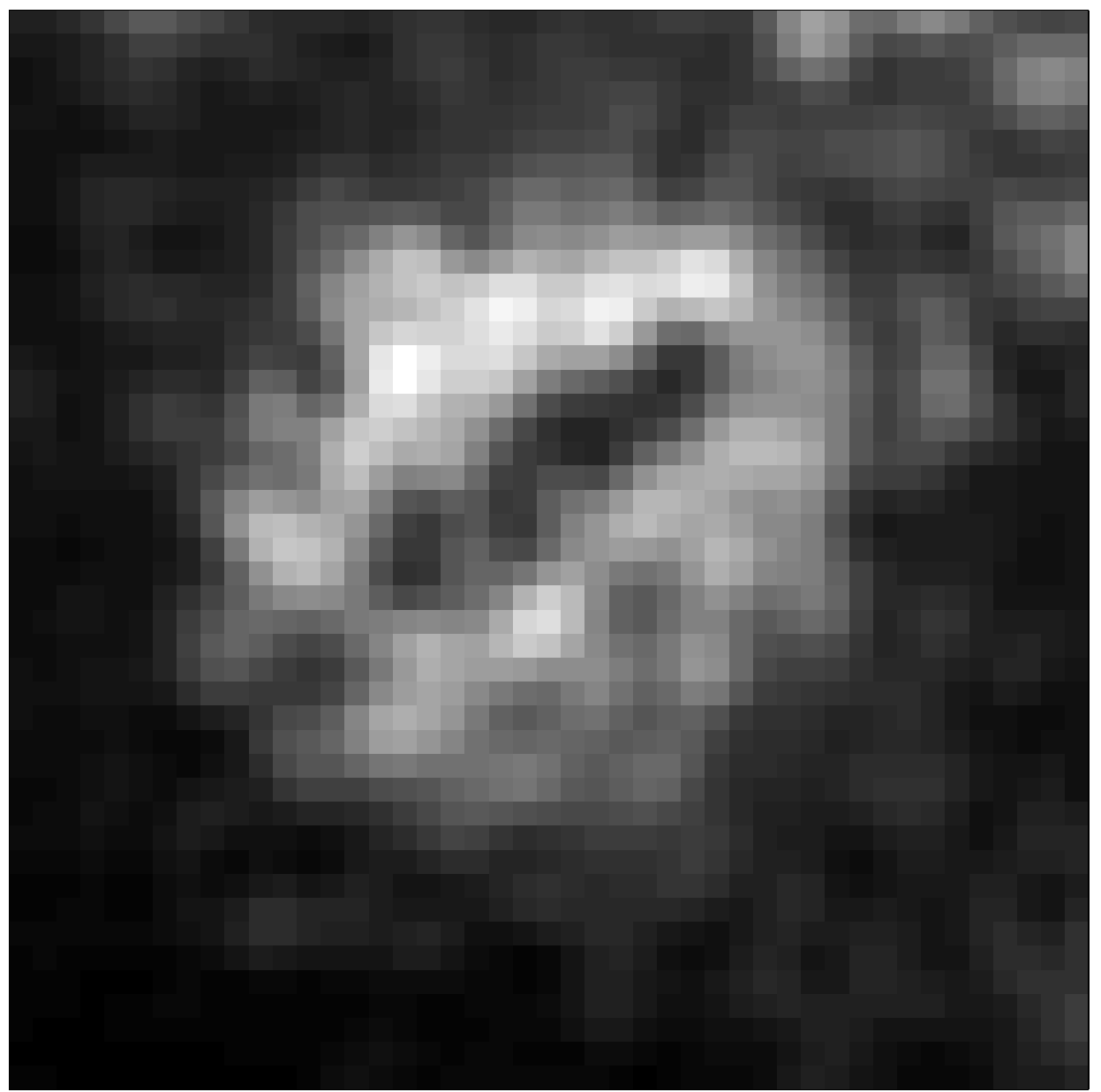}
		 &\includegraphics[width=0.24\textwidth]{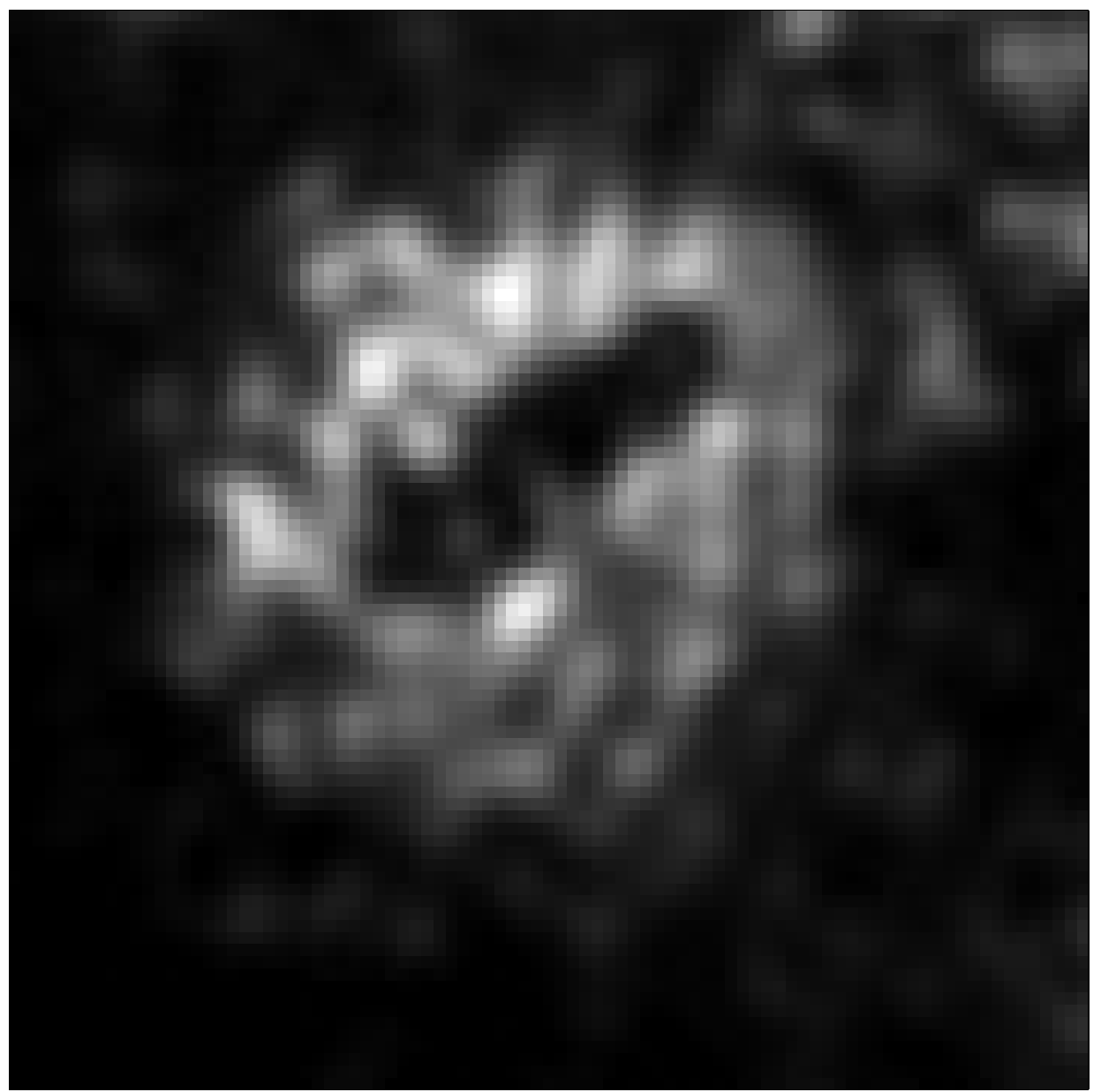}
		 &\includegraphics[width=0.24\textwidth]{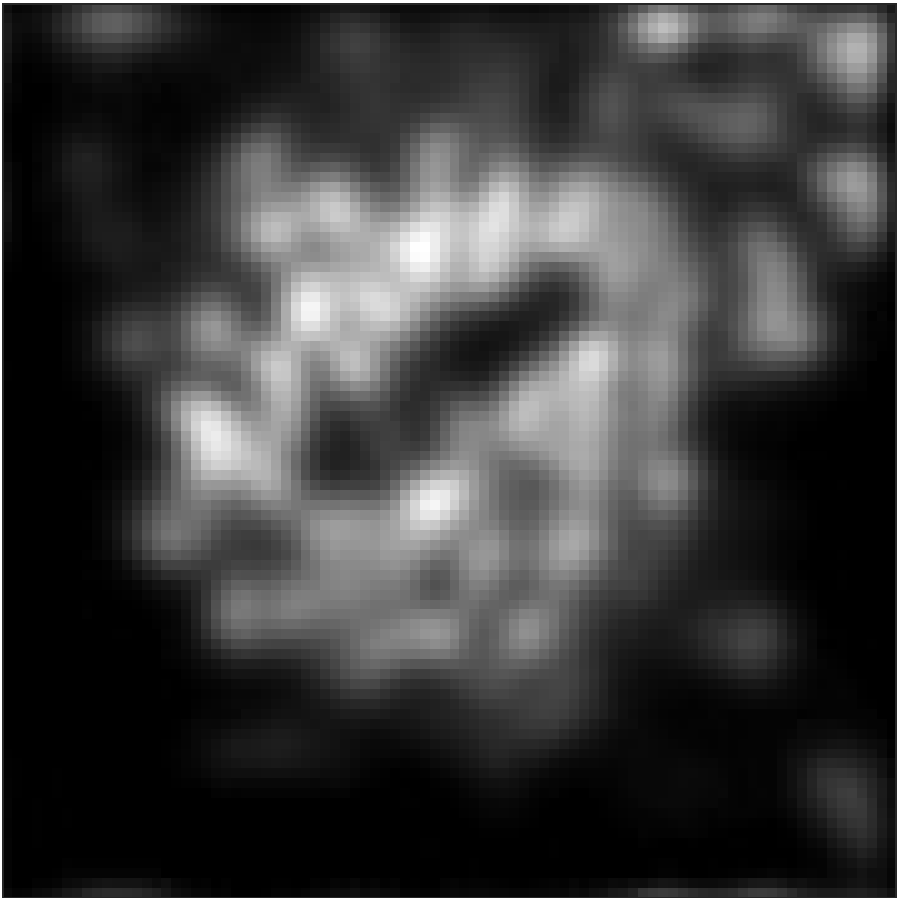}\\

 \small col. 1.) Averaging $\yb_m$ &\small col. 2.) Wiener deconvolution	& \small col. 3.) $\ell_{2,1}\; \hat{\qb}_m$ Std & \small col. 4.) Marginal approach\\

 \\

\end{tabular}  
\caption{\textbf{The images in column $1$-$3$ are partial enlargement of the beads and Podosome images shown in Fig. \ref{fig:realdata}, while the fourth column is obtained by marginal approach presented in Ref. \cite{idier2017super}.} }%

\label{fig.lpqPodosomePart} 
\end{figure*}

\section{Conclusion}\label{sec.conclusion}

In this paper a unified joint reconstruction approach in blind-speckleSIM based on constrained $\ell_{p,q}$ norm minimization of the data is proposed, other prior information of the object (like positivity) can be easily incorporated into the model without big changes of the associated optimization algorithm. Mathematical analysis demonstrates that the joint sparsity of matrix $\Qv$ implies the sparsity assumption of the object. Please note here that in the analysis the statistical prior of speckle patterns are taken into considered, i.e.\ their ensemble average is homogeneous or they are second-order stationary random process, as shown in Eq. \eqref{eq.jointSparsityrhoSparsity1}\eqref{eq.jointSparsityrhoSparsity2}. So it is the sparsity of object together with the statistical prior of speckle patterns that induce the super-resolution imaging in blind-speckleSIM strategy.  The hyper-parameter involves in this model is proportional to the standard variance of noise, thus it is easy to tune. 

This model can examine the performance of different regularizer terms easily. Numerical simulations show that the $\ell_{2,1}$ regularizer is superior in terms of both error and super-resolution in comparison to $\ell_{1,1}$ term. Please note here that this conclusion is drawn under the speckle illumination situation. In cases where conventional sinusoidal patterns are used, $\ell_{2,1}$ regularizer is not a good choice any more as the illumination statistics changed. When $q<1$, the super-resolution information still appears even though the associated primal-dual splitting method can not assure to give the global minimum since the corresponding regularizer term is not convex function any more. In fact reference \cite{Ge2011A} demonstrates that the $\ell_{1/2}$ regularized optimization problem is strongly NP-hard. The binary effect in reconstructions is quite evident in cases $q<1$ as shown in Fig. \ref{fig.lpqnorm2}.

Normally in experiments the inevitable background signal will cause artifacts in reconstruction and reduce the super-resolution in standard SIM. An estimator based on the standard deviation of $\qb_m$ is presented in this paper and it  performs well in both numerical studies and experimental real data without the background estimation and subtraction procedure.

Only the 2D super-resolution problems are considered in this paper. The blind-speckleSIM technique is also compatible with 3D imaging problems of thick samples \cite{jost2015optical}\cite{negash2016improving} and the constrained $\ell_{p,q}$ norm model can also be applied to other inverse problems where the data share a group sparsity structure, such as direction-of-arrival (DOA) estimation \cite{yin2011direction}, photoacoustic microscopy imaging \cite{burgholzer2017super}, and so on.

\appendices
\section{Proximal operator of function $g_1$} \label{appen.lpqNormProximal}
This section focus on the proximal operator of $g_1 = \lVert \bm{\mathfrak{d}} \rVert_{\Gc pq}^q$, whose definition is given by:
\begin{equation}
\begin{aligned}
\text{prox}_{\lambda g_1}(\bm{\mathfrak{x}}) &= \arg \min_{\bm{\mathfrak{d}}} \; \lVert \bm{\mathfrak{d}} \rVert_{\Gc pq}^q + \frac{1}{2\lambda} \lVert \bm{\mathfrak{d}} - \bm{\mathfrak{x}}\lVert_2^2\\
&= \arg \min_{\bm{\mathfrak{d}}} \; \sum_{n=1}^N \Big ( \lVert \bm{\mathfrak{d}}_{\Gc_n} \rVert_p^q +  \frac{1}{2\lambda} \lVert \bm{\mathfrak{d}}_{\Gc_n} - \bm{\mathfrak{x}}_{\Gc_n}\lVert_2^2 \Big)
\end{aligned}
\label{eq.JSPsub2}
\end{equation}

Since the partition of $\bm{\mathfrak{d}}_{\Gc_n}$ is not overlapped, so problem \eqref{eq.JSPsub2} could be decoupled into $N$ independent subproblems. Each subproblem is given below:
\begin{equation}
\label{eq.proxg1Sub}
\arg \min_{\bm{\mathfrak{d}}_{\Gc_n}} \;  \lVert \bm{\mathfrak{d}}_{\Gc_n} \rVert_p^q +  \frac{1}{2\lambda} \lVert \bm{\mathfrak{d}}_{\Gc_n} - \bm{\mathfrak{x}}_{\Gc_n}\lVert_2^2 
\end{equation}

For specific $(p,q)$ pairs, the minimization problem \eqref{eq.proxg1Sub} has the following analytical formula \cite{hu2017group}:
 
\begin{itemize}
	\item for $p=1$ and $q=1$
	\begin{equation}
	\bm{\mathfrak{d}}_{\Gc_n} =  \text{sign}(\bm{\mathfrak{x}}_{\Gc_n}) \circ \max \Big\{  \lvert \bm{\mathfrak{x}}_{\Gc_n} \rvert - \lambda , 0 \Big\}
	\end{equation}
	
	\item for $p=2$ and $q=1$
	\begin{equation}
	\bm{\mathfrak{d}}_{\Gc_n} = \max \Big\{1-\frac{\lambda}{ \lVert \bm{\mathfrak{x}}_{\Gc_n} \rVert_2}, 0 \Big\} \bm{\mathfrak{x}}_{\Gc_n}
	\end{equation}

	\item for $p=2$ and $q = 1/2$ 
	
	\begin{equation}
	\bm{\mathfrak{d}}_{\Gc_n} =
	\begin{cases} 
	\frac{16 \lVert \bm{\mathfrak{x}}_{\Gc_n} \rVert_2^{3/2}\omega_n}{3\sqrt{3}\lambda+ 16 \lVert\bm{\mathfrak{x}}_{\Gc_n} \rVert_2^{3/2}\omega_n} \bm{\mathfrak{x}}_{\Gc_n}, & \lVert \bm{\mathfrak{x}}_{\Gc_n} \rVert_2 > \frac{3}{2}(\lambda)^{2/3} \\      \bm{0} \;\text{or} \; \frac{16 \lVert\bm{\mathfrak{x}}_{\Gc_n}\rVert_2^{3/2}\omega_n}{3\sqrt{3}\lambda+ 16 \lVert\bm{\mathfrak{x}}_{\Gc_n} \rVert_2^{3/2}\omega_n} \bm{\mathfrak{x}}_{\Gc_n}, & \lVert \bm{\mathfrak{x}}_{\Gc_n} \rVert_2 = \frac{3}{2}(\lambda)^{2/3} \\ \bm{0}, & \lVert \bm{\mathfrak{x}}_{\Gc_n} \rVert_2 < \frac{3}{2}(\lambda)^{2/3}
	\end{cases}
	\end{equation}
	
	with 
\begin{equation*}
\begin{aligned}
        \omega_n =  \cos^3(\frac{\pi}{3}-\frac{\psi_n}{3}) \\
	\psi_n = \arccos \bigg(\frac{\lambda}{4} \Big(\frac{3}{\lVert\bm{\mathfrak{x}}_{\Gc_n} \rVert_2}\Big)^{3/2} \bigg)
\end{aligned}
\end{equation*}

	\item for $p=2$ and $q = 2/3$
	
	\begin{equation}
	\bm{\mathfrak{d}}_{\Gc_n} =
	\begin{cases} 
	\frac{3 \eta^4}{2\lambda+3 \eta^4}\bm{\mathfrak{x}}_{\Gc_n},  & \lVert \bm{\mathfrak{x}}_{\Gc_n}\rVert_2 > 2(\frac{2}{3}\lambda)^{3/4}\\
	0 \; \text{or}\; \frac{3 \eta^4}{2\lambda+3 \eta^4}\bm{\mathfrak{x}}_{\Gc_n}, & \lVert \bm{\mathfrak{x}}_{\Gc_n}\rVert_2 = 2(\frac{2}{3}\lambda)^{3/4} \\
	0, & \lVert \bm{\mathfrak{x}}_{\Gc_n}\rVert_2 < 2(\frac{2}{3}\lambda)^{3/4}
	\end{cases}
	\end{equation}
	with 
	\begin{equation*}
         \begin{aligned}
 \eta = \frac{1}{2} \Bigg(\lvert a \rvert + \sqrt{\frac{2\lVert \bm{\mathfrak{x}}_{\Gc_n}\rVert_2}{\lvert a \rvert}-a^2} \Bigg)\\
	a = \frac{2}{\sqrt{3}}\big(2\lambda \big)^{1/4} \Big(\cosh\big( \frac{\phi(\bm{\mathfrak{x}}_{\Gc_n})}{3}\big) \Big)^{1/2} \\
 \phi(\bm{\mathfrak{x}}_{\Gc_n}) = \arccosh\Big( \frac{27\lVert\bm{\mathfrak{x}}_{\Gc_n} \rVert_2^2}{16 \big(2\lambda \big)^{3/2}}\Big)
\end{aligned}
	\end{equation*}

\end{itemize}

\section*{Acknowledgment}
I would like to acknowledge Thomas Mangeat in Universit\'e de Toulouse for offering the raw data corresponding to Fig \ref{fig:realdata} and the partial financial support from the GdR 720 ISIS and China Scholarship Council.

\ifCLASSOPTIONcaptionsoff
  \newpage
\fi




\bibliographystyle{ieeeji}
\bibliography{biblpq}




\end{document}